\pdfoutput=1
\documentclass[prx,aps,12pt,nofootinbib,eqsecnum]{revtex4-2}

\usepackage{lmodern} 
\usepackage[utf8]{inputenc} 
\usepackage[T1]{fontenc} 
\usepackage{microtype} 
\usepackage[english]{babel}

\usepackage{color,graphicx}
\graphicspath{{Images/}}

\usepackage{amsmath,dsfont,url,amssymb,mathrsfs}
\usepackage{slashed,cancel}
\usepackage[usenames,dvipsnames]{xcolor}
\usepackage[colorlinks=true, linktocpage=true, linkcolor=BrickRed, citecolor=BrickRed, urlcolor=RoyalBlue]{hyperref}
\usepackage{mdframed}
\usepackage{comment}
\usepackage{braket}
\usepackage{caption}
\usepackage{subcaption}

\def \d {\partial}

\newcommand{\SDiff}{\text{SDiff}}

\newcommand{\G}{\mathcal{G}}
\renewcommand{\H}{\mathcal{H}}
\newcommand{\sF}{\mathscr{F}}
\newcommand{\sG}{\mathscr{G}}
\newcommand{\sH}{\mathscr{H}}

\newcommand{\cA}{\mathcal{A}}

\newcommand{\cJ}{\mathcal{J}}
\newcommand{\g}{\mathfrak{g}}
\newcommand{\h}{\mathfrak{h}}
\newcommand{\so}{\mathfrak{so}}
\newcommand{\Ad}{\text{Ad}}
\newcommand{\ad}{\text{ad}}
\newcommand{\R}{\mathbb{R}}
\newcommand{\x}{\mathbf{x}}
\renewcommand{\k}{\mathbf{k}}
\newcommand{\p}{\mathbf{p}}
\newcommand{\q}{\mathbf{q}}
\newcommand{\y}{\mathbf{y}}
\renewcommand{\v}{\mathbf{v}}

\newcommand{\E}{\mathbf{E}}
\newcommand{\A}{\mathbf{A}}
\newcommand\<{\langle}
\renewcommand\>{\rangle}
\newcommand{\sign}{\text{sign}}
\newcommand{\n}{\mathbf{n}_\theta}
\newcommand{\s}{\mathbf{s}_\theta}
\newcommand{\su}{\mathfrak{su}}

\DeclareMathOperator{\Tr}{Tr}

\newcommand{\mb}[2]{\{\!\!\{ #1, #2 \}\!\!\}}

\begin{document}
\frenchspacing

\title{Postmodern Fermi Liquids}
\author{Umang Mehta}
\affiliation{Kadanoff Center for Theoretical Physics, University of Chicago, Chicago, Illinois 60637, USA}

\date{\today}
\begin{abstract}
We present, in this dissertation, a pedagogical review of the formalism for Fermi liquids developed in \cite{main:2022} that exploits an underlying algebro-geometric structure described by the group of canonical transformations of a single particle phase space. This infinite-dimensional group governs the space of states of zero temperature Fermi liquids and thereby allows us to write down a nonlinear, bosonized action that reproduces Landau's kinetic theory in the classical limit. Upon quantizing, we obtain a systematic effective field theory as an expansion in nonlinear and higher derivative corrections suppressed by the Fermi momentum $p_F$, without the need to introduce artificial momentum scales through, e.g., decomposition of the Fermi surface into patches. We find that Fermi liquid theory can essentially be thought of as a non-trivial representation of the Lie group of canonical transformations, bringing it within the fold of effective theories in many-body physics whose structure is determined by symmetries. We survey the benefits and limitations of this geometric formalism in the context of scaling, diagrammatic calculations, scattering and interactions, coupling to background gauge fields, etc. After setting up a path to extending this formalism to include superconducting and magnetic phases, as well as applications to the problem of non-Fermi liquids, we conclude with a discussion on possible future directions for Fermi surface physics, and more broadly, the usefulness of diffeomorphism groups in condensed matter physics. Unlike \cite{main:2022}, we present a microscopic perspective on this formalism, motivated by the closure of the algebra of bilocal fermion bilinears and the consequences of this fact for finite density states of interacting fermions.
\end{abstract}

\maketitle

\newpage
\begin{center}
    \hspace{0pt}
    \vfill
        \thispagestyle{empty}
        \textit{To all neurodivergent people, known or unknown,\\
        among whom I finally found a sense of community.}
    \vfill
\end{center}

\newpage
{\hypersetup{linkcolor=MidnightBlue}
\tableofcontents
}

\newpage
\acknowledgments

It was the summer of 2008, about a month before the beginning of the school year, and I had just got back home with my backpack full of new textbooks for class 9. The nerd that I was (and still am), all I could think about on the way back home was the excitement of getting to open and read the books that I had just bought; the curious side of me was just excited to absorb all the knowledge I could from them while the competitive side was daydreaming about having preemptive answers to all the questions that my teachers would later ask in class.

Having already been mesmerized by science and mathematics from the year before, my hands were drawn to the physics textbook, since it had the best colour palette between itself, chemistry, biology, and maths. I picked up the book, opened the cover, and, energized by that new-book-smell, flipped the pages right past the first chapter on measurement and experimentation to the chapters on linear motion and Newton's laws.

In no time I reached the section on the second law of motion, and noticed a footnote that described the inaccuracy of the linear relationship between momentum and velocity at speeds close to the speed of light. The words `special theory of relativity' were mentioned and before I knew it, two whole years had passed with me having read every online resource I could possibly find about special and general relativity and non-Euclidean geometry, convinced that quantum mechanics was not real because ``Einstein didn't believe in it''. It was in that initial spark of interest that I knew that I wanted to pursue a career in theoretical physics, as unorthodox as something like that would be in the culture I grew up in.

I was fortunate enough to have found abundant support for my unusual career choice from my parents Rita and Bharat Mehta, and late grandparents Jaya and Kantilal Mehta, for whom my education always took highest priority. I shall forever be grateful to them for providing me the environment and encouragement to nurture my passion for physics. My father, in particular, has made it a point to read every single paper that I have published, even when it makes no sense to him, and vehemently insists that I send each draft to him for his growing collection, and I will always be glad that my work will, at the very least, be read and appreciated by one person who I admire.

I found my first mentor in Shiraz Minwalla at the Tata Institute for Fundamental Research (TIFR), whose wise words I will always carry with me. He instilled in me the courage I needed to not shy away from difficult problems and even enjoy the often long and tedious calculations that accompany them, to the point where I now get excited at the prospect of taking on such challenges. Shiraz's advice was an important contributor to overcoming the many instances of impostor syndrome that I experienced upon being thrown into the melting pot of all the tremendously talented individuals that I encountered throughout my Ph.D. But most importantly, it was on his suggestion that I found my advisor.

I couldn't have asked for a better advisor than Dam Thanh Son. I switched from high energy to condensed matter physics upon joining the University of Chicago, and if it was not for his guidance, I would have had a much harder time with the transition. In him I found the perfect mentor whose advising style fit with my learning style like pieces of a jigsaw puzzle. Son's visionary foresight is what ultimately lead to the content in the rest of this thesis and I can only hope to be able to replicate that in the future.

I owe a lot to my unofficial mentor, Luca V. Delacr\'etaz, from whom I learned various lessons ranging from the most benign yet consequential tricks to make Mathematica compute integrals when it is being stubborn, to the valuable philosophy behind effective field theory. Luca is and always will be a role model to me for my career and mentorship goals.

My Ph.D. experience would not have been half as incredible as it was if not for the extremely friendly and welcoming environment that my office-mates cultivated. I'm grateful to Alex Bogatskiy, Harvey Hsiao, Kyle Kawagoe, Carolyn Zhang, Yuhan Liu, Yi-Hsien Du, Ruchira Mishra, Ege Eren and Davi Costa for all the wonderful times we had together in our little corner office, for all the insightful discussions that helped me grow as a physicist. I also apologize to them for likely being one of the most disruptive and distracting office-mates that they have encountered.

Everyone at the Kadanoff Center for Theoretical Physics has been pleasantly affable and never once did I feel like I was not welcome by the professors, postdocs and other graduate students. My thesis committee members, Michael Levin, Jeffrey Harvey, and Woowon Kang, were instrumental in making me think deeply about my work and understand it from various different perspectives. The Center has only become more social over the last six years and as much as I'm looking forward to the next step in my career, it saddens me to have to leave behind my wonderful colleagues and the University of Chicago.

Lastly, and perhaps most importantly, I am deeply indebted to my found family, Timothy Hoffman, Claire Baum, and Alex Bogatskiy, (and Bowie Hoffman -- Tim's adorable little pupper) with whom I developed a bond so strong I cannot imagine any force that can break it. Between the Ph.D., the pandemic, and personal setbacks, the last few years have been tumultuous and my friends stood by me with all the love and support for which I was often too afraid to ask. Even on our various rock-hounding vacations we couldn't help but discuss physics and I treasure the precious memories we made along the way.

It was thanks to their support that I persisted through the most prominent milestone of my life -- the day that I discovered that I am neurodivergent. A part of me always knew that I was different but until then I did not have the resources or the labels that I needed to look at it under a positive light. The online neurodivergent community played a major role in this shift of perspective and I am eternally grateful to have found the community and support network built by empathetic neurodivergent strangers who likely will never truly see the scale of the fruits of their efforts. I hope to pay it forward by continuing to advocate for my fellow neurodivergent people. With this discovery, my life came full circle to the realization that theoretical physics has always been a so-called ``special interest'' for me -- a common characteristic of the neurodivergent mind -- and will continue to hold that status for the foreseeable future. I owe my passion for physics to my neurodivergence and therefore also a large part of my happiness.


\newpage
\section{Introduction}

From metals to neutron stars, superconductors to nuclear plasmas, phases of matter described by Fermi surfaces and their instabilities are proliferous. The question ``\textit{What are the different possible ways that interacting fermions can behave at macroscopic scales?}'' is as easy to pose as it is difficult to answer. The possibilities are endless and ever-growing and stand tall and sturdy as a counterpoint to the traditional reductionist-constructivist hypothesis in physics \cite{Anderson1972more_is_diff}. To even begin to answer this question, a broad organizing principle is required.

One such organizing principle is obtained by counting the number of emergent low energy degrees of freedom that govern the behaviour of such systems. The notion of an energy gap helps categorize many-body systems into three possible classes: gapped, gapless and `very gapless'.

\textit{Gapped} systems do not have any propagating, low energy degrees of freedom. The degrees of freedom here are instead topological in nature and are described by topological quantum field theories\footnote{A new class of these that are not described by conventional topological field theories have recently been discovered and are collectively called `fracton models' \cite{Vijay2015fractons,Pretko2017fractons,seiberg2020fracton}. For a review, see \cite{nandkishore2019review,pretko2020review}.}. \textit{Gapless} systems have a finite number of propagating low energy degrees of freedom. These often describe critical points in phase diagrams or boundaries of topologically nontrivial gapped phases.

\textit{`Very gapless'} systems on the other hand have infinitely many low energy degrees of freedom. In particular, the density of states at zero energies is finite. Systems with extended Fermi surfaces are the canonical example of such phases, where low energy excitations can be hosted anywhere on the Fermi surface. Within the realm of Fermi surface physics, a classification of the possible phases of matter is still elusive, largely due to the many possible instabilities that Fermi surfaces can have. One suitable starting point for getting a picture of the various possibilities is to take a free Fermi gas and turn on interactions between the fermions, allowing them to scatter off of each other.

The interactions between fermions can then be put into one of two boxes: short range and long range. Short range interactions are usually mediated by gapped modes. At low energies these can effectively be thought of as point-like interactions between fermions with corrections to this description that do not significantly alter the physical picture. This is the realm of Fermi liquid theory (and its instabilities), one of the pillars of modern condensed matter physics, first developed by Landau \cite{Landau:1956zuh} in a classic 1956 paper. Landau's key insight was that short range interactions in most situations do not dramatically alter the spectrum of excitations of a free Fermi gas. The excitations of the interacting theory are then very similar to free fermions, and thus the notion of a quasiparticle was born. Landau's Fermi Liquid Theory (LFLT), the \textit{classical} formalism for describing Fermi liquids, can perhaps be called the first example of an \textit{effective theory} - a low energy description of a system that is insouciant to microscopic details whose effects are captured by a comparatively small number of parameters\footnote{I thank Luca V. Delacr\'etaz for this succinct description of effective theories.}.

Despite being rather successful at describing the physics of dense, interacting fermions, LFLT stood out among a plethora of other effective descriptions in many body physics as one of the few theories that was not formulated in the language of the renormalization group (RG) and was classical\footnote{Pun intended.} in nature, being described by an equation of motion rather than an action or a Hamiltonian. Progress along these lines was made only in 1990 in \cite{Benfatto:1990zz}, which was then formalized in \cite{Shankar:1990,Polchinski:1992ed} into the \textit{modern} formalism.

The \textit{effective field theory} (EFT) obtained from this analysis can be simplified at the cost of losing locality in space \cite{Altshuler1994patchNFL,Nayak:1993uh,Nayak:1994ng}, so it is not a genuine EFT in that the tower of irrelevant corrections to the scale invariant fixed point cannot be systematically listed, for example through an expansion in spatial and temporal derivatives. An alternate route to a local EFT for Fermi liquids was inspired by the idea of bosonization and pioneered in \cite{Haldane:1994,CastroNetoFradkin:1994,Houghton:2000bn}. But this approach also suffer from the same issue, in that it is unclear how one would construct and classify irrelevant corrections to the scale invariant fixed point. These \textit{contemporary} formalisms are hence also incomplete and in need for further refinement.

Long range interactions, on the other hand, are often mediated by gapless degrees of freedom which cannot be ignored (i.e., integrated out) at any energy scale, and it becomes important to keep track of the additional gapless modes alongside the excitations of the Fermi surface. This can alter the physics of the Fermi surface in ways that are hard to predict, since such interactions often tend to be strong. A celebrated, now solved example of this is the electron-phonon problem \cite{migdal1958,eliashberg1960}, which accounts for the resistivity and superconducting instability of conventional metals\footnote{For recent work on the breakdown of the Migdal-Eliashberg theory of electron-phonon interactions, see \cite{esterlis2018e-ph_breakdown,chubukov2020eliashberg_breakdown}.}.

A more violent example of such an interaction is presented in a class of phases dubbed \textit{non-Fermi liquids} (NFL) (see, e.g., \cite{sslee2017nfl_review} and references therein for a review). The gapless mode that couples to the Fermi surface in these examples is usually either the critical fluctuation of an order parameter or a gauge field in appropriate spatial dimensions. Such interactions trigger an instability of the Fermi surface and the fate of the RG flow is one of the biggest open problems in condensed matter physics. The list of unanswered questions ranges from describing the phase of the end point of the RG flow (metallic NFL or Mott insulator or unconventional superconductor) to developing effective descriptions of the various possibilities and understanding how they compete with one another.

Answers to these questions are crucial from an applied physics perspective since the most common occurrence of NFL physics is in high-temperature superconductivity \cite{keimer2015highTc,phillips2022strange} observed in various different layered materials such as cuprates. In many of these materials, the superconducting dome hides a quantum critical point where the metal undergoes a magnetic phase transition, the order parameter fluctuations of which couple to the Fermi surface and drive the instability to a superconductor. The ultimate goal for NFL physics would be to understand the mechanism that causes high temperature superconductivity in order to be able to engineer materials which could enhance this mechanism and raise the critical temperature of the superconducting phase to larger values, possibly even to room temperature.

From a theoretical standpoint, Fermi and non-Fermi liquids provide a unique playground to explore unconventional RG flows. Almost all tractable RG flows in physics are between two scale invariant fixed points that have no inherent scales. Fermi and non-Fermi liquids, however, enjoy scale invariance despite the presence of an intrinsic scale -- the Fermi momentum $p_F$ -- and understanding the RG flow from one to the other hence necessarily requires a broadening of the notion of RG as well as that of a `scale'. Unconventional RG flows have been gaining interest across various disciplines ranging from the study of fractonic and exotic theories \cite{shirley2019fracton_rg,wang2019fracton_rg,you2021fracton,gorantla2021uv/ir,lake2022fracton_rg} to machine learning \cite{MLscaling_obs,MLscaling_expl} and even information theory and neuroscience \cite{Koch-Janusz2018infoRG,Kline2022infoRG}, and it is likely that Fermi surface physics can serve as a useful launchpad for generalizing the notion of RG beyond its rigid framework and conventional metanarrative.

A fundamental bottleneck to understanding the physics of non-Fermi liquids is the lack of an EFT description for Fermi liquids. Since the scaling behaviour of an NFL can differ dramatically from that of a Fermi liquid, irrelevant corrections to any effective theory of a Fermi liquid can have important consequences for the NFL. A classification of irrelevant corrections to Fermi liquid theory with definite scaling properties, which is missing from the literature so far, would thus hugely benefit the search for an effective description for NFLs.

This is precisely the aim of the \textit{postmodern} formalism developed in \cite{main:2022} and expounded upon in this thesis. We find that LFLT is secretly governed by the geometry of a rather large Lie group -- that of canonical transformations of a single-particle phase space. This constrains the structure of the effective theory for Fermi liquids rigidly enough to be able to construct higher order corrections to the contemporary approaches as well as classify their scaling behaviour. The geometric structure underlying the postmodern formalism also allows us to systematically identify and impose symmetries as well as couple to gauge fields.

Such diffeomorphism groups are not only important for Fermi liquid theory, but also present themselves as a useful tool across other disciplines in condensed matter physics, such as quantum Hall states, lattices of charged monopoles or superfluid vortices and even skyrmions in ferromagnets \cite{main2021vpd,main2021noncomm}, suggesting that diffeomorphism groups have the potential to broadly understand and constrain the properties of various many-body phases.

The rest of this dissertation is organized as follows: in section \ref{sec_review} we review the various historic approaches to Fermi liquid theory and comment on the benefits and drawbacks of each of them. In section \ref{sec_overview} we summarize the postmodern formalism and provide an overview that is stripped off of most technical details for simplicity. In section \ref{sec_mic_ham} we develop the Hamiltonian formalism for Fermi liquids, which is then turned into an action formalism in section \ref{sec_coadj_ac}. Section \ref{sec_coadj_ac} also presents how this action encodes spacetime, gauge, and emergent symmetries, as well as how it simplifies the calculation of correlation functions in Fermi liquids. Section \ref{sec_nfl} then explores how the postmodern formalism can be used as a stepping stone towards perturbative NFLs. In section \ref{sec_extensions} we then switch gears to present different possible generalizations of the postmodern formalism that account for internal symmetries, conventional superconductivity, and large momentum processes. Finally, we conclude in section \ref{sec_conclusion} with an outlook on the various potential applications of the postmodern formalism.


\newpage
\section{Review and history of Fermi liquid theory}\label{sec_review}

We begin by reviewing the various approaches to describing Fermi liquids that have been developed over the last century. This discussion is by no means exhaustive, and we will differ to relevant references for more details.


\subsection{``Classical Fermi liquids'': Landau's kinetic theory}
\begin{figure}[h]
    \centering
    \includegraphics[width=10cm]{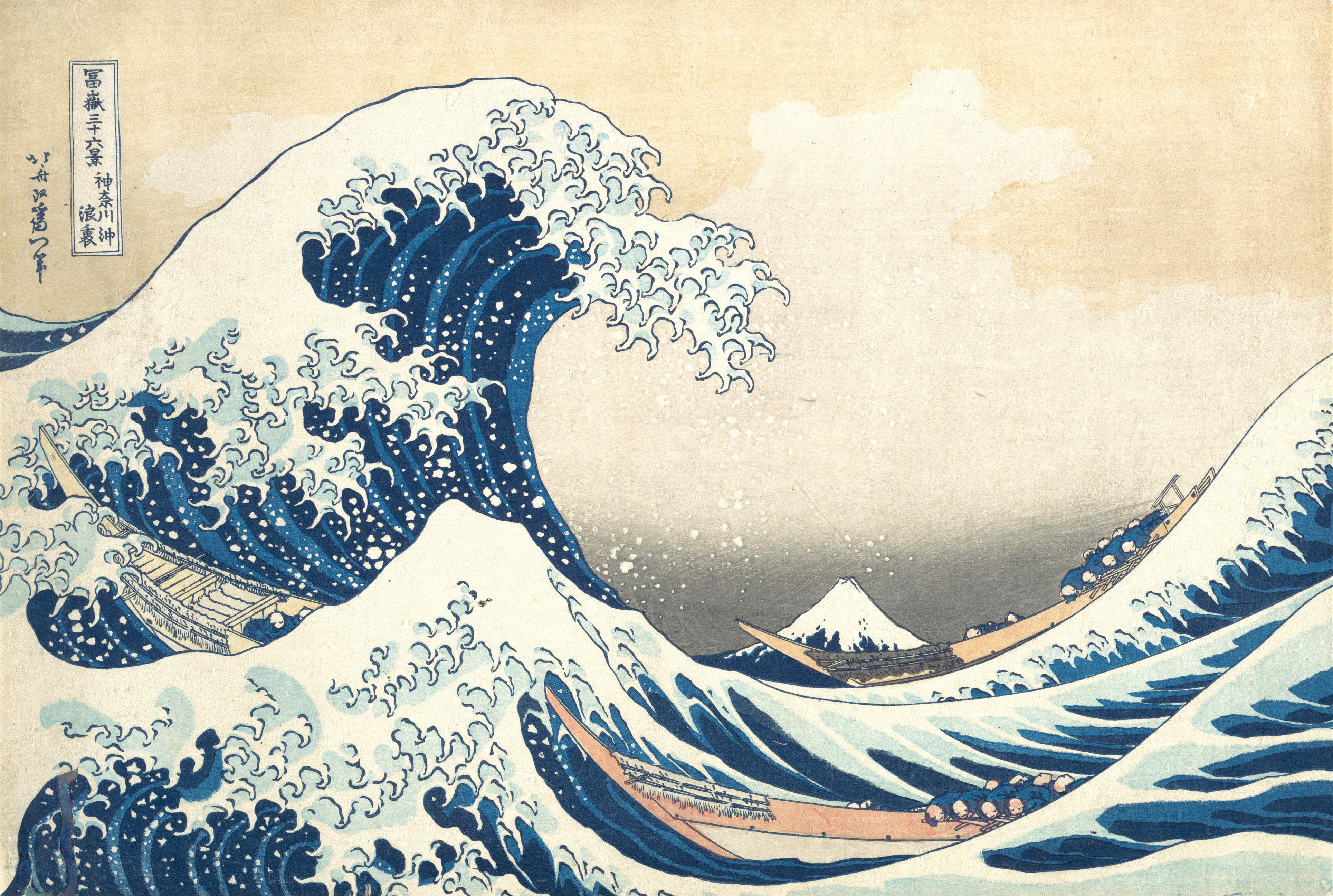}
    \caption{Hokusai's rendition of a propagating mode in kinetic theory.}
    \label{fig_great_wave}
\end{figure}

The very first description for Fermi liquids was proposed by Landau in the form of a kinetic equation. Consider first a gas of non-interacting fermions. Owing to Pauli's exclusion principle, its ground state at zero temperature is described by a occupation number function in momentum space $f_0(\p) = \Theta(\epsilon_F - \epsilon(\p))$ that takes values 1 or 0. $\epsilon_F$ is the Fermi energy and $\epsilon(\p)$ is the dispersion relation for a single fermion. The solution to the equation,
\begin{equation}
    \epsilon(\p) = \epsilon_F\, ,
\end{equation}
defines the Fermi surface at
\begin{equation}
    |\p| = p_F(\theta)\, .
\end{equation}
If the dispersion relation is invariant under rotations, the Fermi momentum $p_F$ is a constant independent of the angles $\theta$ in momentum space. The dynamics of this system is described by a mesoscopic\footnote{By the word `mesoscopic', we mean a regime where we are concerned with physics at length scales much larger than a characteristic length scale, here $1/p_F$. This allows us to describe quantum particles in a semi-classical description using coordinates $\x$ that label the mesoscopic region of size $1/p_F$ that the quantum particle is localized within, and momentum $\p$ of the particle up to uncertainty.} one-particle distribution function $f(t,\x,\p) = f_0(\p) + \delta f(t,\x,\p)$ that obeys the collisionless Boltzmann equation:
\begin{equation}
    \d_t f + \nabla_\p \epsilon(\p) \cdot \nabla_\x f + \mathbf{F}_\text{ext}\cdot\nabla_\p f = 0\, ,
\end{equation}
where $\mathbf{F}_\text{ext}$ is the external force applied to the free Fermi gas. The dynamics of the free Fermi gas are hence entirely captured by the dispersion relation.

For an interacting Fermi liquid, however, the occupation number at every momentum is not a well-defined quantum number, and we cannot characterize its dynamics using the distribution function.

Landau's argument to work around this issue was the following: suppose we start with the free Fermi gas and turn on interactions adiabatically. Thanks to Pauli exclusion principle, the available phase space for the fermions to scatter to is significantly smaller the closer they are to the Fermi surface initially. The low energy ($E\ll\epsilon_F$) part of the interacting many-body spectrum should be continuously deformable to the spectrum of the free theory. Since the spectrum of the free Fermi gas can be constructed from the building block of a single fermion placed outside but close to the Fermi surface (or a single hole inside), this building block should persist as the interactions are adiabatically turned on and also exist in some ``dressed'' form in the low energy spectrum of the interacting Fermi liquid. The remnant of this building block in the interacting theory is what we call a \textit{quasiparticle}.

In situations where this argument holds, we should have an effective single-particle description for the dynamics of interacting Fermi liquids, analogous to the collisionless Boltzmann equation for free fermions. In fact, Fermi liquids are defined retroactively as fermionic phases of matter where this argument holds. The degree of freedom describing the quasiparticle is then also a distribution function:
\begin{equation}
    f(t,\x,\p) = f_0(\p) + \delta f(t,\x,\p)\, .
\end{equation}
However, since the quasiparticle only exists as part of the spectrum for momenta close to the Fermi surface, the distribution $f$ and the fluctuation $\delta f$ are only well defined in a narrow region $|\p|-p_F \ll p_F$. All that we need in order to describe the low energy dynamics of the interacting Fermi liquid is a dispersion relation $\epsilon_\text{qp}$ for the quasiparticle. This is phenomenologically constructed as follows:
\begin{equation}
    \epsilon_\text{qp}(\x,\p) = \epsilon(\p) + \int \frac{d^dp'}{(2\pi)^d} F(\p,\p') \delta f(\x,\p')\, ,
\end{equation}
where $\epsilon(\p)$ is the free fermion dispersion relation, and $F(\p,\p')$ is a phenomenological function that characterizes the interaction contribution to the energy of the quasiparticle at $\p$ due to quasiparticles at $\p'$. Note that the interaction term in the quasiparticle energy is local in space, which is due to the assumption that any interaction between the quasiparticles is short-ranged.

At the risk of being pedantic, we emphasize again that the quasiparticle energy, the interaction function, and the distribution are well-defined only in a small neighbourhood of the Fermi surface. In other words the $\p$ derivatives of all these quantities are only well-defined at the Fermi surface and constitute the various parameters and degrees of freedom of the effective theory.

We can now postulate a collisionless Boltzmann equation that describes the dynamics of the interacting Fermi liquid:
\begin{equation}\label{eq_landau_kinetic_eq}
    \d_t f + \nabla_\p \epsilon_\text{qp}[f] \cdot \nabla_\x f - \left( \nabla_\x \epsilon_\text{qp}[f] - \mathbf{F}_\text{ext} \right) \cdot \nabla_\p f = 0\, .
\end{equation}
We will refer to this equation as \textit{Landau's kinetic equation.} One crucial difference between the interacting Fermi liquid and the free Fermi gas is that equation \eqref{eq_landau_kinetic_eq} is nonlinear in $\delta f$, while the collisionless Boltzmann equation is linear. The nonlinearity comes from the dependence of the quasiparticle energy on the distribution. This also modifies the dynamics at the linear level, since the interaction results in internal forces $\nabla_\x \epsilon_\text{qp}$ acting on the quasiparticles in addition to any external forces.

Since the interaction function $F(\p,\p')$ is well-defined only near the fermi surface, one often assumes that it only depends on two points on the Fermi surface at the angles $\theta,\theta'$, and an angular expansion of the interaction function defines the so-called Landau parameters,
\begin{equation}
    F(\theta,\theta') \sim \sum_l F_l P^{(d)}_l(\theta,\theta')\, ,
\end{equation}
where $P^{(d)}_l(\theta,\theta')$ form a basis of functions in $d$ dimensions that transform covariantly under the symmetries of the Fermi surface, and $l$ is a label for the representations of those symmetries. For example, for a spherical Fermi surface $l=0,1,2,\ldots$ is an `angular momentum' index, and the basis functions are cosines in $d=2$ and Legendre polynomials of cosines in $d=3$.

From Landau's kinetic equation we can calculate a plethora of physical quantities from thermodynamic properties to correlation functions, in terms of Landau parameters which encode the microscopic interactions. In order to calculate correlation functions for, e.g., the particle number density and current, we can couple the theory to background electromagnetic fields through the Lorentz force $\mathbf{F}_\text{ext} = \E + \v\times\mathbf{B}$.

One finds stability conditions for the theory as lower bounds on $F_l$ which when violated, result in Pomeranchuk instabilities. For certain ranges of the Landau parameters, Fermi liquids also exhibit a collective excitation known as zero sound that propagates faster than the Fermi velocity $v_F = \epsilon'(p_F)$ and is hence distinguishable from the particle-hole continuum $\omega \le v_F |\q|$ (figure \ref{fig_great_wave}). The specific calculations that result in these various properties and more can be found, for example, in \cite{landau_statphys_2,Abrikosov:107441}.

While LFLT describes many aspects of interacting Fermi liquids quite well, it has various drawbacks. Firstly, it is unclear how such a theory would emerge from a microscopic model. Since the kinetic equation is written down `by hand' it is not even clear when one should expect a microscopic model of interacting fermions to be described by LFLT.

Second, being an equation-of-motion based description, LFLT is in effect a classical theory, with the only source of `quantumness' being Pauli exclusion and the Fermi-Dirac distribution that gives the ground state $f_0$ of the theory. In practice this means that the theory is blind to subleading corrections to physical quantities such as correlation functions and thermodynamic properties.

These drawbacks would be at least partially, if not completely be remedied by a field theoretic description - one that is amenable to the renormalization group (RG), unlike LFLT.


\subsection{``Modern Fermi liquids'': Renormalization group}
\begin{figure}[h]
    \centering
    \includegraphics[width=10cm]{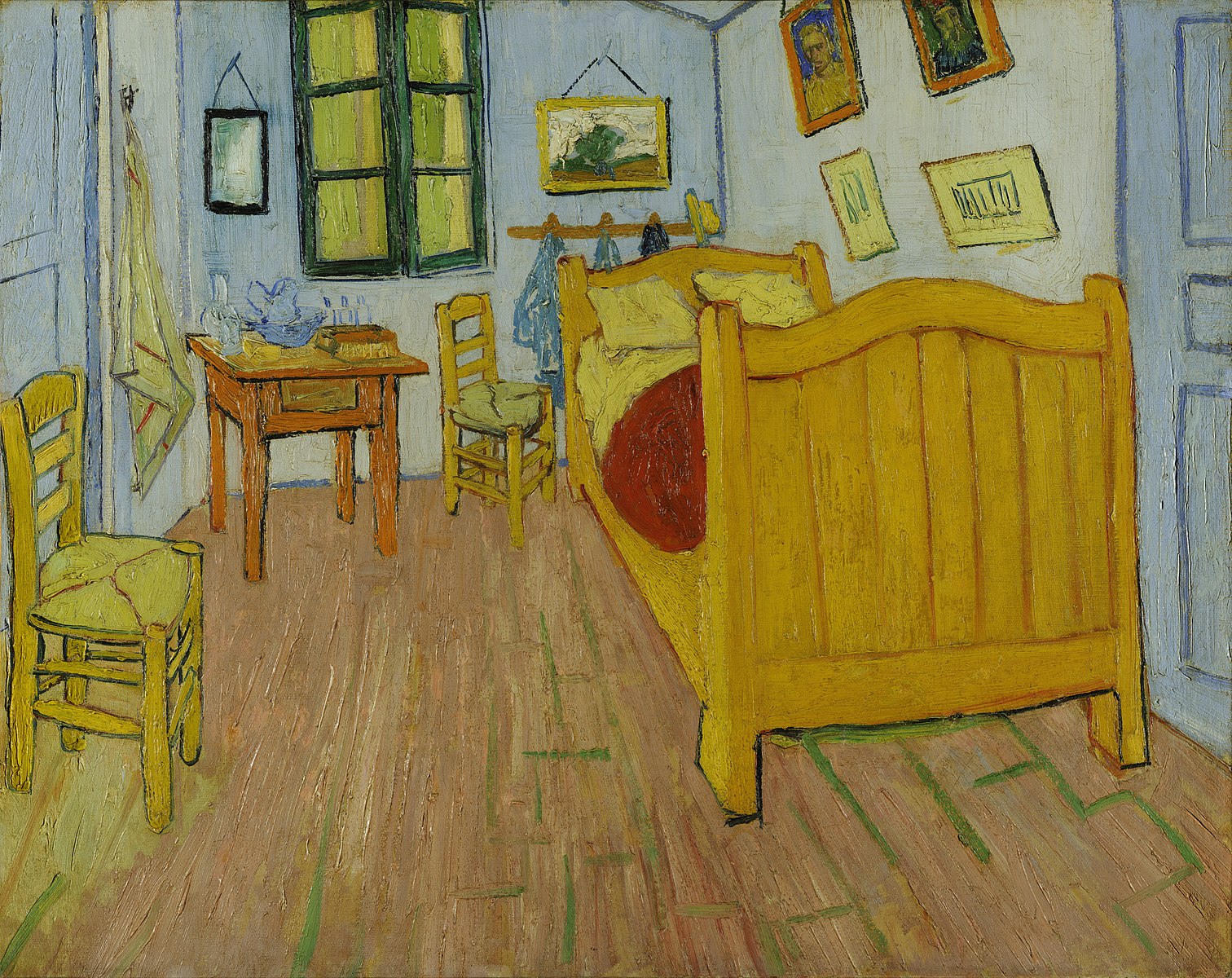}
    \caption{Van Gogh's visualization of scaling towards a (rectangular) Fermi surface.}
    \label{fig_van_Gogh}
\end{figure}

To understand the scaling behaviour of interacting Fermi liquids, we need to pick an RG scheme. The prototypical RG scheme most commonly used in physics, wherein we rescale length to be larger and larger, or equivalently rescale momenta to 0, also shrinks the Fermi surface down to a point! This scheme cannot possibly give physically relevant results since the Fermi surface is an experimentally measurable quantity. We hence need to pick a new scaling scheme.\footnote{It is important to note that in most commonly studied systems in physics such as quantum or statistical field theories, the symmetries of the system uniquely prescribe the RG scheme that can extract universal information from it. Here, however we encounter a system where this is not immediately obvious, so we need to look for other identifiers for the `correct' prescription.}

The most natural RG scheme is one where momenta are rescaled towards the Fermi surface (figure \ref{fig_van_Gogh}). This scheme was introduced in \cite{Benfatto:1990zz,Shankar:1990,Polchinski:1992ed} and is commonly referred to as `Shankar-Polchinski' RG, after the physicists who independently formalized it.

In the spirit of effective field theory, we first identify the low energy degrees of freedom. LFLT tells us that these are fermionic quasiparticles. We define an operator $\psi^\dagger(\p)$ that creates a quasiparticle with momentum $\p$. The annihilation operator $\psi(\p)$ creates a hole in the Fermi sea at the point $-\p$, so that the net momentum of the state with a single hole is $+\p$\footnote{This is different from the usual convention employed in condensed matter physics, where the operator $c(\p)$ creates a hole at the point $\p$, thereby creating a state with momentum $-\p$. We use the less common convention since in our convention, both $\psi^\dagger$ and $\psi$ are Fourier transformed in the same way. This sets a uniform convention for Fourier transforms, allowing us to Fourier transform with impunity without having keep track of sign conventions any more than necessary.}. The free action is given by
\begin{equation}
    \int \frac{dt d^dp}{(2\pi)^d} \psi^\dagger(\p)\left[ i\d_t - (\epsilon(\p) - \epsilon_F) \right] \psi(-\p)\, .
\end{equation}
Each point $\p$ in momentum space can be written as a sum of a vector $\p_F$ on the Fermi surface and another vector $\k$ orthogonal to the Fermi surface at $\p_F$:
\begin{equation}
    \p = \p_F + \k\, , \qquad d^dp = d^{d-1}p_F ~ dk\, ,
\end{equation}
where $d^{d-1}p_F$ is a measure for integrating over the Fermi surface. In our RG scheme, $\p_F$ remain invariant under scaling, while $\k$ get rescaled by a factor of $s\lesssim1$ to $s\k$. The dispersion can be expanded to leading order so that
\begin{equation}
    \epsilon(\p) - \epsilon_F = |\k||\v_F(\p_F)| + \mathcal{O}(k^2)\, ,
\end{equation}
and marginality of the free action requires
\begin{equation}
    [\d_t] = [\k]\, , \qquad [\psi] = -\frac{1}{2}\, .
\end{equation}
We then write down all possible terms allowed by symmetries and analyze their scaling behaviour, both at tree level and at loop level. The leading nontrivial term is a quartic interaction that enables nontrivial $2\rightarrow2$ scattering processes:
\begin{equation}
    \int_t \int_{\p_1\p_2\p_3\p_4} V(\p_{F1},\p_{F2},\p_{F3},\p_{F4}) \psi^\dagger(\p_1) \psi(\p_2) \psi^\dagger(\p_3) \psi(\p_4) \delta(\p_1+\p_2+\p_3+\p_4)\, .
\end{equation}
Immediately, we notice two possibilities for the scaling of the momentum conserving delta function. If the corresponding Fermi momenta $\p_{Fi}$ sum to zero, the delta function scales non-trivially under our RG scheme, while if they do not, the delta function is (approximately) invariant under the scale transformation.

For configurations where $\sum_i \p_{Fi}\ne0$, we find that the quartic term is strictly irrelevant and hence does not change the scale invariant fixed point. For configurations with $\sum_i \p_{Fi} = 0$, on the other hand, the quartic term is marginal. All that remains is find configurations for which the sum vanishes, and check whether loop corrections change the scaling behaviour of the terms corresponding to the relevant configurations.

\begin{figure}[t]
    \centering
    \begin{subfigure}[b]{0.35\textwidth}
        \centering
        \includegraphics[width=\textwidth]{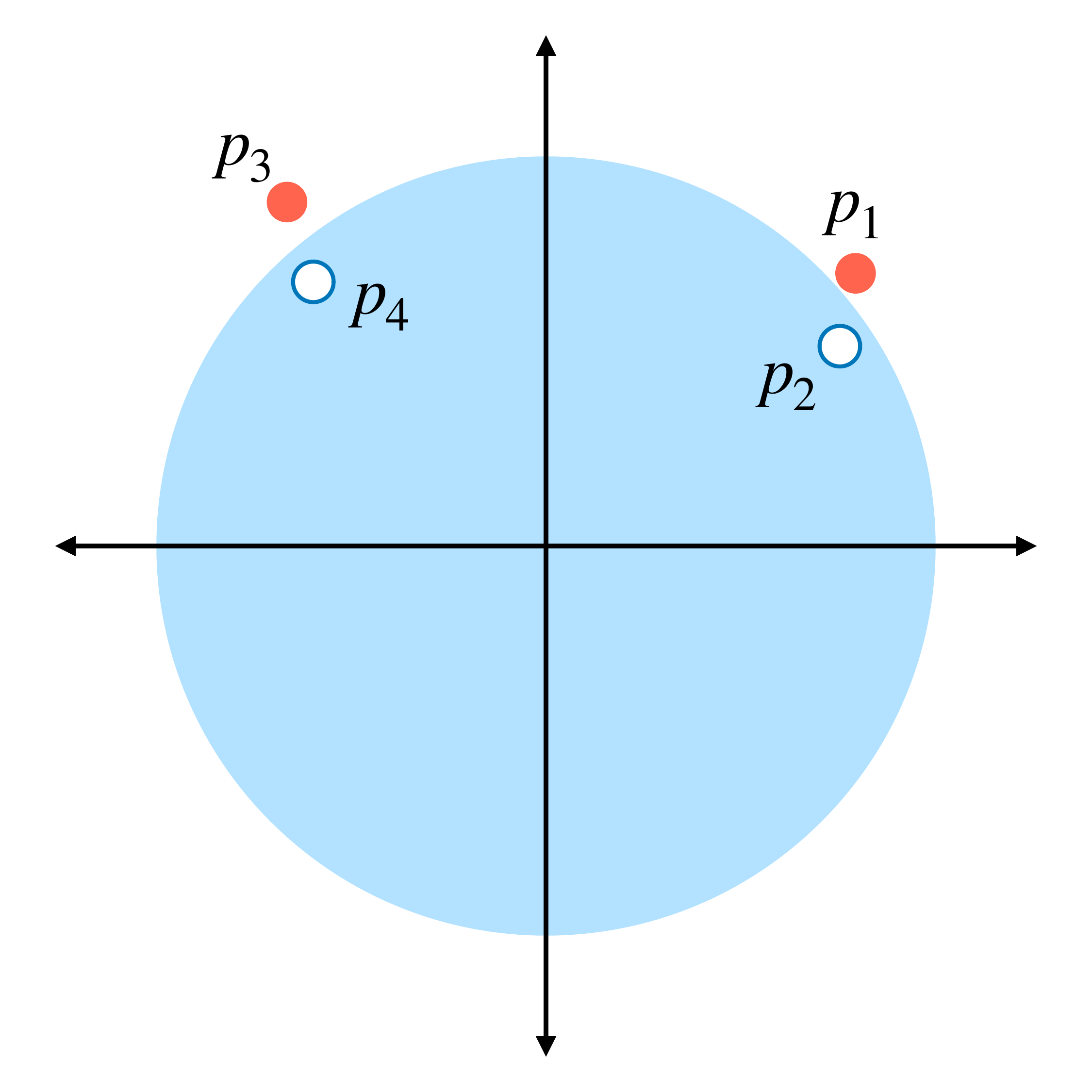}
        \caption{Forward scattering}
        \label{fig_forward_scat}
    \end{subfigure}
    \hspace{1.25cm}
    \begin{subfigure}[b]{0.35\textwidth}
        \centering
        \includegraphics[width=\textwidth]{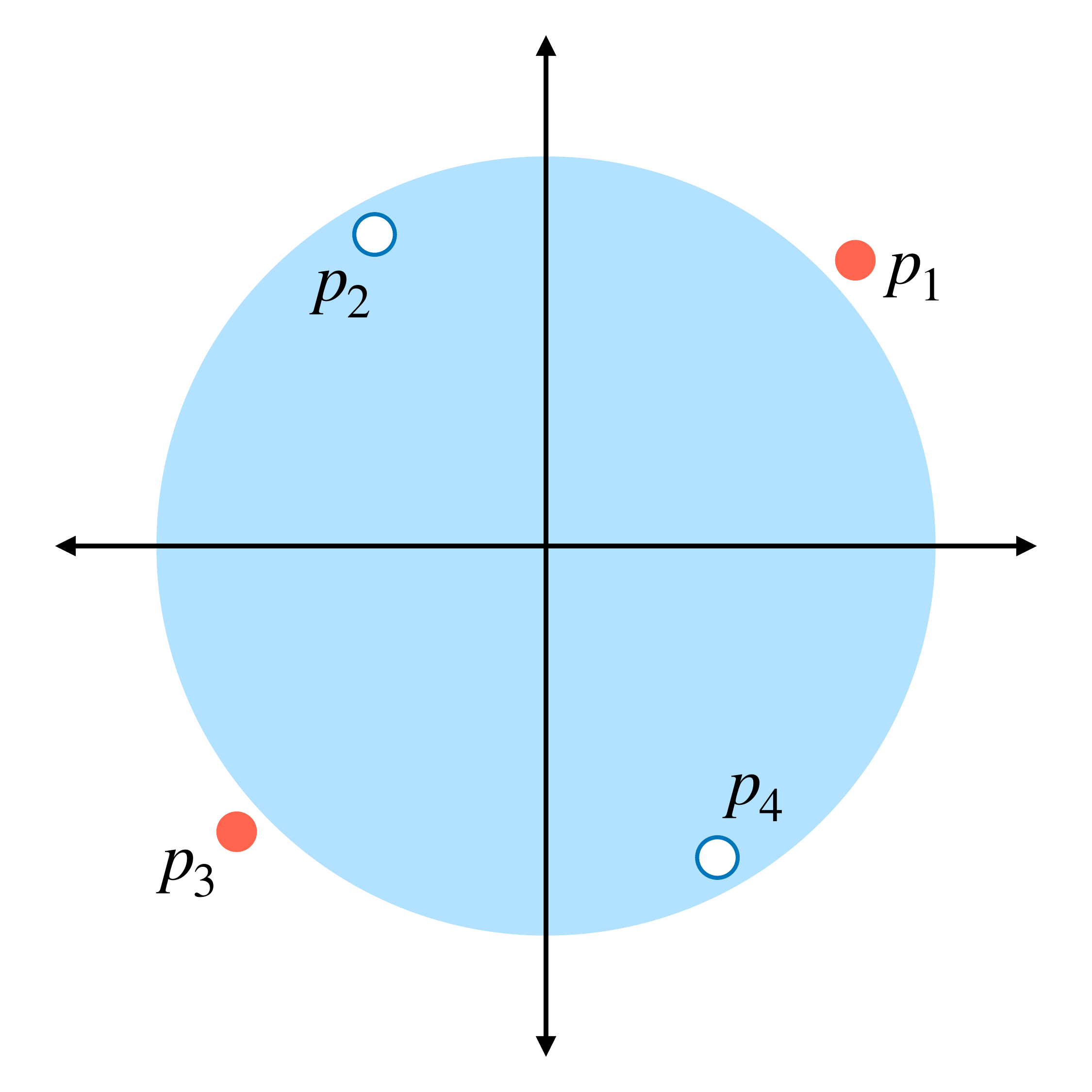}
        \caption{BCS channel}
        \label{fig_bcs_scat}
    \end{subfigure}
    \caption{Scattering configurations for marginal interactions at tree level.}
    \label{fig_scat}
\end{figure}

Consider for instance $d=2$ with a circular Fermi surface. There are two distinct classes of configurations with $\sum \p_F = 0$:
\begin{equation}
    (\p_{F2} = -\p_{F1}, ~ \p_{F4} = -\p_{F3})\, ; \qquad (\p_{F3} = -\p_{F1}, ~ \p_{F4} = -\p_{F2})\, .
\end{equation}
The solution with $\p_{F4} = -\p_{F1}$ is just the first solution with the hole momenta exchanged. The first class of solutions characterize forward scattering, i.e., incoming particles leave with nearly the same or exchanged momenta. These correspond to particle hole pairs with a small net momenta, such as the configuration in figure \ref{fig_forward_scat}. This class of configurations is hence often called the `particle-hole channel'. The form factor $F(\p_{F1},\p_{F3}) = V(\p_{F1},-\p_{F1},\p_{F3},-\p_{F3})$ is the corresponding interaction function.

The second class of solutions has the two particles as well as the two holes align at antipodal points on the Fermi surface respectively, with an arbitrary angle between them, for instance in figure \ref{fig_bcs_scat}. This configuration corresponds to the `Bardeen-Cooper-Schrieffer (BCS) channel'. The interaction form factor $g(\p_{F1},\p_{F2}) = V(\p_{F1},\p_{F2},-\p_{F1},-\p_{F2})$ for this is independent of the forward scattering interaction, except in one special configuration with $\p_{F3}=\p_{F2}=-\p_{F1}$ which imposes a constraint $F(\p_F,-\p_F)$ = $g(\p_F,-\p_F)$. The marginal quartic terms can then be written schematically as
\begin{equation}
    \int_{\p_1\p_3} F(\p_{F1},\p_{F3}) [\psi^\dagger\psi\psi^\dagger\psi]_\text{ph}(\p_1,\p_3) + \int_{\p_1\p_2} g(\p_{F1},\p_{F2}) [\psi^\dagger\psi\psi^\dagger\psi]_\text{BCS}(\p_1,\p_2)\, .
\end{equation}
Both interactions are marginal at tree level, but a one-loop calculation shows that while forward scattering remains marginal, the BCS interaction becomes relevant if the coupling is attractive and irrelevant if the coupling is repulsive. Hence, attractive couplings in the BCS channel trigger a superconducting instability that destroys the Fermi surface.

The forward scattering interaction is just the interaction function in LFLT, but the BCS coupling is one to which LFLT is blind. The inclusion of the pairing instability is the most important advantage of the RG approach over LFLT, and exemplifies the power of effective field theory.

However, this approach still has its limitations. Ideally in an EFT, any isolated term that can be written from symmetry requirements has a fixed scaling dimension which can be calculated simply by adding the scaling dimensions of its constituents --- a principle known as power counting. But as we saw above, understanding the scaling properties of the quartic term was a significantly more complicated task than that, and becomes even more complicated in higher dimensions where the number of possible configurations with $\sum \p_F=0$ is even larger. This procedure becomes all the more gruesome for Fermi surfaces of more complicated geometry such as those for conduction electrons in metals.

In general, any given term in this EFT that can be written from invariance under symmetries does not have a fixed scaling dimension and additional work needs to be done to decompose it into a sum of terms that do. Even then one can find constraints relating one term to another in special cases, such as the configuration $\p_{F1} = -\p_{F2} = -\p_{F3} = \p_{F4}$ where the exactly marginal forward scattering coupling is identical to the marginally relevant or irrelevant BCS coupling. These constraints need to be kept track of by hand and do not immediately follow from any symmetry principle. Instead, the forward scattering -- BCS constraint is a consequence of hacing to decompose a single local operator into different scattering channels that are scaling covariant, but at the cost of an added redundancy.

Furthermore, while coupling LFLT to background gauge fields was a straightforward task, it is much less obvious how one couples this EFT to background gauge fields, given that the EFT lives in momentum space, where no standard minimal coupling procedure exists.

Two remedies for the former issue have been considered, which we will collectively refer to as the `contemporary' formalism, which we review next. Alternate functional RG schemes for Fermi surfaces which hope to capture physics beyond Shankar-Polchinski RG have also recently been developed in \cite{Borges2023funcRG,ma2023funcRG}.


\pagebreak
\subsection{``Contemporary Fermi liquids'': Patch theory and traditional bosonization}
\begin{figure}[h]
    \centering
    \includegraphics[width=7cm]{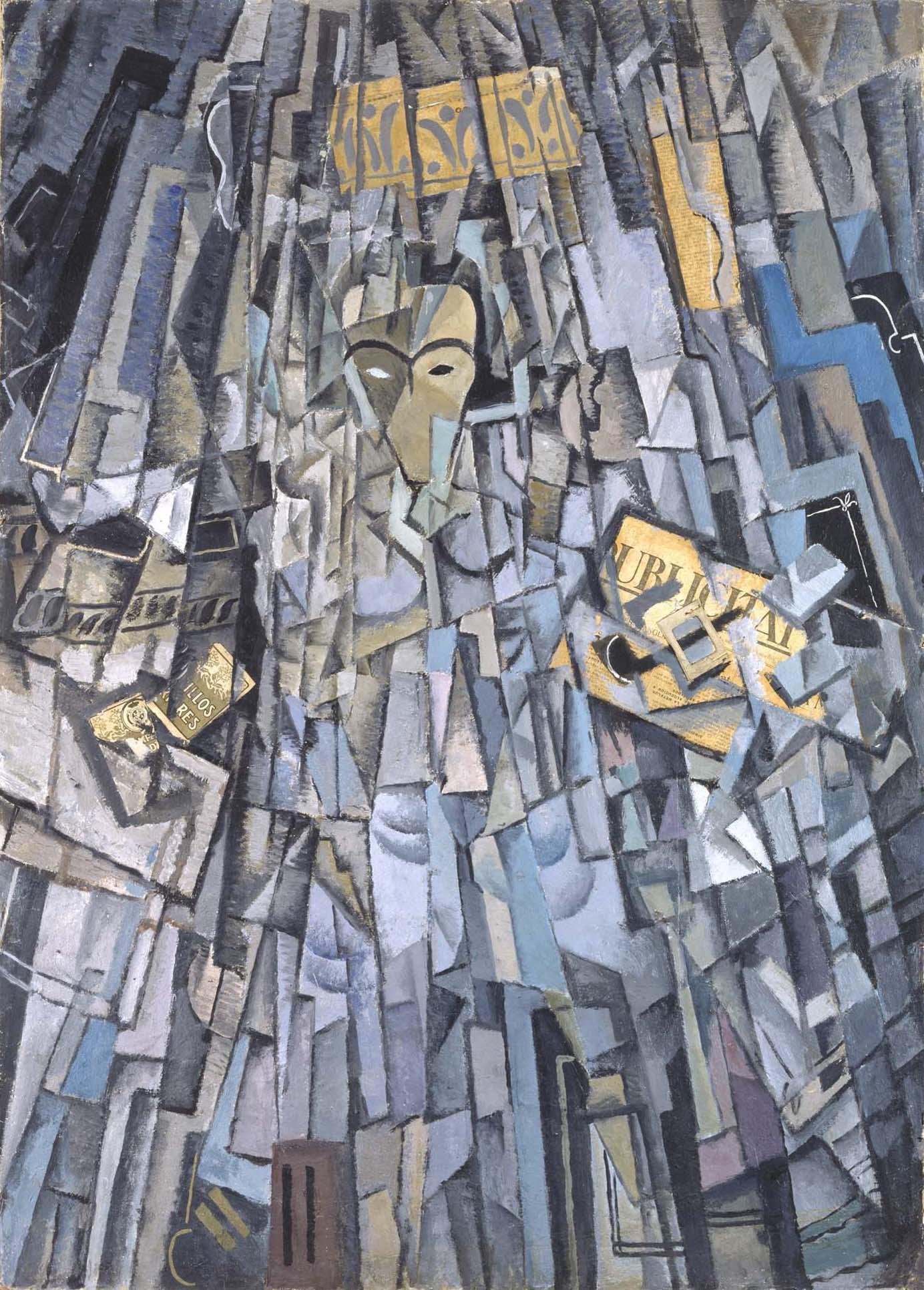}
    \caption{Dali's self-portrait under a patch decomposition.}
    \label{fig_dali}
\end{figure}

One of the key takeaways of the Shankar-Polchinski RG scheme is that, barring BCS interactions, particle-hole pairs have a significant impact on low energy physics only when they are sufficiently close to each other in momentum space (compared to $p_F$). This suggests that one potential workaround to the issue of interactions not having fixed scaling dimensions is the following: we can discretize the Fermi surface to a number of patches of the same size, labelled by a discrete index $\eta$ (figure \ref{fig_dali}), and subsequently separate interactions into intra-patch and inter-patch scattering.

The free fermion action Fourier transformed back to coordinate space can be written as a sum over patches,
\begin{equation}
    S = \sum_\eta \int d^{d-1}x_\parallel \int dt dx_\perp \Psi_\eta^\dagger ~ (x_\perp) \left( \d_t + v_{F\eta} \d_{x_\perp} \right) \Psi_\eta(x_\perp)\, ,
\end{equation}
where $x_\perp$ is a coordinate that is Fourier-conjugate to $\k$, the momentum vector orthogonal to the Fermi surface, $\x_\parallel$ are coordinates conjugate to the transverse directions within a patch, and $\Psi_\eta$ is the fermion on each patch defined by
\begin{equation}
    \psi(\x) = \sum_\eta e^{i\p_{F\eta}\cdot\x} \Psi_\eta(x_\perp)\, ,
\end{equation}
up to normalization. This is simply a collection of chiral fermions at each patch. Intra-patch scattering terms live within a single patch $\eta$, while inter-patch scattering terms couple two different patches $\eta\ne\eta'$. If we restrict our attention to a single patch $\eta_0$, the effect of the latter is simply a logarithmic renormalization of the field strength of $\Phi_{\eta_0}$ as well as its dispersion relation, so inter-patch interactions can be ignored. Intra-patch coupling can be analyzed in the usual way under rescaling of momenta toward the Fermi surface, transverse to the patch. Since the width of the patch is not rescaled in this procedure, the number of patches does not change under rescaling.


\subsubsection{Fermionic patch theory}\label{sec_fermion_patch}

The patch theory in the Shankar-Polchinski RG scheme has an important drawback. Discretizing the Fermi surface makes it so that each patch is effectively flat at low energies. To see this, consider the leading irrelevant correction to the quadratic action, which comes from the curvature of the Fermi surface within the patch,
\begin{equation}
    S = \int d^{d-1}x_\parallel \int dt dx_\perp \Psi^\dagger ~ (x_\perp) \left( \d_t + v_F \d_{x_\perp} + \frac{\kappa}{2} \nabla_\parallel^2 \right) \Psi(x_\perp)\, ,
\end{equation}
where we have dropped the patch index $\eta_0$. Since $\x_\parallel$ does not scale under the Shankar-Polchinski RG scheme, the curvature $\kappa$ scales to zero and we lose crucial information about the shape of the Fermi surface.
\begin{figure}[t]
    \centering
    \includegraphics[width=12cm]{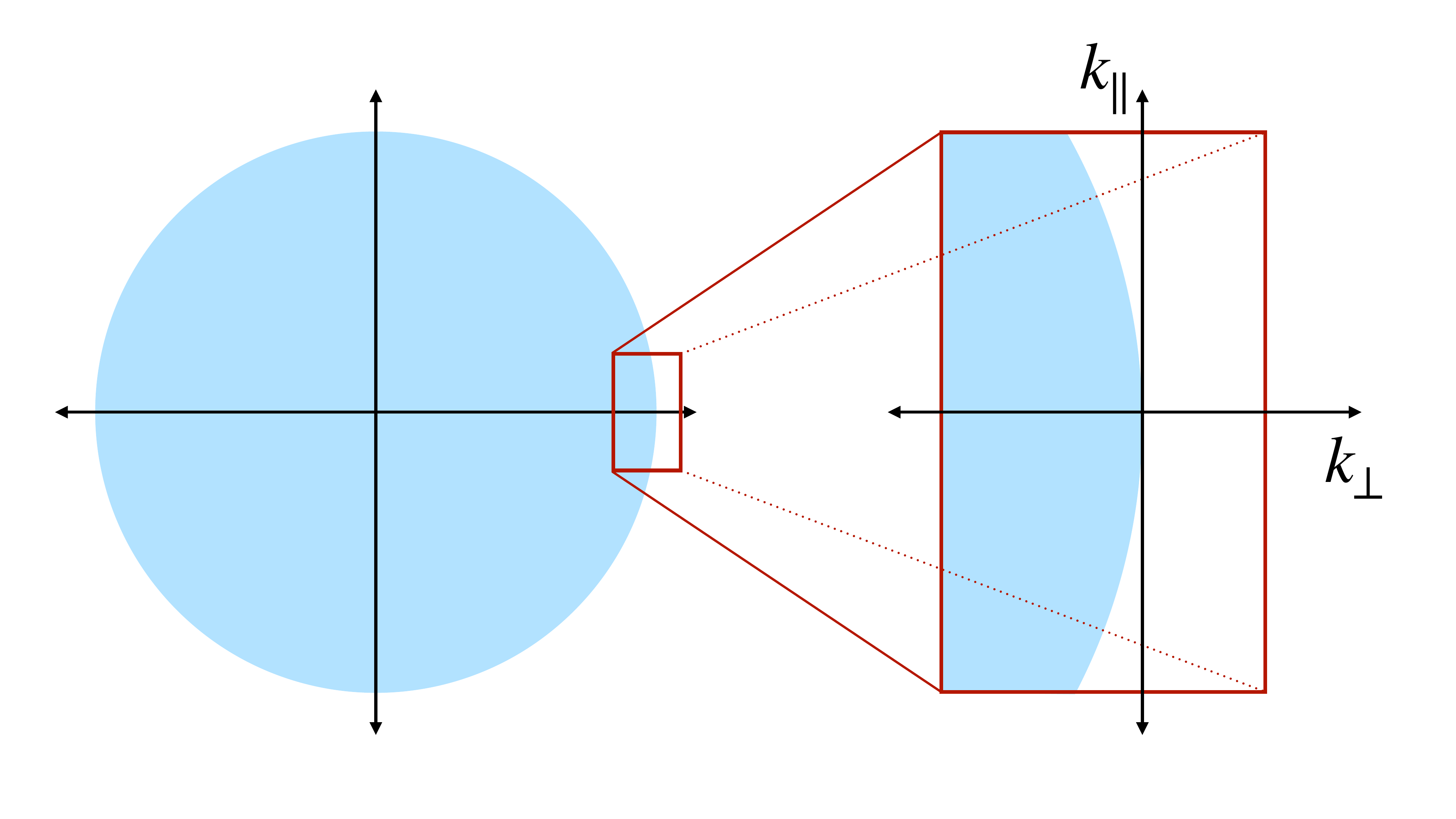}
    \caption{A single Fermi surface patch}
    \label{fig_patch}
\end{figure}

An alternate RG scheme that is more suitable to the patch description \cite{Nayak:1993uh,Nayak:1994ng}(see, e.g., \cite{sachdev_book} for a pedagogical description) and preserves the curvature of the Fermi surface is one where the coordinates $\x_\parallel$ scale like $(x_\perp)^{1/2}$. The curvature term is now scale invariant under this scale transformation, at the expense of the width of the patch scaling down to zero, resulting in a proliferation of the number of patches at the scale invariant fixed point. But if we are only concerned with the low energy properties of fermions within a single patch, we can ignore this drawback. As far as I am aware, so systematic analysis of the consequences of the proliferation of the number of patches exists in the literature, and in particular it is unclear whether this blow up modifies the RG flow of a single patch in any significant way.

One can show that intra-patch scattering from contact interactions under patch scaling is strictly irrelevant in all dimensions, which provides some evidence for the stability of Fermi liquids. Inter-patch couplings can at most logarithmically renormalize the field strength of the patch fermion and the Fermi velocity, and are often ignored. The only interactions that can modify the RG flow are then those that are mediated by a gapless mode. Fermionic patch theory is hence often used as an effective description for non-Fermi liquids, since it provides an RG scheme where other interactions between patch fermions can be safely ignored, in favour of interactions mediated by the gapless mode which couples most strongly to patches that are tangential to its momentum \cite{polchinski1994nfl,Altshuler1994patchNFL}.

Fermionic patch theory has a few more drawbacks. Firstly, in restricting the theory to a single patch, we loose locality in position space. Secondly, single-patch theory cannot accomodate BCS interactions either, which raises questions about the validity of RG flows derived from it. The usual expectation and/or hope is that the NFL fixed point obtained from patch theory would have its own superconducting instability, which would lead it to a superconducting fixed point with the same universal properties as the infrared (IR) fixed point of the physical RG flow without restricting to patches. Lastly, patch theory can only be used for understanding RG flows, but not for calculating physical quantities such as transport properties, for which we need to sum over all patches and be mindful about the proliferation of patches in the IR. Furthermore, the resistance of the Shankar-Polchinski EFT to gauging persists in fermionic patch theory as well.

Additionally, even though fermionic patch theory has attractive properties under RG and simplifies the calculation of scaling dimensions for various operators, the scaling behaviour of correlation functions calculated from patch theory is still not transparent. Various cancellations among diagrams can occur \cite{PhysRevB.58.15449,Metzner1997FermiSW} that alter the IR scaling form of the correlation functions and invalidate power counting arguments. We will discuss this in more detail in section \ref{sec_response} and demonstrate how the postmodern formalism resolves this difficulty.


\subsubsection{Bosonization of patch fermions}

Another approach that starts with the description in terms of patchwise chiral fermions but tries to preserve locality in position space is inspired by bosonization in 1+1d \cite{1dbos_review}. This approach was developed independently by Haldane \cite{Haldane:1994} and by Castro-Neto and Fradkin \cite{CastroNetoFradkin:1994}, and further developed by Houghton, Kwon and Marston \cite{Houghton:2000bn}. Since each patch fermion is a 1+1d chiral fermion, it can be independently bosonized into a collection of chiral bosons to give the following effective action:
\begin{equation}\label{eq_multidimbos_action}
    S = - p_F^{d-3} \sum_\eta \int dt d^dx ~ (\p_{F\eta}\cdot\nabla_\x \phi_\eta) \left( \d_t + v_{F\eta}\p_{F\eta}\cdot\nabla_\x  \right) \phi_\eta\, .
\end{equation}
Although this formalism is local in position space, it suffers from the same drawback as patch theory under Shankar-Polchinski scaling --- it cannot accomodate nonlinear-in-$\phi_\eta$ corrections from Fermi surface curvature and the dispersion relation. This has serious consequences, since even though the nonlinear corrections are irrelevant in Shankar-Polichinski scaling, they contribute at leading order to various higher point correlation functions, which traditional bosonization sans higher order corrections incorrectly suggests would vanish. For instance, the particle number density in traditional bosonization is linear in $\phi$, and since the action is quadratic in $\phi$, the density $(n>2)$-point functions calculated from this action are strictly zero, which certainly is not the case even for free fermions.

In order to solve this issue, various authors appealed to a more algebro-geometric picture underlying the interpretation of Fermi liquid theory as describing the dynamics of droplets in phase space \cite{Stone:1989,Das:1991uta,Dhar:1992rs,Dhar:1993jc,Khveshchenko:1993ug,khveshchenko:1995} similar to quantum Hall droplets on the lowest Landau level in the plane \cite{Iso:1992aa,Karabali:2003bt,Polychronakos:2004es}. This approach is an early precursor to the postmodern formalism described in this dissertation.


\newpage

\section{Postmodern Fermi liquids: A conceptual overview}\label{sec_overview}
\begin{figure}[h]
    \centering
    \includegraphics[width=10cm]{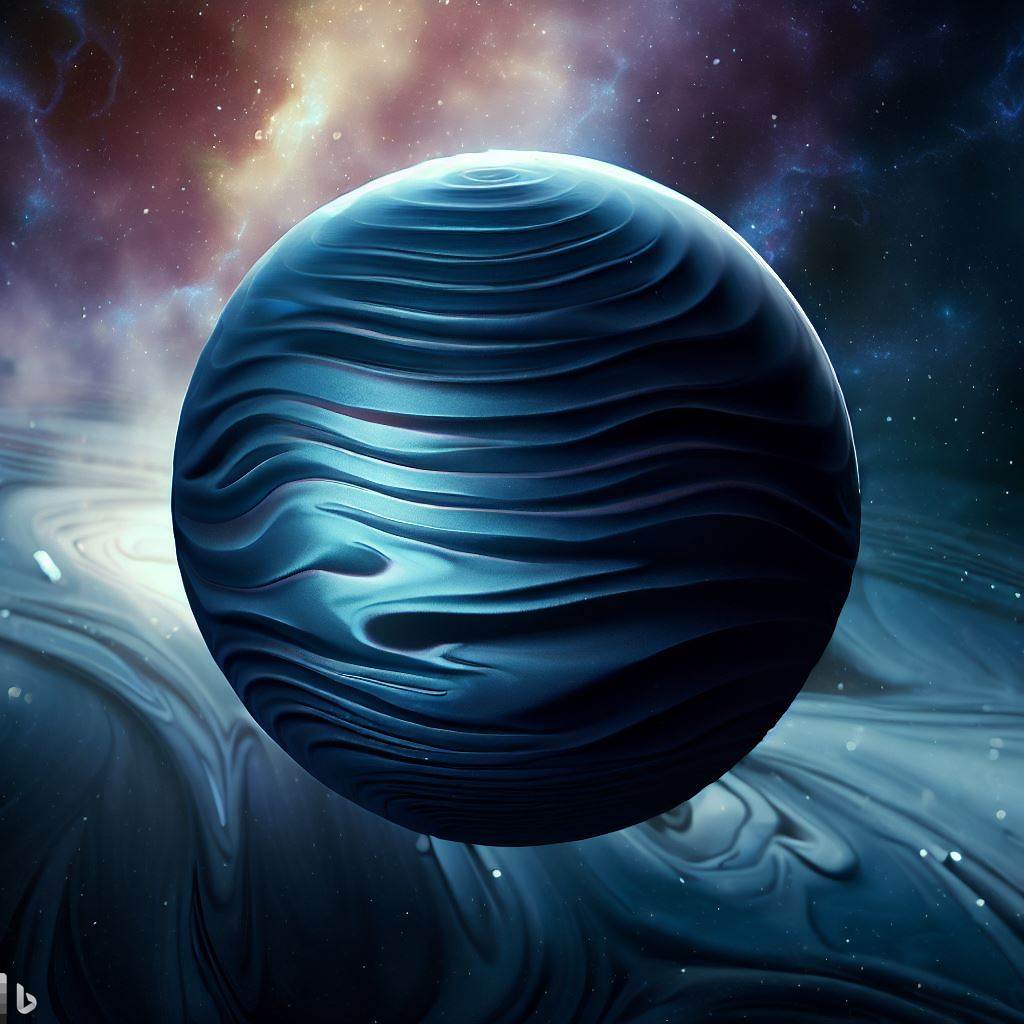}
    \caption{An artificial intelligence's impression of postmodern Fermi liquid theory.}
    \label{fig_wobbly_fermi}
\end{figure}

The starting point for our theory is the observation that the operator algebra constructed from microscopic fermions $\psi(\x)$ has a sub-algebra that is closed under commutators. This is the algebra of operators spanned by (anti-Hermitian) charge 0 fermion bilinears (see section \ref{sec_mic_ham} for details and precise definitions),
\begin{equation}\label{eq_generator_old}
    T(\x,\y) \sim i \psi^\dagger(\x) \psi(\y)\, .
\end{equation}
For theories whose Hamiltonian can be written entirely in terms of these bilinears, the closure of the sub-algebra guarantees that we can restrict our attention to the dynamics of operators in this sub-algebra in the Heisenberg picture, or classes of states distinguished only by expectation values of such operators in the Schr\"odinger picture.

What remains is to find a convenient parametrization for this large space of operators, or equivalently, for the dual space of of states, and figure out how to identify states with Fermi surfaces, to which the next two sections are dedicated. While this is straightforward in principle, some assumptions and approximations need to be made to make it useful in practice. These will be elucidated in the following section.

Conveniently, the question of how to parametrize a Lie algebra and its dual space has a well-established answer in mathematical literature, known as the coadjoint orbit method \cite{Kirillov_book,Wiegmann:1989hn,Alekseev88coadj}. This method was historically developed as a procedure for finding representations of Lie groups, but can also be interpreted as a means of setting up a dynamical system on a Lie group in the Hamiltonian formalism, and then turning that Hamiltonian formalism into an action. The Hamiltonian/action describe time evolution on the Lie algebra, which in our case is the space of fermion bilinears, in the Heisenberg picture, or equivalently on its dual space, which is the space of states, in the Schr\"odinger picture\footnote{Quantization of this action then gives representations of the Lie group under consideration.}.


\subsection{The Lie algebra of fermion bilinears}\label{sec_overview_liealg}

Fermion bilinears $T(\x,\y)$ form a basis for our Lie algebra, which we will call $\g$. A general element of this algebra is a linear combination,
\begin{equation}
    O_F \equiv \int d^dx d^dy ~ F(\x,\y) T(\x,\y) \sim i \int d^dx d^dy ~ F(\x,\y) \psi^\dagger(\x) \psi(\y)\, ,
\end{equation}
where $F(\x,\y)$ is a generic function of two variables. It will be more convenient for us to work with the Wigner transform of the generators:
\begin{equation}
    T(\x,\p) \equiv \int d^d y ~ T \left( \x + \frac{\y}{2}, \x - \frac{\y}{2} \right) e^{i \p \cdot \y}\, ,
\end{equation}
in which case, a general element of the Lie algebra,
\begin{equation}\label{eq_gen_bilin}
    O_F \equiv \int \frac{d^dx d^dp}{(2\pi)^d} F(\x,\p) T(\x,\p)\, ,
\end{equation}
is characterized instead by a function $F(\x,\p)$ of coordinates and momenta instead. The function $F(\x,\p)$ can be thought of as the components of the Lie algebra vector $O_F$, with $\x,\p$ being indices. Since we have already picked a preferred basis for $\g$, we will often refer to the the function $F$ itself as the Lie algebra vector by a slight abuse of terminology.

Using the anti-commutation relations for the fermion creation and annihilation operators, one can show that the commutator of two Lie algebra vectors corresponding to functions $F(\x,\p)$ and $G(\x,\p)$ takes the following form:
\begin{equation}
    [O_F, O_G] = O_{\mb{F}{G}}\, ,
\end{equation}
where the operation in the subscript of the right hand side is the Moyal bracket of two functions,
\begin{equation}\label{eq_moyalbracket}
    \mb{F}{G}(\x,\p) \equiv 2 ~ F(\x,\p) \sin \left( \frac{\overleftarrow{\nabla}_\x \cdot \overrightarrow{\nabla}_\p - \overleftarrow{\nabla}_\p \cdot \overrightarrow{\nabla}_\x}{2} \right) G(\x,\p)\, .
\end{equation}
Note that up until this point, all of our formulas are exact. So far we are working in the full quantum theory, despite the simultaneous occurrence of both position and momentum. This is essentially achieved by a quantization scheme that is different from but equivalent to canonical quantization, known as Weyl quantization (or deformation quantization for more general phase spaces).

Our Lie algebra can hence be characterized as the set of all functions of a single-particle phase space, equipped with the Moyal bracket,
\begin{equation}
    \g_\text{Moyal} \equiv  \left( \{ F(\x,\p) \}; \mb{.}{.} \right)\, .
\end{equation}
We will refer to this as the Moyal algebra or the Weyl algebra\footnote{The Weyl algebra is actually a subalgebra of the Moyal algebra, consisting of only polynomial functions.}. The associated Lie group consists of the exponents of the bilinear operators $e^{\mathcal{O}_F}$. The coadjoint orbit method can be applied directly to the Moyal algebra to yield a formal action that would in principle exactly describe Fermi surfaces, but this action is unwieldy in practice, owing to the fact that the Moyal bracket in equation \eqref{eq_moyalbracket} is only defined in a power series in phase space derivatives, with convergence of the power series having been established only for limited classes of functions \cite{waldmann2019convergence}.

To ameliorate this issue, we can consider a truncation of the Moyal algebra to leading order in the series expansion, which gives the Poisson bracket,
\begin{equation}
    \begin{split}
        \mb{F}{G} &= \{ F, G \} + \mathcal{O}(\nabla_\x,\nabla_\p)^3\, ,\\
        \{ F, G \} &\equiv \nabla_\x F \cdot \nabla_\p G - \nabla_\p F \cdot \nabla_\x G\, ,
    \end{split}
\end{equation}
providing an approximate, semi-classical, action-based description of Fermi liquids via the coadjoint orbit method applied to the truncated Lie algebra of the set of functions of a single-particle phase space, equipped with the Poisson bracket instead of the Moyal bracket,
\begin{equation}
    \g = ( \{ F(\x,\p) \}; \{ . , . \} )\, .
\end{equation}
We will refer to this as the Poisson algebra. Importantly, this is the only truncation of the Moyal algebra that preserves the Jacobi identity. We emphasize that the Poisson algebra is \textit{not} a sub-algebra of the Moyal algebra, but rather a truncation of the Lie bracket.

The Poisson algebra has a useful physical interpretation that can be assigned to it: it is the Lie algebra of infinitesimal canonical transformations of the single-particle phase space. A typical element $F(\x,\p)$ of the Poisson algebra generates a canonical transformation in the following way: we can define new coordinates,
\begin{equation}
    \begin{split}
        \x' &= \x - \nabla_\p F\, ,\\
        \p' &= \p + \nabla_\x F\, .
    \end{split}
\end{equation}
We can verify that the transformed coordinates $\x',\p'$ are canonical pairs. This transformation can be understood as Hamiltonian evolution for infinitesimal time under the Hamiltonian $F(\x,\p)$, and we can also verify that the commutator of two such infinitesimal transformations parametrized by functions $F(\x,\p)$ and $G(\x,\p)$ is an infinitesimal transformation parametrized by the Poisson bracket $\{F,G\}(\x,\p)$. The quickest way to see this is to note that the infitesimal transformation is generated by the phase space vector field:
\begin{equation}
    X_F = \nabla_\x F \cdot \nabla_\p - \nabla_\p F \cdot \nabla_\x\, ,
\end{equation}
and then evaluating the commutator of two vector fields $[X_F, X_G]$ viewed as differential operators acting on test functions. It is not hard to see that
\begin{equation}
    [X_F,X_G] \cdot K(\x,\p) = X_{\{F,G\}} \cdot K(\x,\p)\, ,
\end{equation}
for any function $K(\x,\p)$.

The corresponding Lie group is naturally that of canonical transformations under finite time. For each element $F(\x,\p) \in \g$ of the Poisson algebra, we will define the exponent map, denoted by $\exp$ that associates with $F$ the canonical transformation $U$ obtained by time evolving under $F$ for unit time. The set of all such $U$'s is the group of canonical transformations that we are concerned with (known in the math literature as the group of Hamiltonian symplectomorphisms),
\begin{equation}
    \G \equiv \{ U = \exp F ~|~ F \in \g \}\, .
\end{equation}
Note that the exponent map $\exp F$ from the Lie algebra to the Lie group is different from the point-wise exponential of the function $e^F = 1 + F + F^2/2 + \ldots$. To avoid confusion, we will restrict ourselves to using $\exp$ for the Lie-algebra-to-Lie-group exponent map instead of writing it as $e^F$.

The truncation of the Moyal algebra to the Poisson algebra is subtle and requires some more scrutiny. We will revisit this in section \ref{sec_mic_ham} and clarify the consequences of this truncation, including a discussion on which properties this approximation succesfully captures and which ones it misses out on.

Having understood the operator algebra of concern, we now move on to describing the corresponding space of states that we will be interested in.


\subsection{The space of states}\label{sec_states_overview}

In any quantum mechanical system, states are described by density matrices $\rho$, which can be thought of as linear maps acting on operators to give the expectation value of the operator in the chosen state,
\begin{equation}
    \rho[\mathcal{O}] \equiv \< \mathcal{O} \>_\rho = \Tr(\rho \mathcal{O})\, .
\end{equation}
In principle, if we have access to every operator in the theory, each state is uniquely determined by the list of expectation values of every operator in that state. But since we are only concerned with the subalgebra of charge-neutral fermion bilinears, we inevitably end up being unable to distinguish all microscopic states from each other, but instead are restricted to equivalence classes of microscopic states, where equivalence is established by requiring identical expectation values of all fermion bilinears.

A typical representative of any such equivalence class can be described as follows. Having chosen the basis $T(\x,\p)$ for the space of fermion bilinear, we can pick a dual basis to it, which we will denote by operators $W(\x,\p)$, which have the orthogonality property:
\begin{equation}
    \Tr\left( W(\x',\p') T(\x,\p) \right) = \delta(\x-\x') (2\pi)^d \delta(\p-\p')\, .
\end{equation}
A representative of the equivalence class of states can be expanded in this dual basis with the `coefficients' given by a function of $\x,\p$,
\begin{equation}
    \rho_f = \int \frac{d^dx d^dp}{(2\pi)^d} f(\x,\p) W(\x,\p)\, .
\end{equation}
In this state, the expectation value of a bilinear operator $O_F$ simplifies to
\begin{equation}\label{eq_op_expect}
    \Tr(\rho_f O_F) = \int \frac{d^dxd^dp}{(2\pi)^d} f(\x,\p) F(\x,\p) \equiv \< f, F \>\, .
\end{equation}

Naturally, this set of equivalence classes is the set of linear maps from $\g_\text{Moyal}$ to $\mathbb{C}$, also known as the dual space of $\g_\text{Moyal}$, which we will denote by $\g^*$.
\begin{equation}
    \begin{split}
        \g^* &\equiv \{ f(\x,\p) \}\, ,\\
        f[F] &\equiv \< f, F \> \equiv \int \frac{d^dxd^dp}{(2\pi)^d} f(\x,\p) F(\x,\p)\, ,
    \end{split}
\end{equation}
where the second line defines the action of the linear map $f$ on an element $F$ of $\g_\text{Moyal}$. Note that the dual space is independent of the Lie bracket. Hence, the Moyal algebra and the Poisson algebra share the same dual space $\g^*$.

Ordinarily in physics, vector spaces and their dual spaces are not distinguished between, since they are isomorphic to each other for finite dimensional vector spaces. However, for our purposes we find it crucial to make this pedantic distinction, since the Lie algebra and its dual space will take different physical interpretations and consequently will be equipped with different mathematical structures later.

That the expectation values of operators $O_F$ in a state $\rho_f$ can be written in the form of equation \eqref{eq_op_expect} provides the following interpretation for the functions $F(\x,\p)$ and $f(\x,\p)$ in the semiclassical limit: the function $F(\x,\p)$ that characterizes the linear combination of fermion bilinears will be understood as a single-particle observable, while the function $f(\x,\p)$ characterizing the state is the effective single-particle phase space distribution function (or simply the distribution for brevity) that enters the Boltzmann equation. This connection to the Boltzmann equation will become more precise as we develop the Hamiltonian formalism later in section \ref{sec_mic_ham}, whose equation of motion in the semi-classical limit is precisely the collisionless Boltzmann equation (or Landau's kinetic equation for interacting Fermi liquids).

The pairing or innder product $\< f, F \>$ between elements of $\g^*$ and $\g$ is then just the average value of the single-particle observable $F(\x,\p)$ in the distribution $f(\x,\p)$.


\subsection{Schematic overview of the coadjoint orbit method}

Equipped with the Lie algebra $\g$ consisting of single-particle observables and its dual space $\g^*$ consisting of distribution functions, the coadjoint orbit method provides us an algorithm to derive an action for our theory in broadly two steps.

First, we set up a dynamical system describing time evolution on $\g^*$ via a prescribed Hamiltonian. The choice of Hamiltonian must be governed by microscopics as well as principles of effective field theory, especially since we want to describe the theory via the truncated Poisson algebra instead of the exact Moyal algebra. We will see that these considerations allow us to automatically obtain Landau's kinetic equation for interacting Fermi liquids as the equation of motion, along with systematic higher order corrections to Landau's phenomenological theory.

Second, we attempt to Legendre transform the Hamiltonian into an action. Performing this Legendre transform is a highly non-trivial task, since it turns out that we have to restrict our state space $\g^*$ further in order to achieve this. This restriction, however, is natural, since the set of all possible configurations of the distribution function $f(\x,\p)$ is too large of a set to describe sharp Fermi surfaces at zero temperature. We need only consider functions that take values of either 0 or 1, with the boundary between the two values being the Fermi surface. These functions must also have fixed phase space volume due to Luttinger's theorem. It turns out that restricting $\g^*$ to such states is precisely what is needed to Legendre transform the Hamiltonian to an action. This restriction, therefore, is both physically motivated and mathematically necessary, and we will find that Luttinger's theorem is automatically built into our formalism.

Consequently, the postmodern formalism for Fermi liquids essentially describes the dynamics of a fluctuating codimension one surface in phase space whose topology is $\R^d\times S^{d-1}$, i.e. that of a sphere at every point $\x$ (figure \ref{fig_wobbly_fermi}).

The next two sections are devoted to the two respective steps described above, and a survey of the necessary approximations and consequent validity/invalidity of these steps.


\newpage
\section{The operator algebra and the Hamiltonian formalism}\label{sec_mic_ham}

Before developing the Hamiltonian formalism, we first survey the algebra of fermion bilinears more carefully. We will make a small modification to our definition of the generators and define them instead in center of mass and relative coordinates as
\begin{equation}
    T(\x,\y) \equiv \frac{i}{2} \left[ \psi^\dagger\left(\x+\frac{\y}{2}\right) \psi\left(\x-\frac{\y}{2}\right) - \psi\left(\x-\frac{\y}{2}\right) \psi^\dagger\left(\x+\frac{\y}{2}\right) \right]\, .
\end{equation}
Canonical anti-commutation relations for the fermion operators $[\psi(\x),\psi^\dagger(\y)]_+ = i \delta(\x-\y)$ implies that this definition only differs from equation \eqref{eq_generator_old} by a delta function which serves to regulate the coincidence limit $T(\x,0)$. Furthermore, the Hermitian conjugate takes the form,
\begin{equation}
    T^\dagger(\x,\y) = - T(\x,-\y)\, .
\end{equation}The various Fourier transforms of this generator will be useful for later:
\begin{equation}\label{eq_ph_bilin_config}
    \begin{split}
        T(\x,\y) &\equiv \frac{i}{2} \left[ \psi^\dagger\left(\x+\frac{\y}{2}\right) \psi\left(\x-\frac{\y}{2}\right) - \psi\left(\x-\frac{\y}{2}\right) \psi^\dagger\left(\x+\frac{\y}{2}\right) \right]\, ,\\
        T(\q,\p) &\equiv \frac{i}{2} \left[ \psi^\dagger\left(\frac{\q}{2}+\p\right) \psi\left(\frac{\q}{2}-\p\right) - \psi\left(\frac{\q}{2}-\p\right) \psi^\dagger\left(\frac{\q}{2}+\p\right) \right]\, ,\\
        T(\x,\p) &\equiv \int_\y T(\x,\y) e^{i\p\cdot\y} = \int_\q T(\q,\p) e^{-i \q\cdot \x}\, ,\\
        T(\q,\y) &\equiv \int_{\x,\p} T(\x,\p) e^{i\q\cdot\x} e^{-i\p\cdot\y} = \int_\x T(\x,\y) e^{i\q\cdot\x} = \int_\p T(\q,\p) e^{-i\p\cdot\y}\, ,
    \end{split}
\end{equation}
where integrals over momenta $\q$ and $\p$ are defined with an implicit factor of $1/(2\pi)^d$.

Our convention for the fermion annihilation operator $\psi(\k)$ in momentum space is that $\psi(\k)$ is simply the Fourier transform of $\psi(\x)$. When acting on the Fermi surface it creates a state with momentum $\k$. Therefore, it creates a hole at the point $-\k$ in the Fermi sea. This is different from the usual convention in condensed matter physics, where the annihilation operator $c_\k$ is defined so that it creates a hole at the point $\k$, thereby creating a state with total momentum $-\k$.

\begin{figure}[t]
    \centering
    \includegraphics[width=7cm]{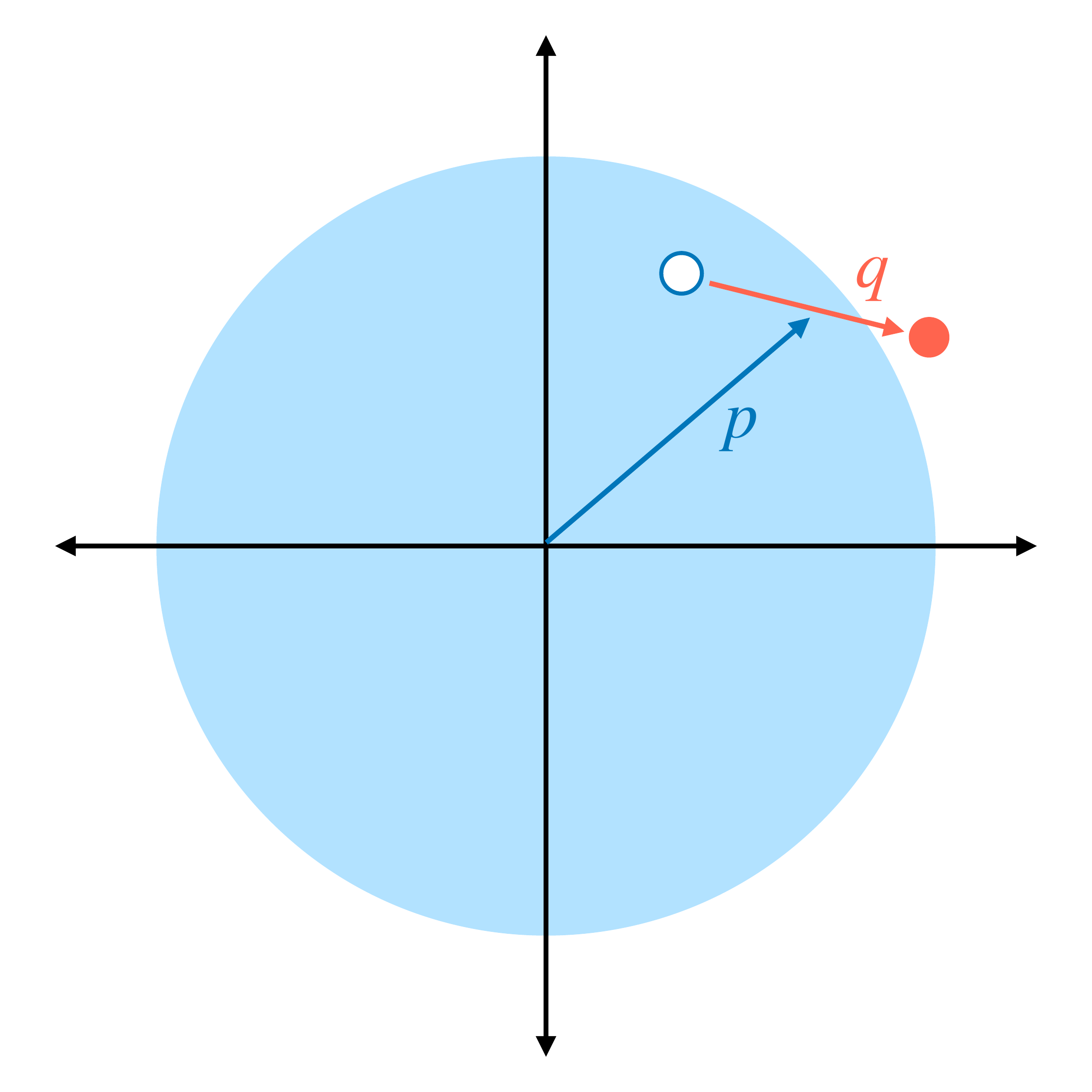}
    \caption{Particle-hole configuration in the parametrization of equation \eqref{eq_ph_bilin_config}.}
    \label{fig_ph_bilin_config}
\end{figure}

It is worth emphasizing that in the notation we have chosen above, $\x$ is the center of mass coordinate of the particle-hole pair described by the fermion bilinear, $\y$ is the relative coordinate or the separation between them. Analogously, given that $\psi(\k)$ creates a hole with momentum $-\k$, the Fourier conjugate $\q$ to the center of mass coordinate $\x$ measures the momentum of the particle-hole pair, which is the difference of the individual momenta of the particle and hole. The Fourier conjugate $\p$ to the separation $\y$ is the average of the individual momenta of the particle and the hole, so the average location of the particle hole pair in momentum space (figure \ref{fig_ph_bilin_config}). We shall restrict ourselves to using this notation convention throughout this thesis, so the arguments of the generator and their specific order should make it clear to which Fourier transform we are referring.

All of the above Fourier transforms are traceless in a fermionic Fock space. Additionally, our definitions imply that $T(\x,\p)$, in particular, is anti-Hermitian,
\begin{equation}
    T^\dagger(\x,\p) = - T(\x,\p)\, .
\end{equation}
The commutator of these generators closes, and we find
\begin{equation}\label{eq_generator_moyal_alg}
    \begin{split}
        [T(\q,\y), T(\q',\y')] &= 2 \sin \left( \frac{\q'\cdot\y - \q\cdot\y'}{2} \right) T(\q+\q',\y+\y')\, ,\\
        [T(\x,\p), T(\x',\p')] &= 2 \sin \left( \frac{\nabla_\x\cdot\nabla_{\p'} - \nabla_{\x'}\cdot\nabla_\p}{2} \right) \left[ \delta(\x-\x') \delta(\p-\p') T(\x,\p) \right]\, .
    \end{split}
\end{equation}
The coefficient functions or differential operators on the right-hand side are the ``structure constants'' of the Lie algebra $\g_\text{Moyal}$, whose typical element is a general linear combination
\begin{equation}
    O_F \equiv \int_{\x\p} F(\x,\p) T(\x,\p)\, ,
\end{equation}
where $F(\x,\p)$ is an arbitrary function, to be thought of a the set of coefficients of the vector $O_F$ in the basis $T(\x,\p)$, with $(\x,\p)$ playing the role of ``incidces'' in this expansion. This results in the Moyal bracket for the commutator of generic linear combinations,
\begin{equation}
    \begin{split}
        [O_F, O_G] &= O_{\mb{F}{G}}\, ,\\
        \mb{F}{G} &= 2 ~ F \sin \left( \frac{ \overleftarrow{\nabla}_\x \cdot \overrightarrow{\nabla}_\p - \overleftarrow{\nabla}_\p \cdot \overrightarrow{\nabla}_\x }{2} \right) G\, .
    \end{split}
\end{equation}
Our generators also obey orthogonality relations:
\begin{equation}
    \begin{split}
        \Tr[T(\x,\p) T(\x',\p')] &= 2 \delta(\x-\x') (2\pi)^d \delta(\p-\p')\, ,\\
        \Tr [T(\q,\y) T(\q',\y')] &= 2 (2\pi)^d \delta(\q+\q') \delta(\y+\y')\, ,
    \end{split}
\end{equation}
where the trace is taken in the fermionic Fock space.

The space of all charge-0 bosonic operators hence forms an infinite dimensional Lie algebra, known as the Moyal algebra. We will restrict ourselves to a class of microscopic Hamiltonians that can be expanded in a polynomial expansion in the generators of this algebra,
\begin{equation}\label{eq_Hmicro}
    \begin{split}
        H_\text{micro} &= \int_\p \epsilon(\p) \psi^\dagger(\p) \psi(-\p)\\
        &+ \int_{\p_1,\p_2,\p_3,\p_4} V(\p_1,\p_2,\p_3,\p_4) \psi^\dagger(\p_1) \psi(\p_2) \psi^\dagger(\p_3) \psi(\p_4) \delta(\p_1 + \p_2 + \p_3 + \p_4)\\
        &+ \mathcal{O}(\psi^\dagger\psi)^3\, ,
    \end{split}
\end{equation}
where $\epsilon(\p)$ is the free particle dispersion, $V(\p_1,\p_2,\p_3,\p_4)$ characterizes $2\rightarrow2$ scattering processes, and so on for higher order terms.


\subsection{Semi-classical truncation of the Moyal algebra}

While the discussion so far has been exact, in practice, using the Moyal algebra can be extremely tedious since the star product and the Moyal bracket are defined as series expansions. A remedy for this is provided by the Poisson truncation discussed in section \ref{sec_overview_liealg},
\begin{equation}
    \begin{split}
        \mb{F}{G} &= \{ F, G \} + \mathcal{O}(\nabla_\x,\nabla_\p)^3\, ,\\
        \{ F, G \} &\equiv \nabla_\x F \cdot \nabla_\p G - \nabla_\p F \cdot \nabla_\x G\, .
    \end{split}
\end{equation}
The Poisson bracket is, in fact, the only truncation of the Moyal bracket that satisfies the Jacobi identity. This truncation, however, comes at a cost, and limits the validity of the theory to regimes where the Poisson bracket is a good approximation to the Moyal bracket. This is only true when
\begin{equation}
    \nabla_\x \cdot \nabla_\p \ll 1\, ,
\end{equation}
which can be rephrased in three other ways by Fourier transforming $\x$ and/or $\p$:
\begin{equation}\label{eq_Poisson_limit}
    \nabla_\x \cdot \nabla_\p \ll 1\, , \quad \Leftrightarrow \quad \q\cdot\y \ll 1\, , \quad \Leftrightarrow \quad \nabla_\x \cdot \y \ll 1\, , \quad \Leftrightarrow \quad  \q\cdot\nabla_\p \ll 1\, .
\end{equation}
Recall that $\x$ corresponds to the center of mass coordinate of a particle-hole pair, $\y$ is the separation, $\q$ measures the net momentum of the particle-hole excitation, and $\p$ is the average of the momenta of the particle and the hole. With these in mind, equation \eqref{eq_Poisson_limit} implies that the Poisson truncation of the Moyal algebra of fermion bilinears is applicable in situations where we have a separation of scales, with $(\x,\q)$ characterizing the long distance or infrared (IR) scale, and $(\y,\p)$ characterizing the short distance or ultraviolet (UV) scale.

In position space, this means that we are restricting ourselves to probing physics at length-scales much larger than the typical separation of a particle-hole pair. In momentum space, a typical particle-hole excitation over a Fermi surface has $|\p|\sim p_F$, and the Poisson truncation is valid for pairs whose net momentum is much smaller than that, i.e.,
\begin{equation}
    |\q| \ll p_F\, .
\end{equation}
The corrections to the Poisson truncation can then be thought of as a derivative expansion with higher derivatives terms being suppressed owing to the fact that
\begin{equation}\label{eq_Poisson_limit_FS}
    \nabla_\x \cdot \nabla_\p \sim \frac{|\nabla_\x|}{p_F} \ll 1\, .
\end{equation}
With this analysis in mind, let us try to understand what consequences the Poisson truncation has for interactions between the fermions. We will consider the quartic term in the microscopic Hamiltonian, which can be written in the following way:
\begin{equation}
    \begin{split}
        H_\text{micro}^\text{int} &= \int_{\q,\p;\q',\p'} V(\q,\p;\q',\p') \psi^\dagger\left( \frac{\q}{2} + \p \right) \psi\left( \frac{\q}{2} - \p \right) \psi^\dagger\left( \frac{\q'}{2} + \p' \right) \psi\left( \frac{\q'}{2} - \p' \right) \delta(\q+\q')\\
        &\simeq \int_{\q,\p;\q',\p'} V(\q,\p;\q',\p') T(\q,\p) T(\q',\p') \delta(\q+\q')\, ,
    \end{split}
\end{equation}
where the symbol $\simeq$ means that we have ignored the quadratic terms generated upon replacing $\psi^\dagger(\k_1)\psi(\k_2)$ with its antisymmetrized version.

The above Hamiltonian characterizes $2\rightarrow2$ scattering processes. In general, the momenta $(\q,\p;\q',\p')$ could take any values allowing for generic scattering configurations on the Fermi surface. However, the semi-classical limit captures those configurations with $|\p|\sim|\p'|\sim p_F$, and $\q,\q'\ll p_F$. This corresponds to particle-hole pairs close to the Fermi surface with small net momentum, such as the configuration in figure \ref{fig_forward_scat}. Higher derivative corrections to the semiclassical limit then systematically account for particle-hole pairs with a larger separation in momentum space.


\subsection{Constructing the Hamiltonian formalism}

To recap the discussion in section \ref{sec_overview}, we find a Lie algebra in the operator algebra, whose generators are fermion bilinears $T(\x,\p)$, whose structure constants can be read off from the commutation relations,
\begin{equation}
    \begin{split}
        [T(\q,\y),T(\q',\y')] &= 2 \sin \left( \frac{\q'\cdot\y - \q\cdot\y'}{2} \right) T(\q+\q',\y+\y')\\
        &= \left( \q'\cdot\y-\q\cdot\y' \right) T(\q+\q',\y+\y') + \mathcal{O}(\q,\y)^3\, .
    \end{split}
\end{equation}
The pair $(\q,\y)$ or its Fourier conjugate $(\x,\p)$ can be thought of as a Lie algebra index. Generic elements of the Lie algebra are linear combinations of the generators,
\begin{equation}
    O_F = \int_{\x,\p} F(\x,\p) T(\x,\p)\, ,
\end{equation}
characterized by functions $F(\x,\p)$. The commutator of two such functions specifies the Lie bracket,
\begin{equation}
    [O_F,O_G] = O_{\mb{F}{G}} = O_{\{F,G\}} + \mathcal{O}(\nabla_\x,\nabla_\p)^3\, ,
\end{equation}
and we can succinctly define the (truncated) Lie algebra as the set of functions of $\x$ and $\p$ equipped with the Poisson bracket:
\begin{equation}
    \begin{split}
        \g &\equiv \{ F(\x,\p) \}\, ,\\
        \{ F, G \} &= \nabla_\x F \cdot \nabla_\p G - \nabla_\p F \cdot \nabla_\x G\, .
    \end{split}
\end{equation}
The corresponding Lie group consists of the set of exponentials $e^{O_F}$ of the operators $O_F$, and in the semi-classical limit takes on the interpretation of canonical transformations $U$ of the single-particle phase space $\R^{2d}$ generated by the function $F$ viewed as a Hamiltonian.
\begin{equation}
    \G \equiv \{ U = \exp F ~|~ F \in \g \}\, .
\end{equation}
We also saw in section \ref{sec_states_overview} that the space of states was given by the dual space $\g^*$, whose elements are also functions $f(\x,\p)$ which are interpreted as quasiprobability distribution functions, which act on elements of the Lie algebra to give the average value of a single-particle observable $F(\x,\p)$ in the state $f(\x,\p)$.
\begin{equation}
    \begin{split}
        \g^* &\equiv \{ f(\x,\p) \}\, ,\\
        \< f, F \> &\equiv \int_{\x,\p} F(\x,\p) f(\x,\p)\, .
    \end{split}
\end{equation}
$\g^*$ is the effective phase space for Fermi liquids and we need to define a Hamiltonian and a Poisson structure on this to get an equation of motion. In order to do so, let us first define the action of the Lie group and Lie algebra on the Lie algebra and its dual space.


\subsubsection{Adjoint and coadjoint representations}

The Lie bracket furnishes a natural action of the Lie algebra on itself, known as the Lie algebra adjoint action:
\begin{equation}
    \begin{split}
        \ad_F ~ &: ~ \g \rightarrow \g\, ,\\
        \ad_F G &\equiv \{ F, G \}\, .
    \end{split}
\end{equation}
This can be exponentiated to obtain an action of the Lie group on the Lie algebra, called the Lie group adjoint action:
\begin{equation}
    \begin{split}
        \Ad_U ~ &: ~ \g \rightarrow \g\, ,\\
        \Ad_{U=\exp F} G \equiv U G U^{-1} \equiv e^{\ad_F} G &= G + \{ F, G \} + \frac{1}{2!} \{ F, \{ F, G \} \} + \ldots\, ~~ .
    \end{split}
\end{equation}
We will often use $UGU^{-1}$ as alternate notation for the adjoint action to make it clear that intuition from quantum mechanics (and matrix Lie groups) applies more or less straightforwardly to our case as well.

The action of the Lie group and Lie algebra on the Lie algebra are called the adjoint representation.

From the above, we can also define the action of the Lie algebra and Lie group on the dual space $\g^*$, known as the coadjoint actions:
\begin{equation}
    \begin{split}
        \ad^*_F, ~ \Ad^*_U ~ &: ~ \g^* \rightarrow \g^*\, ,\\
        \ad^*_F f &\equiv \{ F, f \}\, ,\\
        \Ad^*_{U=\exp F} f \equiv U f U^{-1} \equiv e^{\ad^*_F} f &= f + \{ F, f \} + \frac{1}{2!} \{ F, \{ F, f \} \} + \ldots\, ~~ .
    \end{split}
\end{equation}
Together these define the coadjoint representation.


\subsubsection{Lie-Poisson structure and Hamiltonian}

Next, we need a Poisson structure for functionals of $\g^*$. This requires a bilinear map that takes in two functionals $\sF[f]$ and $\sG[f]$, and spits out a third functional $\sH[f]$ in a way consistent with the product rule as well as with the Jacobi identity. Such a structure is provided by the Lie-Poisson bracket, defined as follows:
\begin{equation}\label{eq_LP_Poisson}
    \{ \sF, \sG \}_\text{LP}[f] \equiv \left\< f, \left\{ \delta\sF|_f, \delta\sG|_f \right\}_\text{Poisson} \right\>\, .
\end{equation}
The above formula is dense, so let us unpack it in a few sentences. $\g^*$ is a vector space. A typical point in this vector space is the function $f(\x,\p)$. Being a vector space, the tangent space $T_f\g^*$ to $\g^*$ at the point $f$ is isomorphic to $\g^*$. Therefore any tangent vector at a point in $\g^*$ can be equivalently thought of as an element of $\g^*$. Analogously, the cotangent space $T^*_f \g^*$ to $\g^*$ at the point $f$ is isomorphic to the space $\g^{**} \cong \g$ that is dual to $\g^*$, which is just the Lie algebra. So cotangent vectors at a point are elements of $\g$.

The variation $\delta \equiv \delta/\delta f$ of a functional $\sF$ is an exterior derivative of a function of $\g^*$. Therefore $\delta \sF$ is a cotangent field on $\g^*$. Its value $\delta\sF|_f$ at the point $f$, being a cotangent vector, is an element of the Lie algebra. The same holds for $\delta\sG|_f$. Since these are both elements of the Lie algebra, i.e., functions of $(\x,\p)$, we can take their Lie bracket, which in our case is the Poisson bracket. The resulting function, when paired with $f$ using our inner product, gives us the value of the Lie-Poisson bracket functional $\{ \sF, \sG \}_\text{LP}$ evaluated at the point $f$.

That the Lie-Poisson bracket obeys the product rule and Jacobi identity follows from the fact that the Poisson bracket itself obeys both.

All that remains is to construct a Hamiltonian functional $H[f]$. Instead of deriving this from the microscopic Hamiltonian in equation \eqref{eq_Hmicro}, we will use effective field theory to write down a Hamiltonian in a systematic expansion. We will assume translation and rotational invariance in the continuum limit, even though the requirement of rotational invariance can be relaxed further to account for materials with more complicated electronic Fermi surfaces.

Our Hamiltonian will take the form of a double expansion, one in nonlinearities in $f(\x,\p)$, and the other in spatial derivatives. The latter will be justified by the semi-classical limit \eqref{eq_Poisson_limit_FS}, since derivatives must be suppressed by the Fermi momentum. To justify the former, we must organize our Hamiltonian in a polynomial expansion in fluctuations around the ground state,
\begin{equation}
    f_0(\p) = \Theta(p_F - |\p|)\, .
\end{equation}
Defining fluctuations around this reference state as
\begin{equation}
    \delta f(\x,\p) \equiv f(\x,\p) - f_0(\p)\, ,
\end{equation}
we can write the most general effective Hamiltonian as follows
\begin{equation}\label{eq_Ham}
    \begin{split}
        H[f] &= \int_{\x\p} \epsilon(\p) f(\x,\p)\\
        &+ \frac{1}{2}\int_{\x\p\p'} F^{(2,0)}(\p,\p') \delta f(\x,\p) \delta f (\x,\p') + \mathbf{F}^{(2,1)}(\p,\p') \cdot \left( \frac{\nabla_\x}{p_F} \delta f(\x,\p) \right) \delta f (\x,\p') + \ldots\\
        &+ \frac{1}{3}\int_{\x\p\p'\p''} F^{(3,0)}(\p,\p',\p'') \delta f(\x,\p) \delta f(\x,\p') \delta f(\x,\p'') + \ldots\\
        &+ ~ \ldots\, ~~ .
    \end{split}
\end{equation}
In the above, $\epsilon(\p)$ is the free fermion dispersion relation and the various coefficient functions $F^{(m,n)}$ parametrize interactions. In our notation, the $m$-index of $F^{(m,n)}$ labels the nonlinearity of the interaction, while the $n$-index labels the number of $\x$-derivatives in that coupling. Of course, there can be multiple independent terms or order $(m,n)$ in which case additional indices are required to distinguish their coefficient functions. The various couplings $(\epsilon,F^{(m,n)})$ are functional analogues of Wilson coefficients in an effective field theory, and we will often refer to them as Wilson coefficients by a slight abuse of terminology, or Wilson coefficient functions if we want to be precise.


\subsubsection{Equation of motion}

Armed with the Lie-Poisson structure \eqref{eq_LP_Poisson} as well as the Hamiltonian \eqref{eq_Ham}, we can write down Hamilton's equation of motion for our system on $\g^*$,
\begin{equation}
    \d_t f = \{ f, H \}_\text{LP}[f]\, .
\end{equation}
The Lie-Poisson bracket can be evaluated from its definition in terms of the Poisson bracket, by using the fact that $\delta f(\x,\p)/\delta f(\x',\p') = \delta(\x-\x')\delta(\p-\p')$ and integrating by parts, to obtain
\begin{equation}
    \d_t f(t,\x,\p) + \left\{ f(t,\x,\p), \frac{\delta H}{\delta f(t,\x,\p)} \right\}_\text{Poisson} = 0\, .
\end{equation}
The variation of the Hamiltonian can be calculated straightforwardly, and defines the quasiparticle dispersion relation,
\begin{equation}
    \epsilon_\text{qp}[f] \equiv \frac{\delta H}{\delta f} = \epsilon(\p) + \int_{\p'} F^{(2,0)}(\p,\p') \delta f (t,\x,\p') + \ldots\, ~~ ,
\end{equation}
in terms of which the equation of motion turns into Landau's kinetic equation \eqref{eq_landau_kinetic_eq}:
\begin{equation}\label{eq_ham_eom}
    \d_t f + \nabla_\p \epsilon_\text{qp}[f] \cdot \nabla_\x f - \nabla_\x \epsilon_\text{qp}[f] \cdot \nabla_\p f = 0\, .
\end{equation}
We see that $F^{(2,0)}(\p,\p')$ is simply Landau's interaction function, but we also find an infinite series of higher order corrections to the quasiparticle energy.

The study of the algebra of fermion bilinears, paired with EFT philosophy, hence provides a a formalism that captures LFLT as well as higher derivative corrections to LFLT in a systematic expansion.

Note that the formalism and equation of motion itself applies generally to any state $f(\x,\p)$, irrespective of whether it describes the excitations of a Fermi surface at zero temperature. The only place that the Fermi surface has entered in this discussion so far is in justifying the series expansion of the Hamiltonian \eqref{eq_Ham}. For other systems, a different choice of Hamiltonian should suffice, as long as time evolution in such a system can be described by canonical transformations.


\subsubsection{An alternate route to the Hamiltonian formalism}

An alternate way to arrive at the Hamiltonian formalism described in this section, without relying on the algebra of fermion bilinears, is the following:

Landau's kinetic equation is simply a non-linear modification of the collisionless Boltzmann equation. Time evolution as determined by the collisionless Boltzmann equation not only preserves volume in the single-particle phase space, as shown by Liouville's theorem, but also preserves the symplectic form (or equivalently Poisson brackets) in the single-particle phase space. This implies that any solution $f(t,\x,\p)$ to the collisionless Boltzmann equation can be described as the action of a one-parameter family of canonical transformations, parametrized by $t$, acting on the initial state $f(t=0,\x,\p)$.

The dynamical system described by the collisionless Boltzmann equation is hence equivalent to a dynamical system on the Lie group of canonical transformations, since the solutions to the equations of motion are simply curves on the group manifold. The method described in the above section is a well-established method to formulate dynamical systems on Lie groups \cite{Kirillov_book,ArnoldKhesin}, and hence automatically applies to our case \cite{MarsdenWeinstein:1982}. This formalism requires a prescribed Hamiltonian to describe time evolution, and the most natural one is the double expansion \eqref{eq_Ham}. As we have already seen, this immediately gives us LFLT at the equation of motion.

In \cite{main:2022}, this was the perspective that was primarily presented in the main body, with the connection to fermion bilinears being relegated to the appendices. In this section, we have instead surveyed in detail the more microscopic approach to constructing the Hamiltonian, with the aim to clarify the connection to microscopics as well as expound upon what approximations and assumptions are required at the microscopic level in order to obtain this effective description. While we have largely appealed to EFT philosophy in order to construct the effective Hamiltonian \eqref{eq_Ham}, it remains to see whether it is possible to derive the effective Hamiltonian for certain classes of microscopic Hamiltonians such as the ones in equation \eqref{eq_Hmicro}, using the properties of the fermion bilinear algebra.


\newpage
\section{Effective action from the coadjoint orbit method}\label{sec_coadj_ac}

The second step towards obtaining an action description for Fermi liquids is to Legendre transform the Hamiltonian. Let us briefly described how this is usually achieved for a Hamiltonian system on a general phase space manifold $\Gamma$, equipped with some choice of Poisson brackets. Defining $\d_I$ as a derivative on the phase space manifold, the Poisson bracket of two functions $F$ and $G$ on $\Gamma$ can always be locally written in the following way:
\begin{equation}
    \{ F, G \} = \Pi^{IJ} \d_I F \d_J G\, ,
\end{equation}
where $\Pi^{IJ}$ is an anti-symmetric rank 2 tensor on $\Gamma$, known as the Poisson bi-vector. To switch from a Hamiltonian formalism to an action formalism, we invert the Poisson bivector to obtain a closed, anti-symmetric, non-degenerate symplectic form:
\begin{equation}
    \omega = \Pi^{-1}\, , \qquad \omega_{IJ} \Pi^{JK} = \delta_I^K
\end{equation}
The symplectic form allows us to write down a `$p\dot{q}$' term in the following way: introduce an extra dimension $s\in[0,1]$ in addition to time $t$ so that $s=1$ corresponds to physical time, and use boundary conditions in $s$ so that all degrees of freedom vanish at $s=0$. Let $\phi^I$ be coordinates on phase space, i.e., the phase space degree of freedom. The $p\dot{q}$ term is then given by
\begin{equation}
    \int dt \int_0^1 ds ~ \omega (\d_t \phi, \d_s \phi) = \int dt \int_0^1 ds ~ \omega_{IJ} \d_t \phi^I \d_s \phi^J\, ,
\end{equation}
with an additional spatial integral involved if $\phi^I$ are fields in space\footnote{The symplectic form is closed ($d\omega=0$) by definition, or as a consequence of the Jacobi identity for the Poisson bracket. This implies that the $\p\dot{q}$ term is independent of the choice of ``bulk'' extension.}. The Legendre transform of the Hamiltonian $H[\phi]$ is then
\begin{equation}
    S = \int dt \int_0^1 ds ~ \omega (\d_t \phi, \d_s \phi) - \int dt H[\phi]\, .
\end{equation}
This entire construction relies on the ability to invert the Poisson bi-vector. However, this invertibility is, in general, not guaranteed by the definition of the Poisson bracket, and when it is not, we cannot find an action that gives the same equation of motion without changing the phase space either by finding a description in terms of different degrees of freedom or by eliminating redundant ones. This is the case for the Hamiltonian formalism described in section \ref{sec_mic_ham}, so the Legendre transformation is not as straightforward as we could have hoped for. Before describing the remedy for this barrier, let us first revisit the microscopic description of the space of states from section \ref{sec_states_overview}.


\subsection{Fermi surface states and their excitations}

To recap the discussion in section \ref{sec_states_overview}, the space of states $\g^*$ is given by the vector space dual to the algebra of fermion bilinears. These are equivalence classes of density matrices that cannot be distinguished by the expectation values of fermion bilinears. A typical representative of such an equivalence class is characterized by the distribution function $f(\x,\p)$ in the following way:
\begin{equation}
    \rho_f = \int_{\x,\p} f(\x,\p) W(\x,\p)\, ,
\end{equation}
where $W(\x,\p)$ is the basis dual to $T(\x,\p)$, defined by
\begin{equation}
    \Tr[W(\x,\p) T(\x',\p')] = \delta(\x-\x') (2\pi)^d \delta(\p-\p')\, .
\end{equation}
The expectation value of a general operator $O_F = \int_{\x\p} F(\x,\p) T(\x,\p)$ in the state $\rho_f$ can be written as
\begin{equation}
    \< O_F \>_{\rho_f} = \Tr[\rho_f O_F] = \int_{\x,\p} F(\x,\p) f(\x,\p) = \< f, F \>\, ,
\end{equation}
and the distribution function $f_\rho(\x,\p)$ that represents any given state $\rho$ itself can be obtained from the state as
\begin{equation}
    f(\x,\p) = \left\< T(\x,\p) \right\>_\rho\, .
\end{equation}
Of course, this is generically true for any (pure or mixed) state, not just states with a Fermi surface. The distribution functions corresponding to these correspond to a subset of $\g^*$.

Consider, for instance, a spherical Fermi surface with Fermi momentum $p_F$. The state that describes is a pure state obtained by filling every momentum within the spherical Fermi surface with a fermion,
\begin{equation}
    \ket{\text{FS}} = \prod_{|\k|\le p_F} \psi^\dagger(\k) \ket{0}\, ,
\end{equation}
where $\ket{0}$ is the vacuum. It is straightforward to show using fermion anticommutation relations that
\begin{equation}
    f_0(\p) = \braket{\text{FS}|T(\x,\p)|\text{FS}} = \frac{1}{2} \sign(p_F - |\p|)\, .
\end{equation}
For later convenience, let us define instead the distribution function of a state as
\begin{equation}
    f(\x,\p) = \< T(\x,\p) \>_\rho + \frac{1}{2}\, ,
\end{equation}
so that
\begin{equation}
    f_0(\p) = \Theta(p_F-|\p|)\, ,
\end{equation}
is the occupation number function for a spherical Fermi surface\footnote{This shift is equivalent to saying that the distribution is defined by the expectation value of the Wigner transform of $\psi^\dagger(\x_1)\psi(\x_2)$, instead of it anti-Hermitian part.}. This shift also ensures that the integral used to define the pairing $\< f, F \>$ converges for states with a sharp Fermi surface, since the domain of integration is effectively bounded in momentum space.

Excitations on top of the Fermi surface take the form of particle-hole pairs, which are created by the action of fermion bilinears on $\ket{\text{FS}}$. A state with a single particle hole excitation is then given by
\begin{equation}
    \ket{\k_1;\k_2} \equiv \psi^\dagger(\k_1)\psi(-\k_2) \ket{\text{FS}}\, .
\end{equation}
Fermion anticommutation relations ensure that this state is different from $\ket{\text{FS}}$ only if $\k_1 \notin \text{FS}$ and $\k_2\in \text{FS}$. Antisymmetrizing over the particle and the hole to regulate the coincidence singularity $\k_1\rightarrow \k_2$, and Wigner transforming allows us to write such states in an alternate basis:
\begin{equation}
    \ket{\x;\p} \equiv T(\x,\p) \ket{\text{FS}}\, .
\end{equation}
In the semi-classical limit, where $|\nabla_\x|\ll \p \sim p_F$, the state $\ket{\x;\p}$ is interpreted as a particle hole pair created at the point $\p$ on the Fermi surface, locally in a mesoscopic region of size $1/p_F$ at the position labelled by the spatial coordinate $\x$. The momentum $\p$ has no relation to the net momentum $\q$ of the particle-hole pair, and only labels on which `patch' of the Fermi surface the particle-hole pair lives.

Another equivalent basis that will be more convenient for us is that of coherent states defined as
\begin{equation}
    \ket{F(\x,\p)} \equiv e^{\int_{\x\p} F(\x,\p) T(\x,\p)} \ket{\text{FS}}\, ,
\end{equation}
whose distribution function is given by the following:
\begin{equation}
    f_F(\x,\p) = f_0(\p) + \mb{F}{f_0} + \frac{1}{2!} \mb{F}{\mb{F}{f_0}} + \ldots\, ~~ .
\end{equation}
This is just the coadjoint action of $F(\x,\p)$ on $f_0(\p)$ in the Moyal algebra! The set of unitary operators $U_F=e^{\int F T}$ form the corresponding group and we find that particle-hole coherent states of a Fermi surface is obtained by the action of all possible group transformations on the spherical Fermi surface. This applies to the parametrization of the states in terms of their distribution functions as well, in that the distribution function for a particle-hole coherent state is obtained by acting on the spherical Fermi surface distribution with a group transformation.

\begin{figure}[t]
    \centering
    \includegraphics[width=17cm]{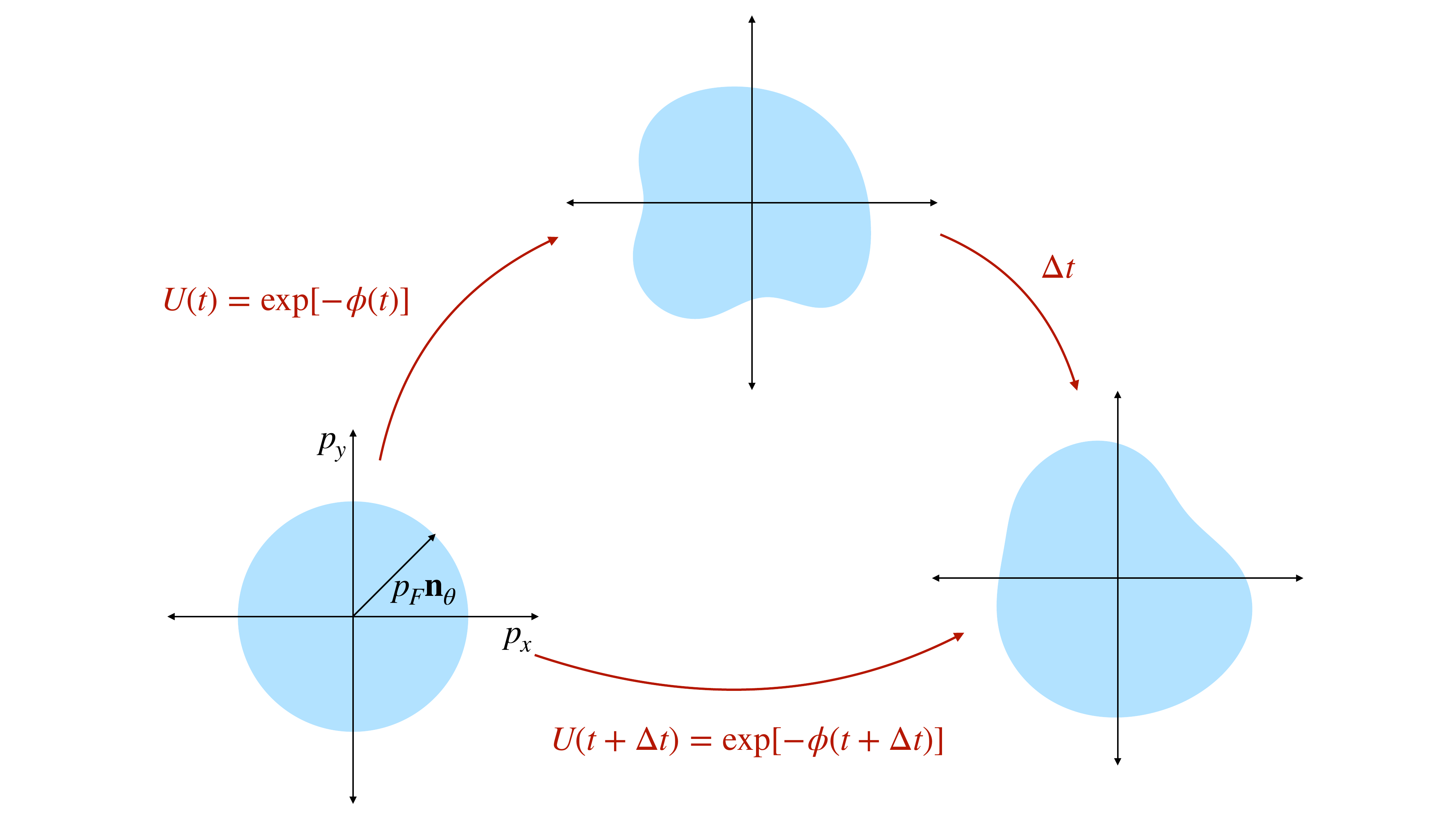}
    \caption{Fermi surface states from canonical transformations}
    \label{fig_fermi_surface}
\end{figure}

In the semi-classical limit, the Moyal brackets are replaced by Poisson brackets and the semi-classical distribution function for a coherent state is given by
\begin{equation}
    f_F = \Ad^*_{\exp F} f_0 = f_0 + \{ F, f_0 \} + \frac{1}{2!} \{ F, \{ F, f_0 \} \} + \ldots\, ~~ ,
\end{equation}
which is interpreted as the action of the canonical transformation $U=\exp F$ on the spherical Fermi surface state. An intuitive picture for this is the following: take all the points within the Fermi surface. The canonical transformation $U$ maps each one of these to a new point. Being a smooth coordinate transformation, this preserves the proximity of points and transforms the initial spherical swarm of points into a new shape that is topologically equivalent to a filled sphere (see figure \ref{fig_fermi_surface}). The precise shape of boundary of this region can be parametrized by a function $p_F(\x,\theta)$, where $\theta$ are angular coordinates in momentum space. We then have
\begin{equation}
    f_F(\x,\p) = \Theta(p_F(\x,\theta) - |\p|)\, ,
\end{equation}
which is entirely characterized by a shape in phase space. The space of states for particle-hole excitations is then just the space of closed surfaces in phase space \cite{CastroNetoFradkin:1994}.

This space of states is described mathematically by what is called a coadjoint orbit, which we define below.


\subsubsection{Coadjoint orbits and the Kirillov-Kostant-Souriau form}
As we saw above, the space of states relevant for zero temperature Fermi surface physics is not all of $\g^*$, but a subset of it consisting of functions that take values $1$ or $0$ separated by a closed surface. This restriction is formally achieved by picking a reference state, $f_0(\p)$ in our case, and acting on it via all possible canonical transformations. Canonical transformations act on $\g^*$ via the coadjoint action, so the set generated from this procedure is known as the coadjoint orbit of $f_0$:
\begin{equation}
    \mathcal{O}_{f_0} \equiv \{ f=\Ad^*_U f_0 \in \g^* ~|~ U \in \G \}\, .
\end{equation}
Two different canonical transformations acting on the same reference state can indeed generate the same element of the coadjoint orbit, owing to the fact that there is a nontrivial subgroup that leaves $f_0$ invariant, called the stabilizer subgroup of $f_0$, which we will denote by $\H$.
\begin{equation}
    \begin{split}
        \H &\equiv \{ V \in \G ~|~ \Ad^*_V f_0 = f_0 \}\\
        &= \{ V = \exp \alpha ~|~ \alpha \in \g, ~ \ad^*_\alpha f_0 = 0 \}\, .
    \end{split}
\end{equation}
So the canonical transformations $U$ and $UV$ create the same state from $f_0$, since
\begin{equation}
    \Ad^*_{UV} f_0 = UV f_0 (U_V)^{-1} = U (V f_0 V^{-1}) U^{-1} = \Ad^*_U f_0\, .
\end{equation}
Each state $f$ in the coadjoint orbit is hence represented by a left coset $U\H$, and the coadjoint orbit is then the left coset space,
\begin{equation}
    \mathcal{O}_{f_0} \cong \G/\H\, .
\end{equation}
Since every element of the coadjoint orbit is related to every other by canonical transformations, we find an important result for time evolution under any Hamiltonian $H[f]$. Infinitesimal time evolution occurs by the action of the infinitesimal canonical transformation $\delta H|_f \in \g$, while finite time evolution occurs by exponentiating the sequence of infinitesimal canonical transformations, which itself is a canonical transformation. Therefore, time evolution takes an initial state to another state in the \textit{same coadjoint orbit} as the initial state.

The coadjoint orbit $\mathcal{O}_{f_0}$ is hence preserved by time evolution, and can hence be thought of as a reduced phase space for Fermi liquids. The Hamiltonian and Lie-Poisson structure can both be restricted to the coadjoint orbit with complete consistency, and the entire Hamiltonian formalism can be defined solely for $\mathcal{O}_{f_0}$ instead of all of $\g^*$.

Unlike the Lie-Poisson structure for $\g^*$, however, the Lie-Poisson structure restricted to $\mathcal{O}_{f_0}$ is invertible, and permits the definition of a closed, non-degenerate symplectic form, known as the Kirillov-Kostant-Souriau (KKS) form. Being a 2-form, it is defined by its action on a pair of vectors tangent to the coadjoint orbit at any given point.

Consider the point $f\in\mathcal{O}_{f_0}$. Since the coadjoint orbit is a submanifold of $\g^*$, the tangent space $T_f \mathcal{O}_{f_0}$ to $\mathcal{O}_{f_0}$ at the point $f$ is a subspace of the tangent space $T_f\g^*$ to $\g^*$. Tangent vectors of $\g^*$ can be thought of as elements of $\g^*$, so defining the KKS form amounts to defining its action $\omega_\text{KKS}(g,k)$ on any two arbitrary functions $g,k\in \g^*$ which are tangent to $\mathcal{O}_{f_0}$.

It can be shown that the tangents $g$ and $k$ at the point $f$ can be obtained from the coadjoint action of two Lie algebra elements $G,K\in\g$ on $f$ (see, for instance, \cite{ArnoldKhesin}), i.e.,
\begin{equation}
    \ad^*_G f = g, \qquad \ad^*_K f = k\, .
\end{equation}
$G$ and $K$ are not uniquely determined by $g$ and $k$ respectively, but rather representatives of equivalence classes of Lie algebra elements. The KKS form is then defined in terms of $G$ and $K$ as follows:
\begin{equation}
    \omega_\text{KKS}(g,k) \equiv \left\< f, \{G,K\}_\text{Poisson} \right\>\, .
\end{equation}
The pairing of the Poisson bracket with $f$ makes it clear that any other choice of representative of the equivalence classes of $G$ and $K$ respectively gives the same answer, using the fact that if $G$ and $G'$ are two elements of the same equivalence class, then $\ad^*_{G-G'} f = 0$. To show that the KKS form is closed, note that the differential $d\omega_\text{KKS}$ acts on three instead of two tangents, and it is not difficult to show that
\begin{equation}
    d\omega_\text{KKS} (g,k,l) = \left\< f, \{ \{ G, K \}, L \} \right\> + \text{cyclic permutations}\, ,
\end{equation}
where $L\in \g$ is such that $\ad^*_L f = l \in \g^*$. The right hand side then vanishes due to the Jacobi identity.

Armed with the Kirillov form, we can formally write down an action for Fermi liquids in terms of the field $f\in\mathcal{O}_{f_0}$, which looks like
\begin{equation}\label{eq_formal_action}
    \begin{split}
        S_\text{FL}[f] &= S_\text{WZW}[f] - \int dt ~ H[f]\, ,\\
        S_\text{WZW}[f] &= \int dt \int_0^1 ds ~ \omega_\text{KKS} \left( \d_t f, \d_s f \right)\, ,
    \end{split}
\end{equation}
where $S_\text{WZW}$ is the Wess-Zumino-Witten (WZW) term, $H[f]$ is the Hamiltonian in equation \eqref{eq_Ham}, and $f$ obeys the following boundary conditions on the $(t,s)$-strip:
\begin{equation}
    f(t,s=1) = f(t)\, , \qquad f(t,s=0) = 0\, .
\end{equation}


\subsection{The Wess-Zumino-Witten term and the effective action}

The action \eqref{eq_formal_action}, while exact (in the semi-classical limit corrected by the derivative expansion) is written in a rather formal way that cannot really be used for calculations. In order to make it more useful, we need to find a convenient parametrization of the coadjoint orbit. The simplest one is obtained directly from the definition of the orbit, i.e., by acting on the reference state $f_0$ by all possible canonical transformations, generated by the field $-\phi(\x,\p) \in \g$. In this parametrization, the field $\phi(\x,\p)$ is our degree of freedom. The minus sign is conventional and chosen for later convenience.

Elements $f(\x,\p)$ of the coadjoint orbit can be parametrized as follows:
\begin{equation}
    \begin{split}
        f_\phi (\x,\p) = \Ad^*_{\exp (-\phi)} f_0 &= f_0 + \{ \phi, f_0 \} + \frac{1}{2!} \{ \phi, \{\phi, f_0 \} \} + \ldots\\
        &= \Theta(p_F-|\p|) + (\n\cdot\nabla_\x \phi) \delta(|\p|-p_F) + \ldots\, ~~ ,
    \end{split}
\end{equation}
where $\n$ is the unit normal to the spherical Fermi surface at the angular coordinates $\theta$ in momentum space.

The stabilizer $\H$ of $f_0$ can be described by its Lie subalgebra $\h$ which corresponds to functions $\alpha(\x,\p) \in \g$ that obey the following condition:
\begin{equation}
    \begin{split}
        \ad^*_\alpha f_0 = \{ \alpha, f_0 \} = 0\, ,\\
        \implies (\n\cdot\nabla_\x \alpha)|_{|\p|=p_F} = 0\, .
    \end{split}
\end{equation}
Consequently, the canonical transformation $\exp \alpha$ leaves $f_0$ invariant,
\begin{equation}
    \Ad^*_{\exp \alpha} f_0 = e^{\ad^*_\alpha} f_0 = f_0\, .
\end{equation}
The equivalence $U\simeq UV$ then leads to an equivalence relation for $\phi$,
\begin{equation}\label{eq_stab_gauge}
    \begin{split}
        \exp(-\phi) &\simeq \exp(-\phi) \exp(\alpha)\, ,\\
        \implies \phi &\simeq \phi - \alpha + \frac{1}{2} \{ \phi, \alpha \} + \ldots\, ~~ ,
    \end{split}
\end{equation}
which allows us to ``gauge fix'' $\phi$ to be independent of the radial momentum coordinate,
\begin{equation}
    \phi = \phi(\x,\theta)\, ,
\end{equation}
where $\theta$ are angular coordinates in momentum space. A suitable choice of $\alpha$ that achieves this, for example, at leading order in the transformation \eqref{eq_stab_gauge}, is
\begin{equation}
    \alpha_\phi(\x,\p) = \phi(\x,\p) - \phi(\x,\theta)|_{|\p|=p_F}\, .
\end{equation}
It is easy to check that $\{ \alpha_\phi, f_0 \} = 0$. While we use a $|\p|$-independent parametrization of our degree of freedom for convenience, any other choice is equally valid and will result in the same physical quantities, with the various choices being related by field redefinitions.

What remains is to write down the WZW term in terms of this field to obtain an action description for Fermi liquids. The definition of the KKS form requires that we find functions $G$ and $K$ such that
\begin{equation}
    \ad^*_G f_\phi = \d_t f\, , \qquad \ad^*_K f_\phi = \d_s f\, .
\end{equation}
Using the fact that $f_\phi = \Ad^*_U f_0 = U f_0 U^{-1}$ where $U = \exp(-\phi)$, we can show that the required functions are\footnote{The simplest way to do this is to pretend that $U,f_0,f$ are all matrices, replace all Poisson brackets with matrix commutators, simplify the expressions and finally replace all commutators back with Poisson brackets.}
\begin{equation}
    G = \d_t U U^{-1}\, , \qquad K = \d_s U U^{-1}\, ,
\end{equation}
so that the KKS form evaluates to
\begin{equation}\label{eq_kks_wzw}
    \begin{split}
        \omega_\text{KKS}(\d_t f, \d_s f) &= \left\< f, \{ \d_t U U^{-1}, \d_s U U^{-1} \} \right\>\\
        &= \left\< f_0, \{ U^{-1}\d_t U, U^{-1} \d_s U \} \right\>\, ,
    \end{split}
\end{equation}
with boundary conditions $\phi(t,s=1)=\phi(t)$ and $\phi(t,s=0)=0$. The above expression can be simplified to a sum of total $s$- and $t$-derivatives, which allows us the write the WZW term as
\begin{equation}
    S_\text{WZW} = \int dt ~ \< f_0, U^{-1} \d_t U \>\, .
\end{equation}
This is a subtle point, since it suggests that the KKS form is necessarily exact, which is not true generally for a Lie group, especially for a coadjoint orbit with non-trivial topology. Since the group of canonical transformations is a diffeomorphism group, the topology of its coadjoint orbits is unknown, and it is unclear whether the KKS form on the coadjoint orbit $\mathcal{O}_{f_0}$ is exact or not.

The expression \eqref{eq_kks_wzw}, on the other hand, is exact, owing to the fact that we are describing a generic point $f$ on the coadjoint orbit as a canonical transformation $U$ acting on the reference state $f_0$. Furthermore, we are restricting ourselves to canonical transformations that are connected to the identity by expressing $U$ as the exponent of a Lie algebra element $-\phi$. This parametrization of the coadjoint orbit is hence incomplete, and only captures the largest possible patch of the coadjoint orbit around $f_0$, missing out on information about disconnected components of the orbit as well as the global topology of the component containing $f_0$.

This choice of parametrization suffices, however, to describe a perturbative expansion around the reference state $f_0$, since all states accessible to such a perturbative expansion necessarily live in a patch around $f_0$, making the choice of the reference state somewhat crucial for this method to work. To account for nonperturbative properties of Fermi liquids, a different parametrization of the coadjoint orbit is required, which we leave to future work.

Finally, we obtain a perturbative action that describes Fermi liquids,
\begin{equation}\label{eq_U_action}
    S_\text{FL} = \int dt ~ \< f_0, U^{-1}\d_t U \> - \int dt ~ H[f_\phi=U f_0 U^{-1}]\, ,
\end{equation}
with $U = \exp(-\phi)$. The action can be expanded order by order in $\phi$, and we will find that higher order terms are suppressed by powers of $p_F$, which takes on the role of the UV cutoff of the theory.

Of course, since this action is just the Legendre transformation of the Hamiltonian \eqref{eq_Ham}, the equation of motion is guaranteed to be equation \eqref{eq_ham_eom}. But this can also be verified directly by varying the action under
\begin{equation}
    U \rightarrow U' = \exp \delta \phi \cdot U\, ,
\end{equation}
with $\delta\phi(t,\x,\p) \in \g$. To linear order in $\delta \phi$, we have
\begin{equation}
    \delta[U^{-1}\d_t U] = U^{-1} (\d_t \delta\phi) U\, , \qquad \delta H[Uf_0 U^{-1}] = \left\< f_\phi, \{ \epsilon_\text{qp}[f_\phi], \delta\phi \}_\text{Poisson} \right\>\, ,
\end{equation}
where $\epsilon_\text{qp}[f] = \delta H/\delta f$ is the quasiparticle energy. This gives us the following result for the variation of the action:
\begin{equation}
    \delta S = - \int dt \left\< \d_t f_\phi + \left\{ f_\phi, \epsilon_\text{qp}[f_\phi] \right\}_\text{Poisson}, \delta \phi \right\>\, ,
\end{equation}
from which we can read off the equation of motion,
\begin{equation}
    \d_t f_\phi + \left\{ f_\phi, \epsilon_\text{qp}[f_\phi] \right\} = 0\, ,
\end{equation}
which is, as expected, identical to equation \eqref{eq_ham_eom}.


\subsection{Symmetries in the postmodern formalism}\label{sec_symmetries}

This geometric perspective for Fermi liquids, in part, powerful because of how it encodes symmetries through the algebra of canonical transformations. We will categorize the symmetries we want to introduce into the formalism into three different groups: spacetime, gauge and internal symmetries. The last of these three requires an extension of the algebra of canonical transformations and will hence be dealt with later in section \ref{sec_spin}.

Let us first discuss some key aspects of how symmetries act in the postmodern formalism, and focus in particular on the unintuitive consequences of the fact that the algebra of canonical transformations is in fact a diffeomorphism algebra as opposed to a global symmetry algebra.

Recall that the coadjoint orbit $\mathcal{O}_{f_0} \cong \G/\H$ is the left coset space of the group of canonical transformations. Therefore every state $f\in \mathcal{O}_{f_0}$ is identical to an equivalence class of canonical transformations under the equivalence relation,
\begin{equation}
    U \simeq UV\, , \quad V \in \H\, .
\end{equation}
The explicit map from $\G/\H$ to $\mathcal{O}_{f_0}$ is given by\footnote{The discussion below equation \eqref{eq_kks_wzw} of the subtlety of not being able to capture every state in the coadjoint orbit does not apply here since we are not requiring $U\in G$ to be the exponent of any Lie algebra element.}
\begin{equation}
    f_U \equiv U f_0 U^{-1}\, .
\end{equation}
Now the group of canonical transformations $G$ can itself act on the coset in one of two different ways, called the left and right actions, respectively given by the transformations
\begin{equation}
    U \xrightarrow{\text{left}} WU\, , \qquad U \xrightarrow{\text{right}} UW\, , \qquad W \in \G\, .
\end{equation}
Both of these induce transformations on the coadjoint orbit as follows:
\begin{equation}
    f_U \xrightarrow{\text{left}} W f_U W^{-1}\, , \qquad f_U \xrightarrow{\text{right}} UW f_0 W^{-1} U^{-1}\, ,
\end{equation}
but only the left action can be naturally and directly written as a transformation of $\G$ on the coadjoint orbit, independent of the choice of reference state $f_0$. Therefore symmetries must act on the coset space via the left action. The right action instead is reserved for transformations by elements $V$ of the stabilizer $\H$, resulting in a \textit{coset redundancy} that is a gauge symmetry of our theory (not to be confused with the gauge symmetry when we couple to background $U(1)$ gauge fields later). The WZW term is invariant under a larger gauge symmetry of all canonical transformations under the right action, since these simply pick out a different reference state to parametrize the coadjoint orbit, but the Hamiltonian breaks this $\G$ gauge symmetry down to a $\H$ gauge symmetry by uniquely picking $f_0$ as the ground state.

Note also that the WZW term is invariant under the left action of every canonical transformation that does not depend on time, since
\begin{equation}
    (WU)^{-1} \d_t (WU) = U^{-1} \d_t U\, ,
\end{equation}
but the Hamiltonian is not. The rule of thumb for imposing symmetries on this theory will be the following:
\begin{itemize}
    \item Identify the subalgebra of canonical transformations that generates the symmetry
    \item If the symmetry being considered is a spacetime symmetry, impose invariance of the action under the transformation $U\rightarrow WU$
    \item If the symmetry in consideration is a gauge symmetry, turn on background fields that make the state $f$ invariant under the transformation $W f W^{-1}$.
\end{itemize}

The last point is unusual and not how we typically gauge a theory, and will be discussed in more detail later. But before imposing any symmetry on our theory, let us describe a global symmetry that does not act on the state $f$, but is instead a consequence of our choice of parametrization of the coadjoint orbit. Recall that we chose to define the canonical transformation $U$ that generates $f$ as the exponent of a Lie algebra element,
\begin{equation}
    U = \exp(-\phi)\, , \qquad \phi(\x,\p) \in \g\, .
\end{equation}
Elements of the Lie algebra have a symmetry built into them, which corresponds to constant shifts\footnote{The more mathematically inclined reader might worry that in order for the pairing $\langle f, F\rangle$ between $\g^*$ and $\g$ to be well-defined, suitable boundary conditions need to be imposed on functions which a constant shift would violate. However, this shift symmetry can be interpreted as a transformation of the boundary conditions to make the pairing well-defined.}:
\begin{equation}\label{eq_global_u1}
    \phi(\x,\p) \rightarrow \phi(\x,\p) + c\, .
\end{equation}
These shifts preserve the action of the canonical transformation on any state, since $f_\phi$ only depends on $\phi$ through its derivatives. While such shifts leave $f$ invariant, they will not leave the WZW term invariant if $c$ is promoted to a function of time, and it is not difficult to show that
\begin{equation}
    \delta S_\text{WZW} = \int dt ~ \< f, \d_t c(t) \> = - \int dt ~ \< \d_t f, c(t) \>\, .
\end{equation}
Noether's theorem then tells us that we must then have
\begin{equation}
    \d_t \int_{\x,\p} f(\x,\p) = 0\, ,
\end{equation}
i.e., the total particle number,
\begin{equation}
    N = \int_{\x,\p} f(\x,\p)\, ,
\end{equation}
is conserved.


\subsubsection{Galilean invariance}

As an example of a spacetime symmetry, let us demonstrate how invariance under Galilean boosts constrains our action. The first step is to identify the subalgebra of canonical transformations that generates Galilean boosts. A typical elements of this algebra is given by the time-dependent function,
\begin{equation}
    B_v = \v\cdot(\p t - m \x)\, ,
\end{equation}
with $W = \exp B_v$ being the corresponding canonical transformation. Under this transformation, we have
\begin{equation}
    f(\x,\p) \rightarrow (\Ad^*_W f)(\x,\p) = f(\x-\v t, \p - m \v  )\, ,
\end{equation}
as can be obtained by observing that the expansion of the coadjoint action takes the form of a Taylor series and then resumming the Taylor series.

Let us first evaluate the constraint on the free fermion action obtained from Galilean invariance. The action can be written as follows:
\begin{equation}
    S_\text{free fermion} = \int dt \left\< f_0, U^{-1} \d_t U \right\> - \int dt \left\< f, \epsilon \right\>\, .
\end{equation}
The WZW term transforms to
\begin{equation}
    \left\< f_0, U^{-1} W^{-1} \d_t (WU) \right\> = \left\< f_0, U^{-1}\d_t U \right\> + \left\< f, W^{-1} \d_t W \right\>\, ,
\end{equation}
while the Hamiltonian term becomes
\begin{equation}
    \left\< W f W^{-1}, \epsilon \right\> = \left\< f, W^{-1} \epsilon W \right\>\, ,
\end{equation}
so the change in the action is given by the following
\begin{equation}
    \delta S = \int dt \left\< f, W^{-1}(\d_t - \epsilon) W - \epsilon \right\>\, .
\end{equation}
Invariance under boosts then requires that
\begin{equation}
    W^{-1} \d_t W = W^{-1} \epsilon W - \epsilon\, ,
\end{equation}
where $W^{-1} \epsilon W = \Ad^*_{W^{-1}} \epsilon = \epsilon(\p+m\v)$ owing to the fact that $W^{-1} = \exp (-B_v) = \exp B_{-v}$. The left hand side can now be expanded using the following formula,
\begin{equation}
    W^{-1} \d_t W = \d_t B_v + \frac{1}{2!} \{ \d_t B_v, B_v \} + \ldots\, ~~ ,
\end{equation}
and compared order by order in $\v$ with the Taylor expansion of the right hand side to obtain the following:
\begin{equation}
    \p = m \nabla_\p \epsilon
\end{equation}
which tells us that the dispersion relation must be quadratic:
\begin{equation}
    \epsilon(\p) = \frac{p^2}{2m} + \text{constant}\, .
\end{equation}
This exactly what is expected for a free fermion with Galilean invariance. Next, we derive the effective mass of Landau quasiparticles by imposing Galilean invariance on the interacting theory truncated to quadratic order in the fluctuation $\delta f = f - f_0$:
\begin{equation}
    H[f] = \int_{\x\p} \epsilon(\p) f(\x,\p) + \frac{1}{2} \int_{\x\p\p'} F^{(2,0)}(\p,\p') \delta f(\x,\p) \delta f(\x,\p') + \mathcal{O}(\delta f^3, \nabla_\x)\, .
\end{equation}
We have already seen that the transformation of the WZW term under a Galilean boost is cancelled by the transformation of a linear-in-$f$ Hamiltonian term with the dispersion $\epsilon = p^2/2m$. Therefore, invariance of the interacting theory can be achieved by demanding invariance of the shifted Hamiltonian:
\begin{equation}
    \tilde{H}[f] = \int_{\x\p} \left( \epsilon(\p) - \frac{p^2}{2m} \right) f(\x,\p) + \frac{1}{2} \int_{\x\p\p'} F^{(2,0)}(\p,\p') \delta f(\x,\p) \delta f(\x,\p') + \mathcal{O}(\delta f^3, \nabla_\x)\, .
\end{equation}
To obtain constraints from boost invariance, it suffices to consider infinitesimal transformations,
\begin{equation}
    f\rightarrow f + \{ B_v, f \} + \mathcal{O}(v^2) = f - \v\cdot (t\nabla_\x + m \nabla_\p) f + \mathcal{O}(v^2)\, ,
\end{equation}
under which the fluctuation transforms as
\begin{equation}
    \delta f \rightarrow - ~ m\v\cdot\nabla_\p f_0 + \delta f - \v\cdot(t\nabla_\x + m\nabla_\p) \delta f\, .
\end{equation}
Note that the transformation of the fluctuation $\delta f$ is inhomogeneous in $\delta f$. In particular, it can reduce the degree of a monomial by up to 1. This results in constraints that mix the various Wilson coefficient functions, so that $F^{(m,n)}$ will be constrained by $F^{(m-1,n)}$.

The transformation of the shifted Hamiltonian under a boost is given by
\begin{equation}
    \begin{split}
        \tilde{H} \rightarrow \tilde{H} &- m\v\cdot\int_{\x\p} \left( \epsilon - \frac{p^2}{2m} \right) \nabla_\p f_0\\
        &+ m\v\cdot\int_{\x\p} \left( \nabla_\p \epsilon - \frac{\p}{m} - \int_{\p'} F^{(2,0)}(\p,\p')\nabla_{\p'} f_0(\p') \right) \delta f(\x,\p)\\
        &+ \mathcal{O}(\delta f^2, \nabla_\x)\, .
    \end{split}
\end{equation}
Rotational invariance kills the term in the first line, while the second line gives us a non-trivial constraint,
\begin{equation}
    \nabla_\p \epsilon - \frac{\p}{m} = \int_{\p'} F^{(2,0)}(\p,\p') \nabla_{\p'} f_0(\p')\, , \qquad ||\p|-p_F| \ll p_F\, .
\end{equation}
The requirement of $\p$ being sufficiently close to $p_F$ comes from the fact that $\delta f$ must be localized near the Fermi surface for a perturbative expansion in $\delta f$ to be valid. It suffices to set $\p$ to a point $p_F \n$ on the Fermi surface and write $\nabla_\p \epsilon|_{p_F} = p_F \n / m^*$, where $m^*=p_F/v_F$ is the effective mass of the quasiparticle. Furthermore, the $\nabla_{\p'} f_0$ term in the integral sets $\p'$ to be on the Fermi surface as well, and we can expand the Landau interaction function in angular channels using rotational covariance. For example, in $d=2$, we write
\begin{equation}
    F^{(2,0)}(p_F\n,p_F\n') = \frac{8\pi^2 v_F}{p_F^2} \sum_{l\ge 0} F_l \cos l(\theta - \theta')\, ,
\end{equation}
to simplify the boost invariance constraint to
\begin{equation}
    p_F \left( \frac{1}{m^*} - \frac{1}{m} \right)\n = - v_F F_1 \n\, .
\end{equation}
Solving for the effective mass in terms of the Galilean boost parameter $m$ and the first Landau parameter $F_1$, we find the known result:
\begin{equation}
    m^* = m(1+F_1)\, .
\end{equation}


\subsubsection{Coupling to $U(1)$ gauge fields}

As mentioned briefly before, the procedure for coupling our theory to background gauge fields is very different from the usual procedure of gauging a global symmetry. A systematic procedure for coupling Fermi liquids to a gauge field has been difficult to achieve in the past owing to the fact that effective theories live in momentum space, and here we present a new approach that provides a solution.

The key observation is that the set of gauge transformations, characterized by functions $\lambda(t,\x)$, forms a subalgebra of infinitesimal canonical transformations. All such functions Poisson-commute with each other, since they do not depend on $\p$, so this subalgebra is abelian. It is not difficult to show that under the canonical transformation $W=\exp\lambda$, we have
\begin{equation}
    (\Ad^*_W f)(\x,\p) = f(\x,\p+\nabla_\x \lambda)\, .
\end{equation}
These then act on the coset representative $U = \exp(-\phi)$ as
\begin{equation}
    U \rightarrow W U\, , \qquad \phi \rightarrow \phi - \lambda + \frac{1}{2} \{ \lambda, \phi \} + \ldots\, ~~ .
\end{equation}
The above transformation makes it clear why the usual procedure of gauging the global $U(1)$ symmetry \eqref{eq_global_u1} by promoting the transformation to depend on space and time is ambiguous when applied to the current theory, since simply promoting the transformation parameter to a function misses out on the nonlinear corrections in the Baker-Campbell-Haussdorff formula. The minimal coupling procedure then is blind to nonlinear couplings to the gauge field as well as contact terms required to ensure gauge invariance.

Naturally, the Fermi liquid action is not invariant under these transformations, so we need to turn on background gauge fields $A_\mu(t,\x)$ that transform under the gauge transformation as
\begin{equation}
    A_\mu (t,\x) \rightarrow W^{-1}(A_\mu - \d_\mu) W = A_\mu (t,\x) - \d_\mu \lambda (t,\x)\, ,
\end{equation}
where $\mu = (t,\x)$ is a spacetime index. The WZW term and the Hamiltonian can be made invariant separately under gauge transformations. Let us start with the WZW term, whose transformation is given by
\begin{equation}
    \begin{split}
        U^{-1} \d_t U  &\rightarrow U^{-1} \d_t U + U^{-1} (W^{-1} \d_t W) U\, ,\\
        \implies \delta_\lambda S_\text{WZW} &= \int dt \left\< f_0, U^{-1} (\d_t \lambda) U \right\>\, .
    \end{split}
\end{equation}
Evidently, making this invariant amounts to modifying it to the following:
\begin{equation}
    S_\text{WZW}[\phi,A_0] = \int dt \left\< f_0, U^{-1} (\d_t - A_0) U \right\>\, ,
\end{equation}
which is now invariant under the simultaneous transformation
\begin{equation}
    U \rightarrow WU\, , \qquad A_0 \rightarrow W^{-1}(A_0 - \d_t) W\, .
\end{equation}

Next, to make the Hamiltonian invariant, it suffices to ensure the invariance of $f$ under gauge transformations by coupling it to the background gauge fields. One can see that the appropriate modification is
\begin{equation}
    f_A(t,\x,\p) \equiv f(t,\x,\p+\A(t,\x))\, ,
\end{equation}
where $\A$ is the spatial part of the gauge field. Since $\x$ does not transform at all under the gauge transformation, the transformation of $\p$ is cancelled by the gauge transformation of $\x$. While $f_A$ is now gauge invariant, its spatial derivatives are not, since
\begin{equation}
    (\nabla_\x f)(\x,\p) \rightarrow (\nabla_\x f)(\x,\p+\nabla_\x\lambda) + \{ \nabla_\x\lambda, f \}(\x,\p+\nabla_\x\lambda)\, .
\end{equation}
But this is straightforwardly remedied by replacing partial derivatives by covariant derivatives:
\begin{equation}
    D_\x f \equiv \nabla_\x f - \{ \A, f \}\, .
\end{equation}
While $f$ transforms covariantly under canonical transformations, the fluctuation $\delta f = f - f_0$ does not, so it is convenient to re-expand the Hamiltonian in $f$ instead of $\delta f$, with modified Wilson coefficient functions $\tilde{F}^{(m,n)}$ that can be related straightforwardly to the original ones in equation \eqref{eq_Ham}. The modified gauge-invariant Hamiltonian is then
\begin{equation}
    \begin{split}
        H_\text{gauged}[f,\A] = H[f_A] &= \int_{\x\p} \epsilon(\p) f(\x,\p+\A)\\
        &+ \frac{1}{2} \int_{\x\p\p'} \tilde{F}^{(2,0)}(\p,\p') f(\x,\p+\A) f(\x,\p'+\A)\\
        &+ \frac{1}{2} \int_{\x\p\p'} \tilde{\mathbf{F}}^{(2,1)}(\p,\p') \cdot (D_\x f)(\x,\p+\A) f(\x,\p'+\A)\\
        &+ \ldots\, ~~ ,
    \end{split}
\end{equation}
and the gauge invariant action can be written as
\begin{equation}\label{eq_u1gauged_action}
    S[\phi;A_0,\A] = S_\text{WZW}[\phi,A_0] - \int dt ~ H_\text{gauged} [f_\phi, \A]\, .
\end{equation}
As a test of the validity of this procedure, let us work out the equation of motion for the gauged action for free fermions and show that it is just the gauged Boltzmann equation. The free fermion action can be written as
\begin{equation}
    S_\text{free}[\phi;A_0,\A] = \int dt \left\< f_0, U^{-1} \left[ \d_t - A_0 - \epsilon(\p-\A) \right] U \right\>\, .
\end{equation}
Under the variation $U\rightarrow \exp \delta\phi \cdot U$, we find
\begin{equation}
    \delta S_\text{free} = - \int dt \left\< \d_t f + \{ f, \epsilon(\p-\A) + A_0 \} , \delta \phi \right\> + \mathcal{O}(\delta\phi^2)\, ,
\end{equation}
which tells us that the equation of motion must take the form,
\begin{equation}
    \d_t f + \{ f, \epsilon(\p-\A) + A_0 \} = 0\, ,
\end{equation}
which, upon expanding the Poisson bracket and defining the group velocity $\v_\p[\A] = \nabla_\p\epsilon(\p+\A)$ reduces to
\begin{equation}
    \d_t f + \v_\p \cdot \nabla_\x f + v_\p^i \d_j A_i \d^j_\p f + \nabla_\x A_0 \cdot \nabla_\p f = 0\, .
\end{equation}
This does not look like the gauged Boltzmann equation, since it is an equation for a distribution function $f(\x,\p)$ that is not gauge invariant, i.e., is evaluated at the canonical momentum $\p$ instead of the gauge invariant momentum $\k = \p + \A$. To bring it to a more familiar form, we make a field redefinition,
\begin{equation}
    f_A(t,\x,\k) = f(t,\x,\k + \A)\, ,
\end{equation}
which turns the equation of motion into the familiar form of the gauged Boltzmann equation with the Lorentz force term:
\begin{equation}
    \d_t f_A + \v_\k\cdot\nabla_\x f_A + \left( \E\cdot\nabla_\k + F_{ij} v_\k^i \d_\k^j \right) f = 0\, ,
\end{equation}
where $v_\k=\nabla_\k\epsilon(\k)$ is the gauge invariant group velocity.


\subsubsection{Emergent symmetries}

Fermi liquids are known to have a tremendously large number of emergent symmetries \cite{Else:2020jln}, corresponding to the conservation of not only the total particle number, but also the particle number at every point on the Fermi surface. This is a consequence of the limited amount of phase space available for quasiparticles to scatter to at low energies. Free fermions have an even larger symmetry group, since the lack of interactions as well as conservation of momentum imply that the occupation number at every momentum is conserved.

These symmetries can be described in the coadjoint orbit formalism as well, by coupling to background gauge fields that make the action invariant under \textit{all} canonical transformations. We begin with the observation that the adjoint and coadjoint action of a general, time-dependent canonical transformation $W = \exp \lambda(t,\x,\p)$ can be written as a coordinate transformation,
\begin{equation}
    \begin{split}
        (\Ad_W F)(\x,\p) &= F(\x^W,\p^W)\, ,\\
        (\Ad^*_W f)(\x,\p) &= f(\x^W,\p^W)\, ,
    \end{split}
\end{equation}
where the transformed coordinates $\x^W$ and $\p^W$ are given by
\begin{equation}
    \begin{split}
        \x^W &= \x + W \nabla_\p W^{-1}\, ,\\
        \p^W &= \p - W \nabla_\x W^{-1}\, .
    \end{split}
\end{equation}
In order to make the action invariant under these, we will turn on background gauge fields in phase space $A_0(t,\x,\p)$, $\A_\x(t,\x,\p)$ and $\A_\p(t,\x,\p)$. $\A_\x$ and $\A_\p$ are the respectively the position and momentum components of the phase space gauge fields. Using $I=(\x,\p)$ to denote a phase space index, we require that the gauge fields transform in the following way:
\begin{equation}
    A_0 \rightarrow W^{-1} (A_0 - \d_t) W\, , \qquad A_I \rightarrow W^{-1}(A_I - \d_I) W\, .
\end{equation}
Unlike $U(1)$ gauge fields, these gauge fields are non-abelian. Making the action invariant under all canonical transformations, however, follows the same steps as for $U(1)$ gauge transformations. The WZW term gets modified to
\begin{equation}
    S_\text{WZW}[\phi;A_0] = \int dt \left\< f_0, U^{-1}[\d_t - A_0] U \right\>\, ,
\end{equation}
which is invariant under the transformation $U\rightarrow WU$ simultaneously with the gauge transformation of $A_0$. To make the Hamiltonian invariant, we look for a gauge invariant modification of the distribution $f$. It is not difficult to see that distribution function evaluated on shifted coordinates,
\begin{equation}
    f_A(\x,\p) = f(\x-\A_\p,\p+\A_\x)\, ,
\end{equation}
does the trick. That this new distribution is gauge invariant can be seen as follows. Define
\begin{equation}
    \tilde{A}_I = W^{-1}(A_I - \d_I) W = A_I(\x^{W^{-1}},\p^{W^{-1}}) - W^{-1} \d_I W\, .
\end{equation}
The transformation of the modified distribution is given by
\begin{equation}
    f_A(\x,\p) \rightarrow f_{\tilde{A}}(\x^W,\p^W) = f \left( \x^W - \tilde{\A}_\p(\x^W,\p^W), \p^W + \tilde{\A}_\x(\x^W,\p^W) \right)\, .
\end{equation}
Now, the gauged transformed $A_I$ evaluated at the transformed coordinates $(\x^W,\p^W)$ can be simplified in the following way:
\begin{equation}
    \tilde{A}_I(\x^W,\p^W) = W \tilde{A}_I(\x,\p) W^{-1} = W [W^{-1}(A_I - \d_I) W] W^{-1} = A_I(\x,\p) + W \d_I W^{-1}\, ,
\end{equation}
so that the arguments of $f$ after the transformation reduce to
\begin{equation}
    \begin{split}
        \x^W - \tilde{\A}_\p(\x^W,\p^W) &= \x + W\nabla_\p W^{-1} - \A_\p(\x,\p) - W\nabla_\p W^{-1} = \x - \A_\p(\x,\p)\, ,\\
        \p^W + \tilde{\A}_\x(\x^W,\p^W) &= \p - W\nabla_\x W^{-1} + \A_\x(\x,\p) + W\nabla_\x W^{-1} = \p + \A_\x(\x,\p)\, .
    \end{split}
\end{equation}
As a result, we find that the modified distribution is indeed gauge invariant:
\begin{equation}
    f_A(\x,\p) \rightarrow f(\x-\A_\p,\p+\A_\x) = f_A(\x,\p)\, .
\end{equation}
Phase space gradients of $f_A$, however, do not transform covariantly under canonical transformations, but covariant derivatives do,
\begin{equation}
    \begin{split}
        D_I f &\equiv \d_I f - \{ A_I, f \}\, ,\\
        (D_I f) &\rightarrow W (D_I f) W^{-1}\, ,
    \end{split}
\end{equation}
which we can then make invariant by evaluating it on shifted coordinates:
\begin{equation}
    (D_I f)_A(\x,\p) \equiv (D_I f)(\x-\A_\p,\p+\A_\x) \rightarrow (D_I f)(\x-\A_\p,\p+\A_\x)\, .
\end{equation}
The Hamiltonian can then be made invariant be re-arranging it in an expansion in $f$ instead of the fluctuation $\delta f = f - f_0$, and replacing the distribution and its derivatives by their invariant counterparts,
\begin{equation}
    \begin{split}
        H_\text{gauged}[f;A_I] \equiv H[f_A] &= \int_{\x\p} \epsilon(\p) f_A(\x,\p)\\
        &+ \frac{1}{2} \int_{\x\p\p'} \tilde{F}^{(2,0)}(\p,\p') f_A(\x,\p) f_{A'}(\x,\p')\\
        &+ \frac{1}{2} \int_{\x\p\p'} \tilde{\mathbf{F}}^{(2,1)}(\p,\p') (D_\x f)_A (\x,\p) f_{A'}(\x,\p')\\
        &+ \ldots\, ~~ ,
    \end{split}
\end{equation}
where $f_{A'}(\x,\p') = f(\x-\A_\p(\x,\p'),\p+\A_\x(\x,\p'))$. The gauge invariant action is given by
\begin{equation}\label{eq_maxgauge_action}
    S_\text{gauged}[\phi; A_0, A_I] = S_\text{WZW}[\phi;A_0] - \int dt ~ H_\text{gauged}[f_\phi;A_I]\, .
\end{equation}

A couple of comments are in order. First, for the case of free fermions, the action can be made independent of $\A_\p$ by a change of integration variables $\x\rightarrow \x+\A_\p, \p\rightarrow\p-\A_\x$ in the Hamiltonian:
\begin{equation}
    \int_{\x\p} \epsilon(\p) f(\x-\A_\p,\p+\A_\x) = \int_{\x\p} \epsilon(\p-\A_\x) f(\x,\p)\, .
\end{equation}
But this does not work for the interacting theory since the various factors of the invariant distribution $f_A$ are evaluated at the same $\x$ but at different momenta $\p,\p',$ etc.

Second, it is tempting to identify $\A_\x$ with a $U(1)$ gauge field and $\A_\p$ with a Berry connection, but this is incorrect due to the fact that they depend on both $\x$ as well as $\p$ and their gauge transformations are non-abelian. The precise encoding of the electromagnetic potentials and the Berry connection in the phase space gauge fields is an interesting question that we leave for future work.

One way to think about these phase space gauge fields is the following. Our theory lives not just in spacetime, but in phase space. Phase space is naturally a noncommutative space owing to the canonical commutation relation,
\begin{equation}
    \{ x^i, p_j \} = \delta^i_j\, .
\end{equation}
Therefore, gauge fields that live in this space are more akin to those in noncommutative field theory (see, e.g. \cite{DouglasNekrasovnoncomm} for a review) than to gauge fields in commutative spacetime. In fact, gauging a global $U(1)$ on a noncommutative space results a nonabelian group of gauge transformations, where the commutator of two gauge transformations is given by the Moyal bracket. Our phase space gauge fields are precisely noncommutative $U(1)$ gauge fields in the Poisson limit.

How does the `maximally gauged' action \eqref{eq_maxgauge_action} encode emergent symmetries? The answer to this question lies in the Ward identity for canonical transformations. The infinitesimal transformation of the phase space gauge fields can be written as
\begin{equation}
    \delta_\lambda A_M = - ~ \d_M \lambda - \{ \lambda, A_M \} + \mathcal{O}(\lambda^2)\, ,
\end{equation}
where $M$ is an index that collectively represents time and phase-space components. The variation of the action under this transformation necessarily takes the form
\begin{equation}
    \delta_\lambda S_\text{gauged} = - \int dt \left\< \cJ^M, \delta_\lambda A_M \right\>\, ,
\end{equation}
thus defining the phase space current $\cJ^M$. The components of this current are given by
\begin{equation}
    \cJ^0 = f\, , \qquad \cJ^{x^i} = f\d_{p_i} \epsilon(\p-\A_\x) + \ldots\, ~~ , \qquad \cJ^{p_j} = 0 + \ldots\, ~~ ,
\end{equation}
where the ellipses denote the contribution of the interacting terms in the Hamiltonian. The Ward identity then becomes
\begin{equation}
    \d_M \cJ^M + \{ \cJ^M, A_M \} = 0\, .
\end{equation}
This takes the form of a (non-)conservation law
\begin{equation}
    \d_\mu \cJ^\mu + \{ \cJ^\mu, A_\mu \} = - \d_{p_i} \cJ^{p_i} - \{ \cJ^{p_i}, A_{p_i} \}\, .
\end{equation}
Let us momentarily turn off the background fields, so that the Ward identity turns into
\begin{equation}
    \d_\mu \cJ^\mu = -\d_{p_i} \cJ^{p_i}\, .
\end{equation}
The source term on the right-hand-side is, in general, non-zero. It can also not typically be written as a total spacetime divergence which prevents us from absorbing it into the spacetime components of the current. This means that even though the Ward identity signifies the conservation of a current in phase space, it does not always reduce to the conservation of a current in space. So the `symmetry' of canonical transformations is not really a global symmetry in that it does not lead to a conservation law. This is just a roundabout way of saying that the action without phase space gauge fields is not invariant under the group of canonical transformations. Rather, the group / algebra of canonical transformations is to be thought of as an organizing principle for the set of operators in Fermi liquid theory.\footnote{This is similar to the Virasoro algebra in $1+1$d conformal field theories, which also does not generally commute with the Hamiltonian of the theory. This analogy between the Virasoro algebra and the algebra of canonical transformations goes even further since minimal models in $1+1$d can be obtained using the coadjoint orbit method to quantize the Virasoro group \cite{witten1988coadj_virasoso}.}

However, despite not being a conservation law, the Ward identity can still be useful for discovering emergent or hidden symmetries. The trivial example is that of free fermions which do not couple to $\A_\p$ at all. The Ward identity for free fermions then looks like a conservation law at every point $\p$ in momentum space,
\begin{equation}
    \d_\mu \cJ^\mu_\text{free}(t,\x,\p) = 0\, ,
\end{equation}
from which we can identify the $U(1)^\infty$ symmetry of the free Fermi gas.

Next, we consider interacting Fermi liquids. Let us look at the leading interaction:
\begin{equation}
    H_\text{int}[f;A] = \frac{1}{2}\int_{\x\p\p'} \tilde{F}^{(2,0)}(\p,\p') f(\x-\A_\p,\p+\A_\x) f(\x-\A'_\p,\p+\A'_\x)\, .
\end{equation}
Its contribution to the momentum space current is given by
\begin{equation}
    \cJ^{p_i}(\x,\p)|_{A=0} = - \frac{\delta H_\text{int}}{\delta A_{p_i}} = \int_{\p'} \tilde{F}^{(2,0)}(\p,\p') (\d_{x^i} f)(\x,\p) f(\x,\p')\, ,
\end{equation}
making the source term in the Ward identity reduce to
\begin{equation}
    \d_\mu \cJ^\mu = - \nabla_\p \cdot \int_{\p'} \tilde{F}^{(2,0)} (\nabla_\x f) f'\, ,
\end{equation}
where we have used $f$ and $f'$ as shorthand for $f(\x,\p)$ and $f(\x,\p')$ to make the expression compact. The source term is evidently neither vanishing nor a total spacetime derivative, so Landau interactions necessarily break the $U(1)^\infty$ symmetry, which should be expected. However, is we now linearize the Ward identity in fluctuations $\delta f = f - f_0$ around the spherical Fermi surface, the Ward identity simplifies to
\begin{equation}
    \d_\mu \cJ^\mu = -\nabla_\x \cdot \nabla_\p \left( \delta f \int_{\p'}\tilde{F}^{(2,0)} f_0(\p') \right) + \mathcal{O}(\delta f^2)\, ,
\end{equation}
and the source term does indeed become a total derivative and can be absorbed into a redefinition of the spatial current,
\begin{equation}
    \cJ^{x^i} \rightarrow \cJ^{x^i} + \d_{p_i} \left( \delta f \int_{\p'}\tilde{F}^{(2,0)} f_0(\p') \right) + \mathcal{O}(\delta f^2)\, .
\end{equation}
Of course, linearization is only justified when the fluctuation $\delta f$ is supported in a small region around the Fermi surface, so the linearized Ward identity can be treated as a conservation law only at points on the Fermi surface. This gives us the well known emergent symmetry of Fermi liquids that corresponds to the conservation of particle number at every point on the Fermi surface, from the linearization of the Ward identity for canonical transformations.

Else, Thorngren and Senthil \cite{Else:2020jln} formalized the study of this symmetry by identifying the symmetry group in $2+1$d as the loop group $LU(1)$ of maps from a circle to $U(1)$ with point-wise multiplication, with a 't Hooft anomaly when coupled to background gauge fields. The current four dimensional $j^M(t,\x,\theta)$ lives in spacetime as well as on the Fermi surface, with $M=t,\x,\theta$. The background gauge field $A_M(t,\x,\theta)$ also lives in the same space and the anomalous conservation law is given by
\begin{equation}
    \d_M j^M = \frac{\kappa}{8\pi^2} \epsilon^{ABCD} \d_A A_B \d_C A_D\, ,
\end{equation}
with $\kappa$ being an integer that evaluates to $\pm 1$ for Fermi liquids. Since this is an emergent symmetry, the background gauge field can be activated against our will, which does in fact happen for Fermi liquids
\begin{equation}
    A_M(t,\x,\theta) = \delta_M^i p_{Fi}(\theta)\, .
\end{equation}
$A_\theta$ is the Berry connection, which we will set to zero. We have seen that the $LU(1)$ symmetry in the absence of background fields emerges as a consequence of linearizing the Ward identity for canonical transformations. Now, let us demonstrate how linearizing the Ward identity also gives the anomaly. For simplicity, we restrict ourselves to free fermions and set $\A_\x = 0$. The free fermion Ward identity reduces to
\begin{equation}
    \begin{split}
        \d_\mu \cJ^\mu + \{ \cJ^\mu, A_\mu \} &= 0\, ,\\
        \d_t \cJ^0 + \d_i \cJ^i + \nabla_\x \cJ^0 \cdot \nabla_\p A_0 &= \nabla_\p \cJ^0 \cdot \nabla_\x A_0\, ,
    \end{split}
\end{equation}
with $A_0(t,\x,\p)$ being the time component of our phase space gauge field. We now expand the current around the spherical Fermi surface,
\begin{equation}
    \cJ^0 = f_0(\p) + \delta \cJ^0\, , \qquad \cJ^i = \delta \cJ^i\, ,
\end{equation}
and linearize the Ward identity in $\delta\cJ^\mu$ and $A_0$ to find that it takes the form,
\begin{equation}
    \d_t \delta \cJ^0 + \d_i \delta \cJ^i = -\delta(|\p|-p_F) (\n\cdot\nabla_\x A_0)\, .
\end{equation}
Integrating over the radial momentum $|\p|$ allows us to identify the Ward identity with the $LU(1)$ anomalous conservation equation by equating
\begin{equation}
    j^0 = \int \frac{pdp}{(2\pi)^2} \delta\cJ^0(t,\x,\p)\, , \qquad j^i = \int \frac{pdp}{(2\pi)^2} \delta\cJ^i(t,\x,\p)\, , \qquad A_0^{LU(1)} = A_0|_{|\p|=p_F}\, ,
\end{equation}
which turns the linearized Ward identity into the $LU(1)$ anomaly,
\begin{equation}
    \d_\mu j^\mu = -\frac{1}{4\pi^2} p_F (\n\cdot\nabla_\x A_0^{LU(1)})\, .
\end{equation}

The same holds for interacting Fermi liquids as well, since as we saw before the linearized source term can be absorbed into a redefinition of the spatial current so that the anomalous conservation law retains the same form as that for free fermions. However, nonlinear corrections to the Ward identity violate both the conservation law in the absence of background fields as well as the anomaly\footnote{This can be seen from the fact that the current has a diamagnetic contribution even for free fermions with $\cJ^{x^i} = f \d_{p_i}\epsilon(\p + \A_\x)$.}.

The algebra of canonical transformations allows us to systematically characterize the violation of the anomalous $LU(1)$ conservation law due to nonlinearities and interactions, the structure of which is somewhat rigidly constrained by the fact that it must descend from a conservation law in phase space.


\subsection{Perturbative expansion and scaling}

So far we have been able to extract a lot of `kinematic' information from the formal action \eqref{eq_U_action} and the algebra of canonical transformations that underlies it, without needing to expand it in the bosonic field $\phi$. In order to calculate correlation functions and understand the renormalization group flow of Fermi liquids, however, we will need to perform the expansion.

We start with the WZW term,
\begin{equation}
    \begin{split}
        S_\text{WZW} &= \int dt \left\< f_0, U^{-1} \d_t U \right\>\\
    &= \int dt \left\< f_0, -\dot{\phi} + \frac{1}{2!} \{ \dot{\phi}, \phi \} - \frac{1}{3!} \{ \{ \dot{\phi}, \phi \}, \phi \} + \ldots \right\>\, ,
    \end{split}
\end{equation}
where $\dot{\phi}$ stands for the time derivative of $\phi$. The first term is a total time derivative and hence vanishes. The second term is quadratic and contributes to the Gaussian part of the action,
\begin{equation}
    S^{(2)}_\text{WZW} = - \frac{p_F^{d-1}}{2} \int_{t\x\theta} ( \n\cdot\nabla_\x\phi ) ~ \dot{\phi}\, ,
\end{equation}
where $\int_{t\x\theta} = \int dt d^dx d^{d-1}\theta / (2\pi)^d$, while the third term is cubic and gives rise to a 3 point vertex for $\phi$,
\begin{equation}
    S^{(3)}_\text{WZW} = - \frac{p_F^{d-2}}{3!} \int_{t\x\theta} (\n\cdot\nabla_\x\phi) \left[ (\s^i\cdot\nabla_\x\phi)\d_{\theta^i}\dot{\phi} - (\s^i\cdot\nabla_\x\dot{\phi})\d_{\theta^i}\phi \right]\, ,
\end{equation}
where $\s^i = \d_{\theta^i} \n$ are tangent vectors on the spherical Fermi surface. We now focus on the Hamiltonian part,
\begin{equation}
    \begin{split}
        S_H[\phi] &= - \int dt ~ H[f_\phi]\, ,\\
        f_\phi &= f_0 - \{ \phi, f_0 \} + \frac{1}{2!} \{ \phi, \{ \phi, f_0 \} \} - \frac{1}{3!} \{ \phi, \{ \phi, \{ \phi, f_0 \} \} \} + \ldots\, ~~ ,
    \end{split}
\end{equation}
with the interacting Hamiltonian from equation \eqref{eq_Ham},
\begin{equation}
    \begin{split}
        H[f] &= \int_{\x\p} \epsilon(\p) f(\x,\p)\\
        &+ \frac{1}{2}\int_{\x\p\p'} F^{(2,0)}(\p,\p') \delta f(\x,\p) \delta f (\x,\p') + \mathbf{F}^{(2,1)}(\p,\p') \cdot \left( \frac{\nabla_\x}{p_F} \delta f(\x,\p) \right) \delta f (\x,\p') + \ldots\\
        &+ \frac{1}{3}\int_{\x\p\p'\p''} F^{(3,0)}(\p,\p',\p'') \delta f(\x,\p) \delta f(\x,\p') \delta f(\x,\p'') + \ldots\\
        &+ ~ \ldots\, ~~ .
    \end{split}
\end{equation}
The higher derivative interaction $\mathbf{F}^{(2,1)}$ is evidently suppressed compared to $F^{(2,0)}$ so we will ignore it for simplicity. The fluctuation $\delta f$ is at least linear in $\phi$, so only the first two lines of the Hamiltonian contribute to the quadratic action,
\begin{equation}
    \begin{split}
        S^{(2)}_H = ~ &- \frac{p_F^{d-1}}{2} \int_{t\x\theta} v_F (\nabla_n\phi)^2\\
        &- \frac{p_F^{d-1}}{2} \int_{t\x\theta\theta'} v_F F^{(2,0)}(\theta,\theta') (\nabla_n\phi)(\nabla_n\phi)'\, ,
    \end{split}
\end{equation}
where $v_F = \epsilon'(p_F)$, $F^{(2,0)}(\theta,\theta') = p_F^{d-1} F^{(2,0)}(p_F \n, p_F \n')/v_F$ is defined to be dimensionless, $\phi' = \phi(t,\x,\theta')$ and $\int_{t\x\theta\theta'}$ is defined with a factor of $(2\pi)^{2d}$ in the denominator. We have also introduced the notation $\nabla_n\phi = \n\cdot\nabla_\x\phi$ for compactness and $(\nabla_n\phi)'$ is the same quantity evaluated at $\theta'$.

For the cubic part of the action, we get contributions from all three lines of the Hamiltonian and we find
\begin{equation}
    \begin{split}
        S^{(3)}_H = ~ &- \frac{p_F^{d-2}}{3!} \int_{t\x\theta} \left( \frac{d-1}{2}v_F + p_F \epsilon'' \right) (\nabla_n\phi)^3\\
        &- \frac{p_F^{d-2}}{2}\int_{t\x\theta\theta'} F_1^{(2,0)}(\theta,\theta') \left[ (\nabla_n\phi)^2(\nabla_n\phi)' + (\theta\leftrightarrow\theta') \right]\\
        &- \frac{p_F^{d-2}}{2}\int_{t\x\theta\theta'} v_F F^{(2,0)}(\theta,\theta') \Big[ \big[ (\nabla^i_s\nabla_n\phi)(\d_{\theta^i}\phi) - (\d_{\theta^i} \nabla_n\phi) (\nabla^i_s\phi) \big](\nabla_n\phi)' + (\theta\leftrightarrow\theta') \Big]\\
        &- \frac{p_F^{d-2}}{3} \int_{t\x\theta\theta'\theta''} F^{(3,0)}(\theta,\theta',\theta'') (\nabla_n\phi)(\nabla_n\phi)'(\nabla_n\phi)''\, ,
    \end{split}
\end{equation}
where $\nabla^i_s\phi = \s^i\cdot\nabla_\x\phi$, and $F_1^{(2,0)}$ is the derivative $\n\cdot\nabla_\p F^{(2,0)}$ evaluated at the Fermi surface and appropriately rescaled to make it dimensionless. $F^{(3,0)}$ has also similarly been evaluated at the Fermi surface and rescaled, and $\epsilon''$ is the second derivative of the dispersion with respect to the radial momentum evaluated at the Fermi surface.

Collecting everything, we can write down the interacting action up to cubic order:
\begin{equation}\label{eq_action_phi}
    \begin{split}
        S = ~ &- \frac{p_F^{d-1}}{2} \int_{t\x\theta} \nabla_n\phi \left( \dot{\phi} + v_F \nabla_n \phi + v_F \int_{\theta'} F^{(2,0)}(\theta,\theta') (\nabla_n\phi)' \right)\\
        &- \frac{p_F^{d-2}}{3!} \int_{t\x\theta} \nabla_n\phi \left[ (\nabla^i_s\phi) (\d_{\theta^i} \dot{\phi}) - (\nabla^i_s\dot{\phi}) (\d_{\theta^i} \phi) \right] + \left( \frac{d-1}{2}v_F + p_F \epsilon'' \right) (\nabla_n\phi)^3\\
        &- \frac{p_F^{d-2}}{2}\int_{t\x\theta\theta'} v_F F^{(2,0)}(\theta,\theta') \Big[ \big[ (\nabla^i_s\nabla_n\phi)(\d_{\theta^i}\phi) - (\d_{\theta^i} \nabla_n\phi) (\nabla^i_s\phi) \big](\nabla_n\phi)' + (\theta\leftrightarrow\theta') \Big]\\
        &- \frac{p_F^{d-2}}{2}\int_{t\x\theta\theta'} F_1^{(2,0)}(\theta,\theta') \left[ (\nabla_n\phi)^2(\nabla_n\phi)' + (\theta\leftrightarrow\theta') \right]\\
        &- \frac{p_F^{d-2}}{3} \int_{t\x\theta\theta'\theta''} F^{(3,0)}(\theta,\theta',\theta'') (\nabla_n\phi)(\nabla_n\phi)'(\nabla_n\phi)''\\
        &+ \mathcal{O}(\phi^4)\, .
    \end{split}
\end{equation}
The first line is the Gaussian part of the action, which includes the Landau parameters $F^{(2,0)}(\theta,\theta')$. The second line is the free fermion contribution to the cubic part of the action, and the remaining three lines are cubic contributions with three independent Wilson coefficient functions.

The quadratic part of the action is almost identical to the action obtained from multidimensional bosonization \eqref{eq_multidimbos_action}\cite{Haldane:1994,CastroNetoFradkin:1994,Houghton:2000bn}, with one crucial difference: the angular coordinates $\theta$ in our case are genuinely continuous variables as opposed to discrete labels for patches on the Fermi surface. Furthermore, the nonlinear and higher derivative corrections that the coadjoint orbit formalism provides can be interpreted as corrections coming from the curvature of the Fermi surface, nonlinearities in the dispersion relation, as well as intra-patch and inter-patch scattering. Since the coadjoint orbit method does not require a discretization of the Fermi surface to begin with, the corrections in our action do not distinguish between intra-patch and inter-patch effects and treat them collectively in an expansion in $\x$ and $\theta$ derivatives.

Note that the cubic terms are suppressed compared to the quadratic ones by a factor of $\nabla_\x/p_F$, owing to the scaling properties of the Poisson bracket described in the discussion before equation \eqref{eq_Poisson_limit_FS}. The expansion in nonlinearities in $\phi$ hence makes our action an effective field theory with a derivative expansion suppressed by the UV cutoff $p_F$.

With the expanded action in hand, we can study its properties under scaling of space $\x\rightarrow s^{-1}\x$ with $s\lesssim1$. In principle we have a choice to make for how $\theta$ scales, e.g., compared to the angle on some external observable. Different choice of this scaling result in different scaling of our theory under RG. The choice that we will make to to leave $\theta$ invariant under scaling. This is consistent with the RG scheme of Shankar and Polchinski \cite{Shankar:1990,Polchinski:1992ed} where they scale all momenta toward the Fermi surface without changing the angle between them.

The quadratic part of the action then tells us that time scales the same way as space, so the dynamical scaling exponent of our theory is $z=1$. The scaling dimension of $\phi$ can be obtained by requiring the quadratic part of the action to be marginal:
\begin{equation}
    [\phi] = \frac{d-1}{2}\, .
\end{equation}
The Landau parameters $F^{(2,0)}(\theta,\theta')$ are marginal as expected. The cubic terms all have an additional factor of
\begin{equation}
    [\nabla \phi] = \frac{d+1}{2}\, ,
\end{equation}
compared to the quadratic terms, which makes them all strictly irrelevant in any number of dimensions, as is necessary for any effective field theory. The same holds for higher order terms in the expansion in $\phi$, as well as interactions.

Note that the interaction terms with Wilson coefficient functions $\epsilon(\p), F^{(m,n)}(\p_1,\ldots,\p_m)$ in the Hamiltonian \eqref{eq_Ham} do not have fixed scaling dimensions. Instead, they characterize a tower of coefficient functions that do have a fixed scaling dimension, given by the various derivatives of $F^{(m,n)}(\p_i)$ with respect to $|\p_i|$ evaluated at the Fermi surface. The first derivative of the dispersion $\epsilon(\p)$ at the Fermi surface is the Fermi velocity and shows up at the quadratic level, whereas the $n$th derivative $\d_p^n\epsilon|_{p_F}$ shows up as a Wilson coefficient at order $\phi^{n+1}$. Each $|\p|$ derivative increases the scaling dimension of the corresponding operator by 1, so that $\d_p^k F^{(m,n)}$ scales in the same way as $F^{(m+l,n+k-l)}$ for all non-negative integer values of $l \le k$.

General observables can be constructed in the EFT from the operator $\phi$ by constructing all possible terms with the required quantum numbers and taking a linear combination of them with arbitrary coefficients that are determined by matching correlation functions with experiments or microscopics, as is common in EFT. Any bosonic operator that is charge neutral can be constructed in this form, while fermionic operators are absent in this EFT.

A special operator is the particle number current, which can be obtained from the gauged action \eqref{eq_u1gauged_action}. The current depends on the Wilson coefficient functions of the theory, but the density is universal since $A_0$ only couples to the WZW term,
\begin{equation}
    \begin{split}
        \rho[\phi](t,\x) &= \frac{\delta S_\text{WZW}}{\delta A_0(t,\x)} = \int_\p f_\phi(t,\x,\p) = \int d^{d-1}\theta ~ \rho[\phi](t,\x,\theta)\, .\\
        \rho[\phi](t,\x,\theta) &= \frac{\delta S_\text{WZW}}{\delta A_0(t,\x,\theta,|\p|=p_F)} = \int \frac{p^{d-1} dp}{(2\pi)^d} f_\phi(t,\x,\theta,p)\, ,
    \end{split}
\end{equation}
where $\rho[\phi](t,\x,\theta)$ is the \textit{angle-resolved density}, which is the density for the emergent LU(1) symmetry. The explicit expression for the total density in terms of expansions in $\phi$ is
\begin{equation}
    \rho - \frac{p_F^d}{d(2\pi)^d} = \frac{p_F^{d-1}}{(2\pi)^d} \int_\theta \left( \nabla_n \phi + \frac{1}{2p_F} \nabla_s^i \left( \d_{\theta^i} \phi \nabla_n \phi \right) + \mathcal{O}(\phi^3) \right)\, ,
\end{equation}
where we have subtracted the background density coming from the spherical Fermi surface and $\int_\theta = \int d^{d-1}\theta$. We will refer to the fluctuation in the density as $\rho$ from here onward and drop the constant background density. Note that the density can be written as a spatial divergence to all orders in phi, since
\begin{equation}
    \int_\x \rho - \rho_0 = \int_{\x\p} f-f_0 = 0\, ,
\end{equation}
where the last equality follows from the fact that corrections to $f_0$ that determine $f$ are all total phase space derivatives. We explicitly calculate the two point and three point density correlation function in the next section and a demonstration of the technical advantages of the postmodern formalism.


\subsection{Linear and nonlinear response}\label{sec_response}

Before calculating correlation functions, we derive a scaling form for $n$-point density correlators from the scaling analysis above as well as the $\phi$-propagator (with Landau parameters set to zero for simplicity),
\begin{equation}\label{eq_phi_propagator}
    \< \phi_\theta \phi_{\theta'} \> (\omega,\q) = i \frac{(2\pi)^d}{p_F^{d-1}} \frac{\delta^{d-1}(\theta-\theta')}{q_n(\omega-v_F q_n)}\, ,
\end{equation}
where $q_n = \n\cdot\q$. The action takes a schematic expansion of the form
\begin{equation}
    \begin{split}
        S \sim p_F^{d-1} &\int_{t\x\theta} \dot{\phi} \left( \nabla\phi + \frac{1}{p_F}(\nabla\phi)^2 + \frac{1}{p_F^2} (\nabla\phi)^3 + \ldots \right)\\
        & + v_F \nabla\phi \left( \nabla\phi + \frac{1}{p_F}(\nabla\phi)^2 + \frac{1}{p_F^2} (\nabla\phi)^3 + \ldots \right)\, ,
    \end{split}
\end{equation}
where we only highlight the dependence of the action on factors that scale. The density has a similar expansion:
\begin{equation}
    \rho \sim p_F^{d-1} \int_\theta \nabla\phi + \frac{1}{p_F}(\nabla\phi)^2 + \frac{1}{p_F^3}(\nabla\phi)^3 + \ldots\, ~~ .
\end{equation}
The nonlinear terms in the action as well as the density activate higher-point correlation functions for the density\footnote{This was anticipated in the context of traditional bosonization in \cite{PhysRevB.52.10877}.} and we can check from the scaling properties of the vertices, the propagator, as well as the nonlinear corrections to the density that all tree level diagrams (e.g. the ones in figure \ref{fig_rrr}) that contribute to the density $n$-point function scale like $q^0$. Their scaling with respect to $p_F$ and $v_F$ can also similarly be determined, and we find the following scaling form for the correlators:
\begin{equation}\label{eq_density_scaling}
    \< \rho(\omega,\q_1) \ldots \rho(\omega,\q_n) \> = \frac{p_F^{d+1-n}}{v_F^{n-1}} g_n \left( \frac{\omega_i}{\omega_j}, \frac{v_F \q_i}{\omega_j} \right) \delta(\Sigma_i \omega_i) \delta(\Sigma_i \q_i) + \mathcal{O}\left( \frac{\omega_i}{v_F p_F}, \frac{\q_i}{p_F} \right)\, ,
\end{equation}
where the subleading corrections come from loops as well as higher-derivative interactions in the Hamiltonian. This scaling form is also apparent from kinetic theory (see appendix F of \cite{main:2022}), but highly counter-intuitive in the fermionic approach where the leading behaviour is given by a single fermion loop with $n$ external legs. Scaling arguments in the Shankar-Polchinski scheme tell us then that a given 1-loop diagram must scale like $p_F^{d-1}/q_\perp^{n-2}$, where $q_\perp$ is the component of the momentum orthogonal to the Fermi surface. This is incorrect and what rescues the calculation is a subtle cancellation that occurs upon symmetrizing the external legs of the diagram \cite{Metzner1997FermiSW,PhysRevB.58.15449}.

Not only can we derive the scaling form of the density correlators from the postmodern formalism, but also determine which Wilson coefficients will contribute to scaling function $g_n$. These are interactions that give vertices of order up to $(\nabla\phi)^n$. For free fermions, these coefficients are
\begin{equation}
    \epsilon^{(m\le n)} = \d_p^m \epsilon|_{p_F}\, ,
\end{equation}
while the interaction functions that contribute to the $n$-point function are given by
\begin{equation}
    F_{(l)}^{(m,k)}|_{p_F} = (\d_p^l F^{(m,k)})|_{p_F}\, , \qquad (m+k+l)\le n\, ,
\end{equation}
where $\d_p$ is a derivative with respect to the one of the radial momenta the $F^{(m,k)}$ depends on.


\subsubsection{Landau damping}

We can now move on to explicit calculations of the density two and three point correlators. For the two point function, the Gaussian action suffices and we only need the linear-in-$\phi$ term in the density operator,
\begin{equation}
    \rho = \frac{p_F^{d-1}}{(2\pi)^d} \int_\theta \nabla_n \phi + \ldots\, ~~ .
\end{equation}
Using the propagator \eqref{eq_phi_propagator} we find that the two-point function evaluates to the following expressions in terms of the hypergeometric function ${}_2 F_1$,
\begin{equation}
    \begin{split}
        \< \rho\rho \>\left(s=\frac{\omega}{v_F|\q|}\right) &= i \frac{p_F^{d-1}}{(2\pi)^d} \frac{1}{v_F} \int d^{d-1}\theta \frac{\cos \theta_1}{\cos\theta_1 - s}\\
        = i \frac{p_F^{d-1}}{(2\pi)^d} \frac{1}{v_F} \frac{\pi^{d/2}}{\Gamma(d/2)} &\frac{2-\delta_{d,1}}{1+|s|}\left[ {}_2 F_1\left( 1, \frac{d+1}{2}; d, \frac{2}{1+|s|} \right) - {}_2 F_1 \left( 1, \frac{d-1}{2}; d-1, \frac{2}{1+|s|} \right) \right]\, ,
    \end{split}
\end{equation}
where $\theta_1$ is one of the angles that parametrize the spherical Fermi surface - the polar angle from the direction of the external momentum $\q$. The $d+1$ loop integrals in the fermionic picture have been replaced by $d-1$ angular integrals. In $d=1$ the answer reduces to the well-known result for a Luttinger liquid:
\begin{equation}
    \< \rho \rho \>(\omega,\q) = -\frac{i}{\pi} \frac{v_F q^2}{\omega^2 - v_F^2 q^2}\, .
\end{equation}
In $d=2$ we recover the expression
\begin{equation}\label{eq_density2pt_2d}
    \< \rho\rho \>(\omega,\q) = \frac{i}{2\pi} \frac{p_F}{v_F} \left( 1- \frac{|\omega|}{\sqrt{\omega^2 - v_F^2 q^2}} \right)\, ,
\end{equation}
with the branch-cut for $|\omega|<v_Fq$ coming from the particle-hole continuum, while in $d=3$ we find \cite{PhysRevB.68.155113}
\begin{equation}
    \< \rho\rho \>(\omega,\q) = \frac{i}{2\pi^2} \frac{p_F^2}{v_F} \left( 1 + \frac{1}{2}\frac{|\omega|}{v_F q} \log \frac{|\omega| - v_F q}{|\omega| + v_F q} \right)\, .
\end{equation}
It is easy to see that the expressions are in agreement with the scaling form \eqref{eq_density_scaling}.


\subsubsection{Cubic response}

Next, we calculate the three point function. Note that even though the nonlinear-in-$\phi$ terms in the Hamiltonian can be set to zero by choosing the interactions appropriately, the nonlinearities in the WZW term cannot be avoided and are a rigid part of the structure of our theory. These have no counterpart in $d=1$ and encode the geometry of the Fermi surface, i.e., its curvature. The density three point function is hence necessarily non-vanishing for $d>1$ for a circular Fermi surface, irrespective of what interactions are turned on.

\begin{figure}[t]
	\begin{equation*}
	\begin{gathered}\includegraphics[width=0.18\linewidth,angle=0]{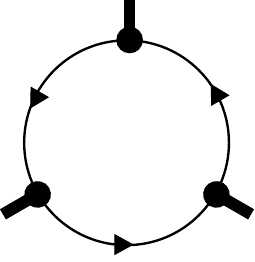}\end{gathered}
	 \quad = \quad
	\begin{gathered}\includegraphics[width=0.18\linewidth,angle=0]{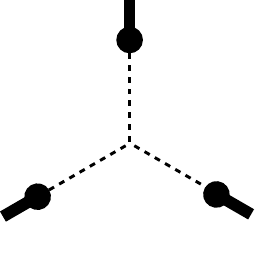}\end{gathered}
	 \quad + \quad
	\begin{gathered}\includegraphics[width=0.18\linewidth,angle=0]{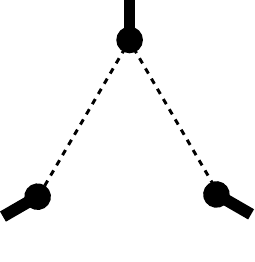}\end{gathered}
	\end{equation*}
\caption{\label{fig_rrr}The density three-point function in fermionic and bosonic descriptions.}
\end{figure}

There are two diagrams that contribute to the density three point function, as shown in figure \ref{fig_rrr}. We will refer to the first of the two as the `star' diagram, and the second as the `triangle' or `wedge' diagram. The latter comes from the quadratic part of the density:
\begin{equation}
    \rho^{(2)} = \frac{p_F^{d-2}}{2(2\pi)^d} \int_\theta \nabla_s^i (\d_{\theta^i}\phi\nabla_n\phi)\, .
\end{equation}
The former is the consequence of the cubic vertices in the action, which can be separated into two distinct terms: the $S_H^{(3)}$ piece obtained from the Hamiltonian and the $S_\text{WZW}^{(3)}$ piece from the WZW term.

The $S_H^{(3)}$ piece is the only one that picks up a contribution from $\epsilon''$, and is given by
\begin{equation}
    \< \rho \rho \rho \>_H = - \frac{p_F^{d-2}}{(2\pi)^d} \left( \frac{d-1}{2}v_F + p_F \epsilon'' \right) \int_\theta \frac{q_n}{\omega-v_F q_n} \frac{q'_n}{\omega'-v_F q'_n} \frac{(q+q')_n}{(\omega+\omega')-v_F (q+q')_n}\, ,
\end{equation}
with $q_n = \n\cdot\q$.

The $S_\text{WZW}^{(3)}$ piece takes the form
\begin{equation}
    \< \rho\rho\rho \>_\text{WZW} = \frac{p_F^{d-2}}{3!(2\pi)^d} \int_\theta \frac{q_n}{\omega-v_F q_n} \frac{q'_{s^i}}{\omega'-v_F q'_n} \d_{\theta^i} \frac{\omega+2\omega'}{(\omega+\omega')-v_F (q+q')_n} + \text{5 perm.}\, ,
\end{equation}
where the permutations are those of the set $\{ (\omega,\q), (\omega',\q'), (\omega'',\q'') \}$ with $\omega''=-\omega-\omega'$ and $\q''=-\q-\q'$ due to conservation of energy and momentum.

Finally, the triangle/wedge diagram evaluates to
\begin{equation}
    \<\rho\rho\rho\>_{\rho^{(2)}} = - \frac{p_F^{d-2}}{2(2\pi)^d} \int_\theta \frac{q_n(q+q')_{s^i}}{\omega-v_F q_n} \d_{\theta^i} \frac{1}{\omega'-v_F q'_n} + \text{5 perm.}\, ,
\end{equation}
and the density three point function is given by the sum of the three expressions
\begin{equation}
    \< \rho\rho\rho \>(\omega,\q;\omega',\q') = \<\rho\rho\rho\>_{\rho^{(2)}} + \<\rho\rho\rho\>_\text{WZW} + \<\rho\rho\rho\>_H\, .
\end{equation}
Each of the terms do indeed have the scaling form \eqref{eq_density_scaling}, as we expected.

While directly matching this expression with the fermion loop in the scaling limit is a highly nontrivial task due to the complexity of the expression evaluated in \cite{Feldman1998,Neumayr1999,Kopper:2001sf} (for a Galilean invariant dispersion relation), we can instead calculate the density 3 point function using kinetic theory for an arbitrary dispersion and show that it matches with the above expressions. This was done in \cite{main:2022} and we refer the reader to appendix F in the paper for details.

Of course this matching should not be unexpected, since the equation of motion for our theory is exactly the kinetic equation, and tree level diagrams reproduce classical physics that is captured by the equation of motion.


\subsection{UV/IR mixing and why it is not all \textit{that} bad}\label{sec_uv/ir}

Since the cubic and higher order terms in our action \eqref{eq_action_phi} are strictly irrelevant, in the deep IR we can focus only on the quadratic part of the action which, for free fermions, is given by
\begin{equation}
    S \sim \int_{t\x\theta} \nabla_n \phi (\dot{\phi} + v_F \nabla_n \phi)\, ,
\end{equation}
The theory has a zero mode which propagates tangent to the Fermi surface. In momentum space, this corresponds to modes
\begin{equation}
    \phi(\omega=0, q_n=0,q_{s^i}, \theta)\, ,
\end{equation}
for all values of the tangential components $q_{s^i}$ of the momentum. In particular, this means that we have low energy modes with indefinitely large momenta (of the order of the cutoff $p_F$) in our EFT, which is the hallmark of UV/IR mixing\footnote{This is similar to fractonic models where exotic symmetries disallow kinetic terms that would suppress large momentum modes at low energies, also resulting in UV/IR mixing.}. This results in UV divergences in loop contributions to correlation functions as well as thermodynamic properties.

For instance, we can calculate the thermal partition function by rotating to imaginary time $t=-i\tau$ and compactifying it on a thermal circle $\tau\in[0,\beta]$,
\begin{equation}
    Z_\text{FL}(\beta) = \det \left[ q_n(-i\omega_k + v_F q_n) \right]^{-1/2}\, ,
\end{equation}
where $\omega_k = 2\pi T k$ are bosonic Matsubara frequencies with $k\in\mathbb{Z}$. The pressure is given by the logarithm of the partition function,
\begin{equation}
    P = \frac{T}{V} \log Z_\text{FL} = -\frac{T}{2} \sum_k \int_{\q,\theta} \log \left[ q_n(-i\omega_k + v_F q_n) \right]\, .
\end{equation}
Since the integrand has a zero mode, the pressure diverges. Nevertheless, we can still extract the scaling form of the pressure with respect to temperature in a hand-wavy manner by writing $\int_\q = \int d^{d-1} q_s \int q_n$. Since the integrand does not depend on $q_{s^i}$, the integral over these components needs to be regulated by some cutoff.

However, the momentum $\q$ is bounded above by a physical cutoff $p_F$, owing to the semiclassical truncation of the Moyal algebra to the Poisson algebra \eqref{eq_Poisson_limit_FS}. This cutoff is not an arbitrary scale that is introduced by hand into low energy physics, but a measurable property of the IR. The integral $\int d^{d-1} q_s$ hence must scale like $p_F^{d-1}$, leaving only
\begin{equation}
    \sum_k \int dq_n \log[q_n(-i\omega_k+v_F q_n)] \sim T\, .
\end{equation}
From this heuristic analysis we surprisingly find the correct scaling form for the pressure:
\begin{equation}
    P \sim p_F^{d-1} T^2\, .
\end{equation}

A similar problem occurs in loop corrections to correlation functions as well, for instance in the 1-loop correction to $\<\rho\rho\>$ and the result is not indifferent to how the integral is cutoff. This suggests that there is a preferred way of introducing the cutoff $p_F$ in loop integrals as well, which needs to be studied more carefully. One possible resolution would be a potential resummation of the Moyal expansion of the theory, which we leave for future work.


\newpage
\section{A road to perturbative non-Fermi liquids}\label{sec_nfl}

One of our main motivations for developing the coadjoint orbit formalism for Fermi liquids was to resolve the drawbacks of Fermi liquid theory that manifest themselves as serious bottlenecks when coupling to a gapless mode and studying the RG flow to a non-Fermi liquid.

This approach to describing non-Fermi liquids as Fermi liquids coupled to a gapless mode is often known as the `Hertz-Millis-Moriya' description \cite{hertz1976,millis1993nfl,Moriya1985} (see \cite{RevModPhys.79.1015} for a review). The upper critical dimension for the coupling to the gapless mode is $d=3$, which makes $d=2$ the most interesting case to study, since there is no extended Fermi surface in $d=1$ and bosonization allows for either exact or perturbative solutions to the $d=1$ problem.

The original approach developed by Hertz in the 1970's was to integrate out the Fermi surface and write down a non-local effective action for the gapless mode. Naturally, this approach is extremely uncontrolled and unreliable. Progress was made after the development of the Shankar-Polchinski RG scheme using both fermionic EFT \cite{polchinski1994nfl,Altshuler1994patchNFL,Nayak:1993uh,Nayak:1994ng,Metlitski:2010pd,Lee:2009epi} as well as traditional bosonization \cite{khveshchenko:1995,PhysRevLett.73.284,Lawler:2006}, but these approaches were also found to be limited owing either to a lack of a systematic expansion \cite{Lee:2009epi,mandal2015uv/ir_nfl,ye2022uv/ir_mfl} for fermionic EFTs or to the incompleteness of the traditional bosonized description. A controlled, systematic expansion is yet to be found and our hope is that the postmodern formalism for Fermi liquids can provide one.

One advantage of a bosonized theory is that some important physical properties of non-Fermi liquids can already be captured from a Gaussian theory. The Gaussian truncation of the EFT \eqref{eq_action_phi} in $d=2$ can be coupled to a bosonic field $\Phi(t,\x)$ through the linearized density,
\begin{equation}
    S^{(2)}_\text{NFL} = - \frac{p_F^2}{8\pi^2} \int_{t\x\theta} \nabla_n\phi \left( \dot{\phi} + v_F \nabla_n\phi \right) - \frac{1}{2} \int_{t\x} \left[ ( \nabla \Phi )^2 + k_0^2 \Phi^2 \right] + \lambda \frac{p_F}{4\pi^2} \int_{t\x} \Phi \int_\theta \nabla_n\phi\, ,
\end{equation}
with the bare mass $k_0^2$ tuned to criticality. The coupling can be generalized to a spin-$l$ harmonic of the Fermi surface by inserting an additional factor of $\cos(l\theta)$.

This action is Gaussian and can hence be exactly solved. The $\Phi$ propagator is Landau damped,
\begin{equation}\label{eq_landaudamped_bosonprop}
    \< \Phi \Phi \>(\omega,\q) = \frac{i}{q^2 + k_0^2 - \< \rho\rho \>(\omega,\q)}\, ,
\end{equation}
with $\<\rho\rho\>$ being the tree level density two point function \eqref{eq_density2pt_2d}. Taking the limit $\omega\ll q$ and tuning the boson mass to criticality by setting $k_0^2 = - p_F \lambda^2/2\pi v_F$, we find
\begin{equation}
    \< \Phi\Phi \>(\omega,\q) \simeq \frac{1}{q^2 - i \frac{p_F \lambda^2}{2\pi v_F^2} \frac{|\omega|}{v_Fq}}\, , \qquad \omega \ll v_F q\, ,
\end{equation}
from which we can read off the dynamical critical exponent:
\begin{equation}
    z = 3\, .
\end{equation}

The temperature scaling of the specific heat can also be calculated from this Gaussian theory from the thermal partition function,
\begin{equation}
    Z_\text{NFL}(\beta) = \int D\phi D\Phi ~ e^{-S_E}\, ,
\end{equation}
where $S_E$ is the Euclidean action obtained by Wick rotating $t=-i\tau$ and putting imaginary time on a circle $\tau\in[0,\beta]$. The partition function can be calculated by first integrating over $\phi$ followed by $\Phi$ and we find that it factorizes into a product of a Fermi liquid contribution and a Landau-damped critical boson contribution,
\begin{equation}
    Z_\text{NFL} = \det \left[ q_n (-i \omega_k + v_F q_n) \right]^{-1/2} \det \left( q^2 + \frac{p_F \lambda^2}{2\pi v_F} \frac{|\omega_k|}{\sqrt{\omega_k^2 + v_F^2 q^2}} \right)^{-1/2}\, ,
\end{equation}
where $\omega_k = 2\pi T k$ are bosonic Matsubara frequencies with $k\in \mathbb{Z}$. The free energy or pressure then also splits up into a sum of a Fermi liquid contribution and a Landau-damped critical boson contribution.
\begin{equation}
    \begin{split}
        P &= \frac{T}{V}\log Z_\text{NFL}\\
        &= -\frac{T}{2} \sum_k \int_{\q,\theta} \log \left[ q_n (-i \omega_k + v_F q_n) \right] - \frac{T}{2} \sum_k \int_\q \log \left( q^2 + \frac{p_F \lambda^2}{2\pi v_F} \frac{|\omega_k|}{\sqrt{\omega_k^2 + v_F^2 q^2}} \right)\, .
    \end{split}
\end{equation}
As discussed in section \ref{sec_uv/ir}, the Fermi liquid contribution in the EFT suffers from UV/IR mixing and needs to be regulated appropriately. Nevertheless we can deduce its scaling form to be $P_\text{FL} \sim p_F T^2$ from a heuristic scaling analysis.

Restricting our attention to low temperatures, the Matsubara sum for the critical boson contribution is dominated in the IR by frequencies of order $\omega_k \sim q^3 \ll v_F q$, allowing us to simplify the integral to
\begin{equation}
    \int_q \log \left( q^2 + \tilde{\lambda}^2\frac{|\omega_k|}{q} \right) = \tilde{\lambda}^{4/3} \frac{|\omega_k|^{2/3}}{2\sqrt{3}}\, ,
\end{equation}
after dropping a temperature-independent UV divergence, and defining $\tilde{\lambda}^2 = p_F \lambda^2/2\pi v_F^2$. The Matsubara sum is also divergent but can be regulated by introducing an exponential $e^{-\varepsilon k}$ in the sum with $\varepsilon>0$ to suppress the large $k$ contribution, expanding for small $\varepsilon$ and then subtracting off divergent pieces to find
\begin{equation}
    \sum_k k^{2/3} \simeq \zeta(-2/3)\, .
\end{equation}
We ultimately find that the critical boson contribution to the pressure evaluates to
\begin{equation}
    P = - \frac{\zeta(-2/3)}{4\sqrt{3}} \tilde{\lambda}^{4/3} T^{5/3}\, .
\end{equation}
At low temperatures, $T^{5/3}$ dominates over $T^2$ and the Fermi liquid contribution to the specific heat can be dropped. Any concerns about UV/IR mixing also vanish with it, since the critical boson contribution does not suffer from UV/IR mixing. The low temperature specific heat of the Gaussian NFL is hence given by
\begin{equation}
    c_V = T \frac{ds}{dT} = T \frac{d^2P}{dT^2} = - \frac{5\zeta(-2/3)}{18\sqrt{3}} \tilde{\lambda}^{4/3} T^{2/3}\, ,
\end{equation}
in perfect agreement with the $T^{2/3}$ scaling of the specific heat found from other approaches \cite{sslee2017nfl_review}.


\subsection{Scaling in non-Fermi liquids\footnote{The results presented in this section are based on ongoing work, soon to appear.}}

The Gaussian truncation of the Fermi liquid EFT is evidently insufficient for a full description of the NFL (see e.g., \cite{PhysRevLett.97.226403}). But now that we know how systematically add corrections to the Gaussian action, we can hope to analyze the theory with the corrections and perform an RG analysis for the coupling to the gapless boson. The bosonized NFL action up to cubic order in arbitrary dimensions $d$, for instance, looks like
\begin{equation}\label{eq_nfl_ac_cubic}
    \begin{split}
        S_\text{NFL}[\phi,\Phi] = &- \frac{p_F^{d-1}}{2(2\pi)^d} \int_{t\x\theta} \nabla_n\phi \left( \dot{\phi} + v_F \nabla_n\phi \right)\\
        &- \frac{p_F^{d-2}}{3!(2\pi)^d} \int_{t\x\theta} \nabla_n\phi \left[ \nabla^i_s\phi \d_{\theta^i}\dot{\phi} - \nabla^i_s\dot{\phi} \d_{\theta^i} \phi \right] + \left[ \epsilon'' + \frac{d-1}{2}\frac{v_F}{p_F} \right] (\nabla_n\phi)^3\\
        &- \lambda \frac{p_F^{d-1}}{(2\pi)^d} \int_{t\x} \Phi \int_\theta \nabla_n\phi + \frac{1}{2p_F} \nabla^i_s (\d_{\theta^i}\phi \nabla_n\phi)\\
        &- \frac{1}{2} \int_{t\x} \Phi \left( -|\nabla|^{1+\epsilon} \right) \Phi\\
        &+ \mathcal{O}(\phi,\Phi)^4\, ,
    \end{split}
\end{equation}
where we have replaced the kinetic term for the critical boson by a non-local term, \'a la Nayak-Wilczek \cite{Nayak:1993uh,Nayak:1994ng}. The bare mass term for the critical boson has also been suppressed for brevity, since it is tuned to criticality anyway.

We can now attempt to understand the scaling properties of this theory. From the tree level propagator \eqref{eq_landaudamped_bosonprop} of the critical boson, it is clear that time must scale with a non-trivial power $z\ne1$ of space. However, requiring every term in the Gaussian part of the action then necessitates that the angles $\theta$ scale with $q$ as well. This can be understood in the following way: ultimately the scaling properties of the actions are to be applied to correlation functions with external momenta. Pick one such external momentum --- $\mathbf{Q}$, and decompose the momentum $\q$ of the fields parallel and perpendicular to the external momentum:
\begin{equation}
    \q = q_\parallel \frac{\mathbf{Q}}{|\mathbf{Q}|} + \q_\perp\, .
\end{equation}
Parametrize the Fermi surface with angles $\theta_i$ such that $\theta_{d-1}$ is the polar angle subtended from the direction of $\mathbf{Q}$ and the rest $\theta_1,\ldots,\theta_{d-2}$ are azimuthal angles for the $(d-2)$-spherical slices of the Fermi surface for a fixed $\theta_{d-1}$. The external momentum $\mathbf{Q}$ couples most strongly to the parts of the Fermi surface that are tangent to it, i.e., at the equator when $\theta_{d-1} \approx \pi/2$. Define $\delta\theta = \theta_{d-1} - \pi/2$. In this parametrization we have
\begin{equation}
    \nabla_n \sim |\q_\perp| + q_\parallel \delta\theta\, .
\end{equation}
Marginality of the quadratic part of the Fermi liquid action (the first line of equation \eqref{eq_nfl_ac_cubic}) then requires
\begin{equation}
    \omega \sim |\q_\perp| \sim q_\parallel \delta\theta\, .
\end{equation}
If we let frequency scale with an arbitrary power (greater than 1) of the parallel momentum,
\begin{equation}
    \omega \sim q_\parallel^z\, ,
\end{equation}
we find that the polar angle must scale toward the equator and the field momentum must scale towards a direction tangential to the Fermi surface (and collinear with the external momentum):
\begin{equation}
    \delta\theta \sim \frac{\omega}{q_\parallel} \sim q_\parallel^{z-1}, \qquad |\q_\perp| \sim \omega \sim q_\parallel^z\, .
\end{equation}
Since the transverse components $\q_\perp$ scale to zero much faster than the parallel component, the parallel component in the IR scales like the magnitude of the field momentum $q_\parallel \sim q$ and the following scaling relations hold:
\begin{equation}
    \omega \sim |\q_\perp| \sim q^z\, , \qquad \delta\theta \sim q^{z-1}\, , \qquad \nabla_n \sim q^z\, , \qquad \nabla^i_s \sim q\, .
\end{equation}
From this we can calculate the scaling dimension of $\phi$,
\begin{equation}
    \phi \sim q^{1+z(d-3)/2}\, ,
\end{equation}
and that of the density,
\begin{equation}
    \rho \sim \int_\theta \nabla_n\phi \sim q^{z(d+1)/2}\, .
\end{equation}
The scaling dimension of the critical boson can be calculated from its kinetic term,
\begin{equation}
    \Phi \sim q^{(zd-\epsilon)/2}\, ,
\end{equation}
and requiring the Gaussian part of the interaction to be marginal sets the dynamical critical exponent:
\begin{equation}
    z = 2+\epsilon\, ,
\end{equation}
which is consistent with $z=3$ for $\epsilon = 1$.

Now let us look at the cubic terms. The cubic part of the Hamiltonian term scales like
\begin{equation}
    \frac{S^{(3)}_H}{S^{(2)}} \sim \nabla_n\phi \sim q^{1 + z(d-1)/2}\, ,
\end{equation}
which is irrelevant for all values of $z>0,d\ge1$, and hence does not contribute to the RG flow in any dimension. The cubic parts of the WZW term as well as the coupling, on the other hand, scale differently:
\begin{equation}
    \frac{S^{(3)}_\text{WZW}}{S^{(2)}} \sim \frac{S^{(3)}_\text{int}}{S^{(2)}} \sim \nabla_s^i \d_{\theta^i} \phi = q^{3 - z(5-d)/2}\, .
\end{equation}
This gives us a set of marginal cubic corrections in the $(d,z)$-plane
\begin{equation}
    z = \frac{6}{5-d} \quad \Leftrightarrow \quad d = 5 - \frac{6}{z}\, , \qquad \begin{tabular}{c|c c c c}
        ~$d$~ & ~$2$~ & ~$3$~ & ~$4$~ & ~$5$~ \\
        \hline
        ~$z$~ & ~$2$~ & ~$3$~ & ~$6$~ & ~$\infty$~
    \end{tabular}\, ~~ ,
\end{equation}
with the corrections being relevant if $z$ is larger at fixed $d$ or $d$ is smaller at fixed $z$. This suggests two possible methods to obtain a perturbative NFL fixed point:
\begin{itemize}
    \item $d=2,z=2-\epsilon$ for small $\epsilon$ (Nayak-Wilczek).
    \item $d=3-\epsilon,z=3$ for small $\epsilon$ (dimensional regularization).
\end{itemize}
The former has the advantage of being technically simpler by virtue of having fewer angles to integrate over, while the latter has the advantage of having a local order parameter and a more traditional and familiar expansion, similar to the perturbative fixed point for the $O(N)$ model. We leave an explicit analysis of both these expansions to future work.


\newpage
\section{Spin and BCS extensions}\label{sec_extensions}

So far we have been exclusively working with spinless fermions and the charge 0 bosonic operators that can be constructed from them. Only a small class of fermion systems fall into this category so a natural extension would be to understand how to include internal symmetries as well as charged operators. The way this is achieved in traditional multidimensional bosonization is by writing a non-abelian patch fermion in terms of a bosonic vertex operator (see e.g., \cite{Houghton:2000bn}),
\begin{equation}
    \psi_i(\eta) \sim e^{i\phi_i(\eta)}\, ,
\end{equation}
where $i$ is an internal index, e.g., spin, and $\eta$ is a discrete label for the patches into which the Fermi surface is decomposed.

There are various issues with this construction. The most immediate objection one could have is the mismatch of operator statistics on both sides. A fermion operator cannot possibly  be written as a bosonic operator. In 1+1d this works in a subtle way since the Bose-Fermi duality is not strictly between the bosonic and fermionic theories, but rather the bosonic theory is dual to the fermionic one with a gauged $(-1)^F$ fermion parity symmetry. An intuitive way of thinking about this is that exchanging operators in 1+1d forces us to pass a coincidence singularity which allows for non-trivial transition functions to enter the exchange statistics of operators, unlike in higher dimensions.

The usual workaround for this is an `engineering' solution which multiplies the bosonic vertex operators by a `Klein factor' $O_\eta$ that obeys anticommutation relations and fixes the mismatch of exchange statistics on both sides. But this solution is unsatisfactory and unsystematic since its not clear whether these factors are supposed to be treated as dynamical quantities (to be integrated over in a path integral) or effectively as transition functions between different patches on the Fermi surface and if the physics of the bosonized theory is independent of the choice of Klein factors.

Secondly, the bosonization prescription ignores the non-abelian nature of the fermion, since the bosonic field $\phi_i$ transforms in the same representation of the internal symmetry as the patch fermion. This is evidently incorrect since nonabelian bosonization requires the addition of WZW terms in one higher dimension \cite{witten1984nonabelian_bos}, and the bosonized field lives in the square of the representation of the patch fermion.

We take an alternate approach to bosonizing Fermi surfaces of non-abelian fermions - one that relies on the algebra of fermion bilinears that can be constructed from the microscopic fermion bilinears.


\subsection{Spinful Fermi surfaces}\label{sec_spin}

Recall that our starting point for the postmodern formalism was the algebra of fermion bilinears. For spin-$1/2$ fermions, the same holds, but the generators of our algebra have additional indices.
\begin{equation}
    \begin{split}
        T_{\sigma\sigma'}(\x,\y) &\equiv \frac{i}{2} \left[ \psi_\sigma^\dagger\left(\x+\frac{\y}{2}\right) \psi_{\sigma'}\left(\x-\frac{\y}{2}\right) - \psi_{\sigma'}\left(\x-\frac{\y}{2}\right) \psi_\sigma^\dagger\left(\x+\frac{\y}{2}\right) \right]\, ,\\
        T_{\sigma\sigma'}(\q,\p) &\equiv \frac{i}{2} \left[ \psi_\sigma^\dagger\left(\frac{\q}{2}+\p\right) \psi_{\sigma'}\left(\frac{\q}{2}-\p\right) - \psi_{\sigma'}\left(\frac{\q}{2}-\p\right) \psi_\sigma^\dagger\left(\frac{\q}{2}+\p\right) \right]\, ,\\
        T_{\sigma\sigma'}(\x,\p) &\equiv \int_\y T_{\sigma\sigma'}(\x,\y) e^{i\p\cdot\y} = \int_\q T_{\sigma\sigma'}(\q,\p) e^{-i \q\cdot \x}\, ,\\
        T_{\sigma\sigma'}(\q,\y) &\equiv \int_{\x,\p} T_{\sigma\sigma'}(\x,\p) e^{i\q\cdot\x} e^{-i\p\cdot\y} = \int_\x T_{\sigma\sigma'}(\x,\y) e^{i\q\cdot\x} = \int_\p T_{\sigma\sigma'}(\q,\p) e^{-i\p\cdot\y}\, .
    \end{split}
\end{equation}
Ignoring the dependence on phase space coordinates, the generators live in the tensor product representation,
\begin{equation}
    \frac{1}{2} \otimes \frac{1}{2} = 0 \oplus 1\, ,
\end{equation}
of the fundamental (spin--1/2) representation of $SU(2)$, which decomposes into a direct sum of the scalar (singlet) and the adjoint (triplet). Therefore an alternate choice of basis for these generators is given by
\begin{equation}
    T^a(\x,\p) = \frac{i}{2} \int_\y \left[ \psi^\dagger\left( \x + \frac{\y}{2} \right)\cdot S^a\cdot \psi \left( \x - \frac{\y}{2} \right) - \text{h.c.} \right] e^{i\p\cdot\y}\, , \qquad a = 0,1,2,3\, ,
\end{equation}
where h.c. stands for hermitian conjugate and $S^0 = 1$ is the identity matrix and $S^i = \sigma^i/2$ are the generators of the Lie algebra $\su(2)$. The generators close under commutation and we have
\begin{equation}
    \begin{split}
        [T^a(\q,\y), T^b(\q',\y')] = 2 \left( i \cos \frac{\q'\cdot\y-\q\cdot\y'}{2} [S^a,S^b]^c + \sin \frac{\q'\cdot\y-\q\cdot\y'}{2} [S^a,S^b]_+^c \right)\\
        \times ~ T^c (\q+\q',\y+\y')\, ,
    \end{split}
\end{equation}
where $[S^a,S^b]^c$ and $[S^a,S^b]_+^c$ are respectively the components of the commutator and anticommutator of the spin generators expanded in the $S^c$ basis. This Lie algebra, which we refer to as the $\su(2)$\textit{-extended Moyal algebra} or the \textit{spin-Moyal algebra} is isomorphic, as a vector space, to the tensor product
\begin{equation}
    \g_\text{spin-Moyal} \cong \left( \mathbb{C} \oplus \su(2) \right) \otimes \g_\text{Moyal}\, ,
\end{equation}
where $\mathbb{C}$ is a one-dimensional complex vector space.

The semi-classical / Poisson limit is the same as before \eqref{eq_Poisson_limit}, and we find that this truncates to the following algebra:
\begin{equation}
    \begin{split}
        [T^0(\q,\y), T^0(\q',\y')] &= (\q'\cdot\y-\q\cdot\y') T^0(\q+\q',\y+\y')\, ,\\
        [T^0(\q,\y), T^i(\q',\y')] &= (\q'\cdot\y-\q\cdot\y') T^i(\q+\q',\y+\y')\, ,\\
        [T^i(\q,\y), T^j(\q',\y')] &= -f^{ijk} T^k(\q+\q',\y+\y')\, ,
    \end{split}
\end{equation}
where $f^{ijk}$ are the structure factors of $\su(2)$. One can check by explicit calculation that these truncated commutators do indeed obey the Jacobi identity. Convoluting with a $0\oplus1$-valued phase space function to define a general linear combination of these generators
\begin{equation}
    O_F \equiv \int_{\x,\p} F^a(\x,\p) T^a(\x,\p)\, ,
\end{equation}
we find that the commutator $[O_F, O_G]$ for two arbitrary $0\oplus1$-valued functions $F^a(\x,\p)$ and $G^a(\x,\p)$ is given by another operator $O_{[F,G]_\text{spin}}$ corresponding to the components
\begin{equation}\label{eq_spin-poisson_algebra}
    \begin{split}
        [F,G]^0_\text{spin} &= \{ F^0, G^0 \}\, ,\\
        [F,G]^k_\text{spin} &= \{ F^0, G^k \} + \{ F^k, G^0 \} - f^{ijk} F^j G^k\, ,
    \end{split}
\end{equation}
We will refer to this Lie algebra as the $\su(2)$\textit{-extended Poisson algebra} or the \textit{spin-Poisson algebra}:
\begin{equation}
    \g_\text{spin-Poisson} \cong \left(\mathbb{C}\oplus\su(2)\right)\otimes\g\, .
\end{equation}
This algebra also has an interpretation in terms of canonical transformations, in a single particle phase-space, except we now allow the action of (infinitesimal) canonical transformations on functions with spin indices to mix with transformations of the spin indices. For a function $O(\x,\p)$, with suppressed internal indices, that transforms under some representation $\rho$ of $\su(2)$, the infinitesimal transformation is given by
\begin{equation}
    \begin{split}
        \x \rightarrow \x' &= \x - \nabla_\p F^0\, ,\\
        \p \rightarrow \p' &= \p + \nabla_\x F^0\, ,\\
        O(\x,\p) \rightarrow O'(\x',\p') &= \left( 1_\rho + F^i \rho(S^i) \right) \cdot O(\x,\p)\, ,
    \end{split}
\end{equation}
where $1_\rho$ is the identity operator in the representation $\rho$. One can see that these transformations are generated by the phase space vector field valued in the representation $\rho$:
\begin{equation}
    X_F = \left( \nabla_\x F^0 \cdot \nabla_\p - \nabla_\p F^0 \cdot \nabla_\x \right) \cdot 1_\rho + F^i \rho(S^i)\, .
\end{equation}
Evaluating the commutator of two such vector fields acting on a spinful test function, we find the Lie bracket \eqref{eq_spin-poisson_algebra} of the spin-Poisson algebra. This perspective also makes it evident that the Lie bracket must obey Jacobi identity, without having to explicitly demonstrate it, since vector fields by definition obey it. The corresponding Lie group of canonical transformations augmented with spin, or \textit{spin-canonical transformations} will be labelled $\G_\text{spin}$.

The dual space $\g^*_\text{spin-Poisson}$ then consists of $(0\oplus1)$-valued distributions corresponding to the expectations values of the fermion bilinears in a given state $\sigma$,
\begin{equation}
    f^a(\x,\p) = \< T^a(\x,\p) \>_\sigma\, ,
\end{equation}
with the ground state distribution for a spherical Fermi surface given by
\begin{equation}
    f^0_\text{gs}(\p) = \Theta(p_F-|\p|)\, , \qquad f^i_\text{gs} = 0\, .
\end{equation}
The $0\oplus1$ decomposition of the fermion bilinears allows for a useful physical interpretation of the various components of the distribution, with $f^0$ being the charge fluctuation and $f^i$ being the spin fluctuation. At first glance, this formalism hence seems to be amenable to a description of spin-charge separation without the need for a parton construction, and perhaps might be able to answer questions about the energetic favourability of spin-charge separation states. We leave a study of this to future work.

The spin-Poisson bracket \eqref{eq_spin-poisson_algebra} points out an important scaling relation between the charge and spin fluctuations. For every term in the commutator $[F,G]^k_\text{spin}$ to scale homogeneously, we need
\begin{equation}\label{eq_spin_scaling}
    f^i \sim \nabla_\x \nabla_\p f^0 \sim \frac{q}{p_F} f^0\, ,
\end{equation}
so the spin fluctuations are suppressed compared to the total charge in the Poisson limit.

The coadjoint orbit action for spinful Fermi surfaces can be computed in the usual way, by first finding the stabilizer $\H_\text{spin}$, which consists of functions $\alpha^a(\x,\p)$ such that
\begin{equation}
    [\alpha,f_0]_\text{spin} = 0\, , \qquad \implies \qquad (\n\cdot\nabla_\x \alpha^a)_{|\p|=p_F} = 0\, .
\end{equation}
This allows us to parametrize the coadjoint orbit $\G_\text{spin}/\H_\text{spin}$ by the degree of freedom
\begin{equation}
    \phi^a(\x,\theta)\, ,
\end{equation}
with a typical state given by
\begin{equation}
    f_\phi = U f_0 U^{-1} = f_\text{gs} - [\phi,f_\text{gs}]_\text{spin} + \frac{1}{2!}[\phi,[\phi,f_\text{gs}]_\text{spin}]_\text{spin} + \ldots\, ~~ , \qquad U = \exp (-\phi)
\end{equation}
We find that the fluctuations of a spinful Fermi surfaces are characterized by twice as many degrees of freedom compared to multidimensional bosonization, which is a fact that is well known in the conventional Fermi liquid approach (see, e.g., \cite{vollhardt2013superfluid}).

The Gaussian part of the EFT can be evaluated in the usual way to find an expression very similar to the spinless case,
\begin{equation}
    S = - \frac{p_F^{d-1}}{2} \int_{t\x\theta} (\nabla_n\phi^a) \left( \dot{\phi}^a + v_F \nabla_n \phi^a + \int_{\theta'} F^{(2,0)}_{ab} (\theta,\theta') (\nabla_n\phi^b)' \right)\, ,
\end{equation}
where we allow interactions $F^{(2,0)}_{ab}$ that can break the $\su(2)$ symmetry. The cubic WZW term, however, has an important difference in the last term:
\begin{equation}
    \begin{split}
        S_\text{WZW}^{(3)} = - \frac{p_F^{d-2}}{3!} \int_{t\x\theta} &\nabla_n \phi^0 \left( \nabla_s\dot{\phi}^0 \d_\theta \phi^0 - \nabla_s\phi^0 \d_\theta\dot{\phi}^0 \right)\\
        &+\nabla_n \phi^i \left( \nabla_s\dot{\phi}^0 \d_\theta \phi^i - \nabla_s\phi^i \d_\theta\dot{\phi}^0 + \nabla_s\dot{\phi}^i \d_\theta \phi^0 - \nabla_s\phi^0 \d_\theta\dot{\phi}^i \right)\\
        &- f^{ijk} (\nabla_n \phi^i) \dot{\phi}^j \phi^k\, .
    \end{split}
\end{equation}
The expansion of the charge density $\rho^0$ in terms of $\phi^0$ remains identical to the spinless case, but the spin density picks up new types of terms:
\begin{equation}
    \rho^i = \frac{p_F^{d-1}}{(2\pi)^d} \int_\theta \nabla_n \phi^i + \frac{1}{2p_F} \nabla_s \left( \d_\theta \phi^0 \nabla_n \phi^i + \d_\theta \phi^i \nabla_n \phi^0 \right) + f^{ijk} \phi^j \nabla_n \phi^k\, .
\end{equation}

The scaling scheme is determined by requiring the quadratic part of the action to be exactly marginal, implying that we need to scale $\phi^i\sim\phi^0$. This causes the $f^{ijk}$ terms in the WZW piece as well as the spin density to scale with an additional factor of $q^{-1}$ compared to the others, making the spin density correlators scale differently compared to their charge density analogues (see section VI B of \cite{main:2022}).


\subsection{Charged fermion bilinears}

While we do not necessarily have access to the patch fermion operator in the postmodern formalism, we can consider charged bilinears of the form,
\begin{equation}
    \begin{split}
        T^{(2)}(\x,\y) &= - T^{(2)}(\x,-\y) \equiv i \psi^\dagger \left( \x + \frac{\y}{2} \right) \psi^\dagger \left( \x - \frac{\y}{2} \right)\, ,\\
        T^{(-2)}(\x,\y) &= -T^{(-2)}(\x,-\y) \equiv i \psi \left( \x + \frac{\y}{2} \right) \psi \left( \x - \frac{\y}{2} \right)\, ,\\
        T^{(0)}(\x,\y) &\equiv \frac{i}{2} \left[ \psi^\dagger\left(\x+\frac{\y}{2}\right) \psi\left(\x-\frac{\y}{2}\right) - \psi\left(\x-\frac{\y}{2}\right) \psi^\dagger\left(\x+\frac{\y}{2}\right) \right]\, ,
    \end{split}
\end{equation}
where the number in the parenthesis in the superscript denotes their charge under the particle number conserving $U(1)$ symmetry. We have defined these operators so that
\begin{equation}
    T^{(q)}(\x,\y)^\dagger = - T^{(-q)}(\x,-\y)\, .
\end{equation}
The various Fourier transforms of these are defined in the usual way, but we write down the momentum space versions here for later use:
\begin{equation}
    \begin{split}
        T^{(2)}(\q,\p) &\equiv i \psi^\dagger \left( \frac{\q}{2} + \p \right) \psi^\dagger \left( \frac{\q}{2} - \p \right) = - T^{(2)}(\q,-\p)\, ,\\
        T^{(-2)}(\q,\p) &\equiv i \psi \left( \frac{\q}{2} + \p \right) \psi \left( \frac{\q}{2} - \p \right) = -T^{(-2)}(\q,-\p)\, ,\\
        T^{(0)}(\q,\p) &\equiv \frac{i}{2} \left[ \psi^\dagger\left(\frac{\q}{2}+\p\right) \psi\left(\frac{\q}{2}-\p\right) - \psi\left(\frac{\q}{2}-\p\right) \psi^\dagger\left(\frac{\q}{2}+\p\right) \right]\, .
    \end{split}
\end{equation}
It turns out that these also close under commutation, and we find the following Lie algebra:
\begin{equation}\label{eq_bcs_moyal_algebra}
    \begin{split}
        [T^{(0)}(\q,\y),T^{(0)}(\q',\y')] &= 2 \sin \left( \frac{\q'\cdot\y-\q\cdot\y'}{2} \right) T^{(0)}(\q+\q',\y+\y')\, ,\\
    [T^{(0)}(\q,\y),T^{(\pm 2)}(\q',\y')] &= i e^{\frac{i}{2}(\q'\cdot\y - \q\cdot\y')} T^{(\pm2)}(\q+\q',\y+\y')\\
    & \qquad \pm i e^{-\frac{i}{2}(\q'\cdot\y - \q\cdot\y')} T^{(\pm2)}(\q+\q',\y-\y')\, ,\\
    [T^{(2)}(\q,\y),T^{(-2)}(\q',\y')] &= 2 \sin \left( \frac{\q'\cdot\y-\q\cdot\y'}{2} \right) T^{(0)}(\q+\q',\y+\y')\\
    &\qquad - 2 \sin \left( \frac{\q'\cdot\y+\q\cdot\y'}{2} \right) T^{(0)}(\q+\q',\y-\y')\, .
    \end{split}
\end{equation}
What remains is to find an appropriate semi-classical truncation of this algebra in order to apply the coadjoint orbit method and obtain an action that would involve three distributions $f^{(q)}(\x,\p)$ as the degrees of freedom, corresponding to the usual occupation number distribution for $q=0$, as well as charged distributions for $q\pm 2$ whose values encode the BCS gap function.

Taking a semiclassical limit in this case is a bit more involved, since the right hand side of the commutators includes operators not only evaluated at $\y+\y'$, but also at $\y-\y'$. From an intuitive perspective, the Poisson limit for the charge 0 section \eqref{eq_Poisson_limit}
\begin{equation}
    |\q| \ll |\p| \sim p_F
\end{equation}
still seems to give the most relevant configurations of both particle-hole pairs as well as Cooper pairs, since in this limit the particle and hole nearly coincide at the Fermi surface, while the particles in, say, $T^{(2)}$ become nearly antipodal\footnote{Recall that in our convention $\psi(\k)$ creates a hole at $-\k$, while $\psi^\dagger(\k)$ creates a particle at $+\k$.} (see figure \ref{fig_bcs_scat}). However, the appropriate truncation of the algebra \eqref{eq_bcs_moyal_algebra} in this limit that also obeys the Jacobi identity needs to be found to ensure that this intuition works quantitatively.

\newpage
\section{Conclusion and Outlook}\label{sec_conclusion}

To summarize, we presented in this dissertation a formalism for the study of Fermi surface physics that is built out of a robust geometric structure underlying a large subalgebra of operators that governs low energy physics in the presence of a Fermi surface. This formalism provides an algorithm to obtain an effective field theory description for Fermi liquids given the internal symmetries of the microscopic fermions. The effective field theory has a rigid structure determined by the geometry of the Fermi surface and a collection of Wilson coefficient functions that parametrize fermion interactions.

Unlike previous approaches, this postmodern formalism systematizes the expansion of low energy properties in a way that makes the scaling behaviour of these properties transparent. The amenability of this EFT to simple power counting arguments exemplifies its usefulness, on top of making diagrammatic calculations simpler compared to earlier approaches. Not only that, the geometric nature of this formalism allows us to identify emergent symmetries in a straightforward manner as well, entirely by analyzing the Ward identity for canonical transformations.

The postmodern formalism opens up many different avenues of exploration, the primary one being a systematic study of non-Fermi liquids. The fact that power counting arguments work even for the non-Fermi liquid theory presented above in section \ref{sec_nfl} is promising, but the theory needs to be studied more carefully to obtain quantitative results.

The ability to include charged fermion bilinears in the algebra opens up the possibility of combining Fermi liquids and conventional superconductors into a parent effective field theory, which could serve as a useful theoretical platform to analyze the competition between non-Fermi liquid and superconducting instabilities of a Fermi liquid, as well as provide a path towards understanding the mechanisms underlying high temperature superconductivity, e.g., in cuprates.

Other, slightly less ambitious, directions include a study of the non-perturbative properties of the postmodern formalism, for instance through an analysis of the topological properties of the coadjoint orbit $\mathcal{O}_{f_0}$ which were largely ignored in the present construction since we were looking for a perturbative expansion around the ground state $f_0$ in the coadjoint orbit. The nonlinear Ward identity might also serve as a powerful non-perturbative constraint, especially if it holds in more general systems beyond Fermi liquids.

One rather curious aspect of the postmodern formalism is that despite the presence of UV/IR mixing, it seems possible to analyze the scaling behaviour of various physical quantities such as the specific heat, since our EFT seemingly comes with a preferred choice of UV cutoff. A more careful exploration of such UV divergences needs to be undertaken to see whether we can obtain quantitative results through a prescription for the cutoff or a resummation of the Moyal expansion. We hope that such an analysis will also shed light on the question of how to deal with UV/IR mixing in other effective theories.

One lesson to take away from this formalism and its ubiquity across other phases of matter is that diffeomorphism groups have an untapped potential to constrain the emergent physics of many-body systems. We hope that this work will serve as a stepping stone towards exploiting this potential further.


\newpage
\appendix

\section{Coadjoint orbit method --- mathematical details}

The coadjoint orbit method \cite{Kirillov_book} is, in principle, a method used to quantize a Lie group, i.e., find irreducible (linear and/or projective) representations of the group. The notion of quantizing a classical dynamical theory is closely tied to finding irreducible representations of its symmetries, as is made evident by the example of a single spin. Classically, a single spin is just some vector of arbitrary length in 3 dimensions whose dynamics are governed by the rotation group $SO(3)$. It is only when we quantize the classical dynamics that we find that the magnitude of the spin must be $\sqrt{l(l+1)}\hbar$ where $l$ is a half-integer or an integer. When $l$ is an integer, we find linear representations of $SO(3)$, while for $l$ a half-integer, the representation is a projective representation of $SO(3)$ which is equivalently a linear representation of $SU(2)$. Of course this distinction between the two groups only occurs once we take into consideration the global topological structure of the Lie groups, since both have identical Lie algebras.

One approach to the coadjoint orbit method is hence to set up a dynamical system that evolves under the action of the Lie group, and then quantize it \cite{Wiegmann:1989hn,Alekseev88coadj}. Quantizing the dynamical system will result in some consistency constraints which will label the irreducible representations of the Lie group.

For the purpose of this draft, we are only interested in the first step: setting up a dynamical system that evolves under the action of canonical transformations. While established methods of quantizing this dynamical system describing semi-classical Fermi liquids should in principle apply, in practice they are rather difficult to implement due to the fact that the Lie group of canonical transformations is an infinite dimensional diffeomorphism group. Therefore, we resort to a more `lowbrow' approach to quantizing the theory like one would any other quantum field theory.

For most of this section, we will keep the discussion rather general, and provide intuition for the results we obtain using the example of a single spin (or equivalently a rigid body in the center of mass frame).

Consider a Lie group $\G$, whose typical element will be represented by the letter $g$. The identity element of the Lie group will be represented as $e$. It is Lie algebra $\g$ consists of left-invariant vector fields $X$ on the Lie group, with the commutator of these vector fields (viewed as differential operators acting on test functions of the Lie group) determines the Lie bracket, which will be denoted by $[~,~]$. Alternately, one can think of the Lie algebra as the tangent space to the Lie group at unity, with the Lie bracket prescribed externally.

For any Lie group we can define an exponent map and its inverse, the logarithm,
\begin{equation}
    \exp: \g \rightarrow \G_e, \qquad \log :\G_e \rightarrow \g\, ,
\end{equation}
which map the Lie algebra to and from the largest possible simply connected patch $\G_e$ of the Lie group that includes the identity. In general, the exponent map is not globally defined, i.e., it is not always possible to take the logarithm of a general Lie group element. $SO(3)$ provides an example of this, since the logarithm of a $\pi$-rotation around any axis does not exist within the Lie algebra (unless the Lie algebra is complexified). Therefore, if we insist upon parametrizing elements of a Lie group as exponents of the elements of its Lie algebra, like we do for the case of canonical transformations, we necessarily lose information about the topological structure of the Lie group.

For $SO(3)$, a Lie group element is a $3\times 3$ orthogonal matrix $O^T O = 1_3 = OO^T$. A Lie algebra element is an antisymmetric $3\times3$ matrix with real components $M^T=-M$, and the exponent map is the literal exponent of the matrix. The group and algebra are both 3 dimensional and a general Lie algebra element can be written as a 3 dimensional vector $\vec{\Omega}$ with real components. Given the usual generators $L_1, L_2, L_3$ of $\so(3)$, the matrix that the vector $\vec{\Omega}$ corresponds to is simply
\begin{equation}
    M_{\vec{\Omega}} = \sum_i \Omega^i L_i\, .
\end{equation}
If we are using $SO(3)$ to describe the configuration space of a single spin or a rigid body, an element of the Lie algebra $\vec{\Omega}$ can be interpreted as an angular velocity. The Lie bracket of two antisymmetric matrices is just the matrix commutator and takes the following form:
\begin{equation}
    [M_{\vec{\Omega}}, M_{\vec{\Omega}'}] = M_{\vec{\Omega}\times\vec{\Omega}'}\, ,
\end{equation}
where $\vec{\Omega}\times\vec{\Omega}'$ is the cross product of the two angular velocities.

Next, for the Lie algebra, we can define its dual space $\g^*$, i.e., the space of linear functions acting on the Lie algebra. A typical element of the dual space will be labelled by lowercase Greek letters:
\begin{equation}
    \begin{split}
        \eta: ~ &\g \rightarrow \R\, ,\\
        \eta[X] &\equiv \langle \eta, X \rangle\, .
    \end{split}
\end{equation}
The angular brackets are standard notation for the action of a dual space element on the Lie algebra element. The dual to any finite dimensional vector space is isomorphic to the vector space itself, but we will maintain the distinction between the two. For $SO(3)$, the dual space also consists of 3 dimensional real vectors $\vec{l}$, which act on Lie algebra elements $\vec{\Omega}$ via the dot product:
\begin{equation}
    \vec{l}[\vec{\Omega}] \equiv \< \vec{l}, \vec{\Omega} \> \equiv \vec{l}\cdot\vec{\Omega}\, .
\end{equation}
Elements of the dual to $\so(3)$ are interpreted as angular momenta of the rigid body, or the orientation of the spin itself. These characterize the state of the spin or the rigid body.

The adjoint action of the Lie algebra on itself is given by the Lie bracket:
\begin{equation}
    \ad_X Y = [X,Y]\, .
\end{equation}
This induces a coadjoint action of the Lie algebra on its dual space, determined uniquely by the requirement
\begin{equation}
    \< \ad^*_X \eta, Y \> = \< \eta, -\ad_X Y \>\, .
\end{equation}
We will avoid rigorous definitions of the Lie group adjoint and coadjoint actions $\Ad_g$ and $\Ad^*_g$ on $\g$ and $\g^*$ respectively since these definitions are somewhat involved, but it suffices to know that these action exist and generalize the definitions via the exponentials of $\ad$ and $\ad^*$ to group elements that cannot be written as exponentials of Lie algebra elements. If the Lie group and its Lie algebra consist of matrices, then the group adjoint and coadjoint actions are just matrix conjugation $gXg^{-1}$ and $g\eta g^{-1}$.

Returning to $SO(3)$, the Lie algebra adjoint and coadjoint actions are given by cross products of vectors,
\begin{equation}
    \ad_{\vec{\Omega}} \vec{\Omega}' = \vec{\Omega} \times \vec{\Omega}'\, , \qquad \ad^*_{\vec{\Omega}} \vec{l} = \vec{\Omega} \times \vec{l}\, ,
\end{equation}
while the group adjoint and coadjoint actions reduce to rotation of 3d vectors:
\begin{equation}
    \Ad_O \vec{\Omega} = O\cdot\vec{\Omega}\, , \qquad \Ad^*_O \vec{l} = O\cdot\vec{l}\, .
\end{equation}

Now, since $\G$ is the configuration space of our system, the phase space is given by the cotangent bundle (which can be shown to be a trivial direct product for any Lie group),
\begin{equation}
    T^*\G \cong \G \times \g^*\, .
\end{equation}
Roughly speaking, $\G$ itself is also a symmetry of our dynamical system, so we can quotient it out to obtain a reduced phase space,
\begin{equation}
    T^*\G/\G \cong \g^*\, ,
\end{equation}
that is isomorphic to the dual space\footnote{More precisely, this is achieved by defining a momentum map $\mu: T^*\G\rightarrow \g^*$ such that the pre-image $\mu^{-1}(0\in\g^*)$ of this map gives us the reduced phase space.}. This is just to say that the configuration of a rigid body in the center of mass frame is effectively determined by its total angular momentum, for the purpose of time evolution.

With $\g^*$ as the reduced phase space for the dynamics of our system, we need only a Poisson structure and a choice of Hamiltonian to obtain Hamilton's equations of motion. The Poisson structure is given by the Lie-Poisson bracket of two functionals $\sF[\eta]$ and $\sG[\eta]$ of $\g^*$:
\begin{equation}
    \{ \sF, \sG \}_\text{LP}[\eta] \equiv \left\< \eta, [ d_\eta \sF, d_\eta \sG ] \right\>\, .
\end{equation}
This definition can be understood as follows: the differentials $d_\eta\sF$ and $d_\eta\sG$ at the point $\eta\in\g^*$ live in the cotangent space to $\g^*$ at the point $\eta$. Since $\g^*$ is a vector space, its cotangent spaces are isomorphic to its dual $\g^{**}$, which is just the Lie algebra $\g$. Since the differentials can be treated as Lie algebra elements, we can take their Lie bracket to obtain a new Lie algebra element. The pairing of $\eta$ with this Lie algebra element defines the value of the functional $\{\sF,\sG\}_\text{LP}$ at the point $\eta$. This can be done for every point $\eta$ to define the Lie-Poisson bracket. It is instructive to write this formula in terms of the structure constants $f^{abc}$ of the Lie group:
\begin{equation}
    \{ \sF, \sG \}[\eta] = \eta_c f^{abc} \d_a \sF \d_b \sG = \Pi^{ab}(\eta) \d_a \sF \d_b \sG\, ,
\end{equation}
where $\d_a$ are derivatives on $\g^*$ in the basis of generators, and $\Pi^{ab}(\eta) = f^{abc}\eta_c$ is the Poisson-bivector. 

For $SO(3)$ recall again that the Lie bracket is the cross product of vectors and the pairing is given by the dot product, so the the Lie-Poisson bracket of two functions of the angular momentum takes the form:
\begin{equation}
    \{ \sF, \sG \}_\text{LP}[\vec{l}] \equiv \vec{l} \cdot \left( \frac{\d \sF}{\d\vec{l}} \times \frac{\d \sG}{\d\vec{l}} \right)\, .
\end{equation}

The choice of Hamiltonian is determined by the dynamical system under consideration. For a rigid body, the natural choice of Hamiltonian is the total rotational energy defined in terms of the inverse of the moment of inertia tensor, with an additional torque term,
\begin{equation}
    H[\vec{l}] \equiv \frac{1}{2} \left( \vec{l}\cdot I^{-1} \cdot \vec{l} \right) - \vec{\tau} \cdot \vec{l}\, .
\end{equation}
The equation of motion is then given by
\begin{equation}
    \dot{\vec{l}} = \{ \vec{l}, H \}_\text{LP}[\vec{l}] = - \left(I^{-1}\cdot\vec{l}\right)\times\vec{l} + \vec{\tau}\, .
\end{equation}
The moment of inertia tensor defines a map from $\g$ to $\g^*$ and vice versa, so that $I^{-1}\cdot\vec{l}=\vec{\Omega}$, the angular velocity, and the equation of motion takes the more familiar form of Euler's equations for a rigid body:
\begin{equation}
    \dot{\vec{l}} ~ + ~ \vec{\Omega}\times\vec{l} = \vec{\tau}\, .
\end{equation}
For a single spin, the Hamiltonian would not have a quadratic term, but there could be an external magnetic field providing the torque, so the equation of motion is identical, except without the moment of inertia term.

The final step is to turn this Hamiltonian into an action, which requires a symplectic form on the reduced phase space, obtained by inverting the Lie-Poisson bivector. However, $\g^*$ does not host a symplectic form, since the Lie-Poisson bivector $\Pi^{ab} = f^{abc}\eta_c$ is not invertible, since $\eta_c$ can be zero! $SO(3)$ once again provides some intuition for this: the dual space for this Lie group is a 3 dimensional vector space. Symplectic forms can only exist on even dimensional manifolds. Therefore it is impossible to define one on the dual space.

However, given that time evolution on $\g^*$ for any choice of Hamiltonian occurs through the action of a one-parameter family of group elements, the space of states in $\g^*$ that are reachable from one another is smaller than $\g^*$. Such a space is called a coadjoint orbit. It is defined as an equivalence class of states $\eta\in\g^*$ such that any two such states are related by the coadjoint action of some group element.

We will avoid the proof here, but it is possible to show that the Lie-Poisson bivector does become invertible when restricted to functions of the coadjoint orbit. The symplectic form hence obtained on a given coadjoint orbit is known as the Kirillov-Kostant-Souriau (KKS) form, and is defined by its action on two vectors $\rho,\sigma$ tangent to a point $\nu$ in the coadjoint orbit in $\g^*$, which can be thought of as elements of $\g^*$,
\begin{equation}\label{eq_kks_general}
    \omega_\text{KKS}(\rho,\sigma)|_\eta \equiv \left\< \eta, [X,Y] \right\>\, ,
\end{equation}
where $X$ and $Y$ are Lie algebra elements such that
\begin{equation}
    \ad^*_X \eta = \rho\, , \qquad \ad^*_Y \eta = \sigma\, .
\end{equation}
$X$ and $Y$ are not uniquely determined by this condition, but it is possible to show that the expression on the right hand side is independent of this ambiguity. The action that reproduces the same equation of motion as the Hamiltonian $H[\eta]$ is then given by
\begin{equation}
    S = \int_0^1 ds \int dt ~ \omega_\text{KKS}(\d_t \eta, \d_s \eta) - \int dt ~ H[\eta]\, ,
\end{equation}
where $s$ is an extra dimension with $s=1$ corresponding to physical time and boundary conditions $\eta(s=0) = 0$.

Consider once again the case of $SO(3)$ whose coadjoint action on $\g^*$ is simply the rotation of an angular momentum vector. Evidently, coadjoint orbits are spheres of fixed radius $|\vec{l}|$, so that the Poisson bivector,
\begin{equation}
    \Pi^{ij}(\vec{l}) = \epsilon^{ijk} l_k\, ,
\end{equation}
becomes invertible on such a sphere, with the inverse given by
\begin{equation}
    (\omega_\text{KKS})_{ij} = \frac{l^k}{l^2} \epsilon_{ijk}\, .
\end{equation}
This is just the rescaled area form on the sphere, which is closed but not exact. While for this case we were able to find an explicit expression for $\omega_\text{KKS}$, this will not necessarily happen in general, and we have to resort to the definition \eqref{eq_kks_general}. The action for a rigid body or a spin is then given by
\begin{equation}
    S = \frac{1}{l^2} \int_0^1 ds \int dt ~ \vec{l} \cdot \left( \d_t \vec{l} \times \d_s \vec{l} \right) - \int dt ~ H[\vec{l}]\, .
\end{equation}
The term obtained by integrating the Kirillov is the familiar WZW term for a spin or a rigid body, and making $\vec{l}$ a local function of space turns it into the Berry phase term for the effective field theory of a ferromagnet.

It is worth pointing out that since the KKS form is not exact, the extra dimension cannot be integrated over unless we work in a perturbative expansion around some fixed ground state angular momentum $\vec{l}_0$ (every closed form is locally exact). However, had we parametrized the coadjoint orbit as the action of exponentiated infinitesimal rotations acting on $\vec{l}_0$ to begin with, we would have found the KKS form to be exact and the WZW term to be a total $s$-derivative. This is what happens in the case of Fermi liquids, and we leave an exploration of the topological structure of the coadjoint orbit to future work.


\newpage
\section{Luttinger liquids from the coadjoint orbit method}

In this section we show that the coadjoint orbit formalism reproduces the bosonized theory of Luttinger liquids. In particular, the mixed anomaly between the emergent chiral $U(1)$ symmetries at the Fermi points can be understood as a linearization of the Ward identity for canonical transformations. Luttinger liquids have been extensively studied in the literature, see in particular Refs.~\cite{Stone:1989,Das:1991uta,Dhar:1992rs,Dhar:1993jc,Khveshchenko:1993ug} for constructions using coadjoint orbits.

We begin with a review of the construction of the bosonized action for Luttinger liquids from the algebra of densities. Fermi `surfaces' in 1+1 dimensions are a collection of discrete points in momentum space. Assuming that the dispersion relation $\epsilon(p)$ is an even function that monotonically increases with positive momentum, the Fermi surface consists of exactly two points at momentum values $p=\pm p_F$. Each Fermi point hosts a chiral mode whose chirality is given by $\text{sgn}[\d_p \epsilon]$. Denoting the chiral modes at the points $+p_F$ and $-p_F$ by the subscripts $R$ and $L$ (for `right' and `left') respectively, the particle number densities obey the following equal time commutation relations
\begin{equation}\label{eq_alg_lin_1d}
    \begin{split}
        [\rho_R(x), \rho_R(x')] &= -\frac{i}{2\pi} \d_x \delta(x-x')\, ,\\
        [\rho_L(x), \rho_L(x')] &= \frac{i}{2\pi} \d_x \delta(x-x')\, ,\\
        [\rho_R(x), \rho_L(x')] &= 0\, .
    \end{split}
\end{equation}
The so-called Schwinger terms on the right-hand side of the first two lines are indicative of the chiral anomalies carried by each chiral fermion. $\rho_{R,L}$ are the charge densities corresponding to two copies of $U(1)$ symmetry, which we will refer to as $U(1)_R$ and $U(1)_L$. The chiral algebra can be realized in terms of bosonic fields $\phi_{R,L}$ by defining the densities as
\begin{equation}\label{eq_Luttinger_rhotophi}
    \rho_R = \frac{1}{2\pi} \d_x \phi_R\, , \qquad \rho_L = -\frac{1}{2\pi} \d_x \phi_L\, .
\end{equation}
The commutators of the densities with the bosonic fields are then
\begin{equation}\label{eq_Luttinger_phirho}
    \begin{split}
        [\phi_R(x), \rho_R(x')] &= -i \delta(x-x')\, ,\\
        [\phi_L(x), \rho_L(x')] &= -i \delta(x-x')\, ,
    \end{split}
\end{equation}
which tells us that the $U(1)_{R.L}$ symmetries are non-linearly realized on the bosonic fields as
\begin{equation}
    \phi_R \rightarrow \phi_R - \lambda_R\, , \qquad \phi_L \rightarrow \phi_L - \lambda_L\, .
\end{equation}
An action that produces the algebra \eqref{eq_Luttinger_phirho} is
\begin{equation}\label{eq_Luttinger_WZW}
\begin{split}
    S 
    &= \frac12 \int dt dx \, \dot \phi_R\rho_R + \dot \phi_L \rho_L \\
    &= -\frac{1}{4\pi} \int dt dx \, \d_x \phi_R \dot \phi_R  - \d_x \phi_L \dot \phi_L \, .
\end{split}
\end{equation}
The factor of $\frac12$ in the first line comes from the fact this is a constrained system: using the appropriate Dirac brackets one recovers the commutation relation \eqref{eq_Luttinger_phirho} as desired. 

This action corresponds to the WZW term in the coadjoint orbit construction. The integral over the Fermi surface angle $\theta$ becomes a sum over two points $\theta = 0,\, \pi$, so that one finds
\begin{equation}
\begin{split}
S_{\rm WZW}
    &=
    -\frac{1}{4\pi}\sum_{\sigma=\pm}\sigma\int dt dx \, \d_x 
    \phi_\sigma  \dot \phi_\sigma \\ 
    &= -\frac{1}{4\pi} \int dt dx \, \d_x \phi_R \dot \phi_R  - \d_x \phi_L \dot \phi_L \, ,
\end{split}
\end{equation}
in agreement with \eqref{eq_Luttinger_WZW}. Nonlinearities in the WZW term, present for any $d>1$, entirely vanish in $d=1$. These nonlinearities are associated with the curvature of the Fermi surface, which explains why they are absent in one dimension. For the same reason, the relation between $\rho$ and $\phi$ \eqref{eq_Luttinger_rhotophi} does not receive nonlinear corrections.

In $d=1$, all nonlinearities in the bosonized description of a Luttinger liquid come from the Hamiltonian, in particular from nonlinearities in the dispersion relation. The Hamiltonian part of the action also produces a term in the quadratic action,
\begin{equation}\label{eq_ungauged_luttinger}
\begin{split}
S^{(2)}
    &=
   -\frac{1}{4\pi}\sum_{\sigma=\pm}\int dt dx \, \d_x 
    \phi_\sigma \left(\sigma \dot \phi + v_F \d_x \phi\right) \\ 
    &= -\frac{1}{4\pi} \int \d_x \phi_R \left( \d_0 \phi_R + v_F \d_x \phi_R \right) - \d_x \phi_L \left( \d_0 \phi_L - v_F \d_x \phi_L \right) \, ,
\end{split}
\end{equation}
which is the well-known Gaussian action for a Luttinger liquid.

\subsection{Chiral anomaly as a linear approximation}\label{sapp_chiralanomaly}

When coupled to background gauge fields, both chiral symmetries are anomalous with opposite anomalies. If $A_\mu^R$ and $A_\mu^L$ are the background fields for the two global symmetries, the anomalous conservation laws are
\begin{equation}
    \begin{split}
        \d_\mu j^\mu_R &= -\frac{1}{4\pi} \epsilon^{\mu\nu}F^R_{\mu\nu}\, ,\\
        \d_\mu j^\mu_L &= \frac{1}{4\pi} \epsilon^{\mu\nu}F^L_{\mu\nu}\, .
    \end{split}
\end{equation}

In the coadjoint orbit formalism, the chiral anomalies appear as a linearized approximation to the invariance of the maximally gauged action \eqref{eq_maxgauge_action} under all canonical transformations. To see this, we begin with the Ward identity for free fermions, that have $\cJ_{p^j} = 0$
\begin{equation}
    \d_\mu \cJ^\mu + \{ \cJ^\mu, A_\mu \} = 0\, . 
\end{equation}
Turning off $A_x$ for simplicity, the conservation law takes the form
\begin{equation}
    \d_0 \cJ^0 + \d_x \cJ^x + \d_x \cJ^0 \d_p A_0 = \d_p \cJ^0 \d_x A_0\, .
\end{equation}
Recall that $\cJ^0$ is simply the phase space distribution $f$. Hence, it has a nonzero expectation value in the ground state
\begin{equation}
    \langle \cJ^0 \rangle = f_0\, .
\end{equation}
If we now linearize the equation around the two Fermi points by writing
\begin{equation}
    \cJ^0 = f_0 + \delta \cJ^0, \qquad \cJ^x = \delta \cJ^x\, ,
\end{equation}
and treat $A_0(t,x,p)$ to be of the same order as $\delta \cJ^\mu$, we find that the equation takes the form
\begin{equation}
    \d_0 \delta \cJ^0 + \d_x \delta \cJ^x = (\d_x A_0^L) \delta(p+p_F) - (\d_x A_0^R) \delta(p-p_F)\, .
\end{equation}
Integrating over either $p>0$ or $p<0$ and using the expressions for the chiral density and current
\begin{equation}
    \begin{split}
        \rho_R = \int_0^\infty \frac{dp}{2\pi} ~ \delta \cJ^0, \qquad j_R &= \int_0^\infty \frac{dp}{2\pi} ~ \delta \cJ^x\, ,\\
        \rho_L = \int_{-\infty}^0 \frac{dp}{2\pi} ~ \delta \cJ^0, \qquad j_L &= \int_{-\infty}^0 \frac{dp}{2\pi} ~ \delta \cJ^x\, ,
    \end{split}
\end{equation}
we find that the Ward identity takes the form of the anomalous conservation laws for the chiral anomalies
\begin{equation}
    \begin{split}
        \d_t \rho_R + \d_x j_R &= - \frac{1}{2\pi}\d_x A_0^R\, ,\\
        \d_t \rho_L + \d_x j_L &= \frac{1}{2\pi}\d_x A_0^L\, .
    \end{split}
\end{equation}
The chiral anomaly is therefore a linear approximation to the non-abelian Ward identity, or a covariant conservation law, around a state with nonzero charge density $\langle \cJ^0 \rangle \ne 0$.
\newpage

\bibliography{postmodern.bib}{}

\end{document}